\documentclass[12pt]{article}
\pdfoutput=1

\usepackage{jheppub}
\usepackage{amsmath,amssymb,euscript,array,mathrsfs}
\usepackage{slashed}

\usepackage{epsfig}

\def\a{\alpha}
\def\b{\beta}
\def\c{\gamma}
\def\d{\delta}
\def\e{\epsilon}

\def\l{\lambda}
\def\m{\mu}
\def\n{\nu}

\def\s{\sigma}

\def\w{\omega}

\def\D{\Delta}

\def\tr{{\rm tr}}

\def\Dbarslash{\,\,{\raise.15ex\hbox{/}\mkern-12mu {\bar D}}}
\def\Dslash{\,\,{\raise.15ex\hbox{/}\mkern-12mu D}}
\def\delslash{\,\,{\raise.15ex\hbox{/}\mkern-9mu \partial}}
\def\delbarslash{\,\,{\raise.15ex\hbox{/}\mkern-9mu {\bar\partial}}}

\def\thalf{\tfrac{1}{2}}

\def\rta{\rightarrow}
\def\tr{{\rm tr}}

\def\Re{{\rm Re}}
\def\MOL{{\textsf{MOL }}}
\def\EXP{{\textsf{EXP }}}
\def\HEL{{\textsf{HEL }}}
\def\TT{{\cal T}}
\def\VV{{\cal V}}
\def\NR{{\rm NR}}
\newcommand{\wh}{\widehat}

\title{\begin{center}{\Large  Lorentz and \textsf{CPT} violation and the
    hydrogen and antihydrogen molecular ions II -- hyperfine-Zeeman spectrum}
\end{center}\vskip0.5cm}

\author{\vskip1cm
\large{Graham M.~Shore} }

\emailAdd{g.m.shore@swansea.ac.uk  }

\affiliation{\vskip0.8cm 
Department of Physics, Faculty of Science and Engineering, Swansea University, 
Singleton Park, Swansea, SA2 8PP, UK
}

\date{\today}

\abstract{Fundamental principles of quantum field theory such as Lorentz invariance,
\textsf{CPT} symmetry and locality may be tested to extremely high precision in
atomic and molecular spectroscopy. The narrow natural linewidth of rovibrational 
states in the hydrogen molecular ion ${\rm H}_2^{\,+}$ and its antimatter counterpart 
$\overline{\rm H}_2^{\,-}$ make these ideal candidates, and give $O(m_p/m_e)$ increased 
sensitivity to Lorentz and \textsf{CPT} violation in the proton sector compared to 
${\rm H}$ and $\overline{\rm H}$ atoms. In a previous paper, we presented a detailed 
analysis of the rovibrational spectrum of ${\rm H}_2^{\,+}$ and $\overline{\rm H}_2^{\,-}$ 
in an effective QFT encoding Lorentz and \textsf{CPT} violation, focusing on 
spin-independent effects. Here, we extend this analysis to include the full hyperfine-Zeeman
spectrum and include spin-dependent Lorentz and \textsf{CPT} violating operators in the 
effective theory. The results demonstrate how constraints on these symmetry-violating
couplings may be extracted from specific rovibrational transitions between
hyperfine-Zeeman states in the presence of an applied magnetic field.

\vskip3cm}

\notoc

\begin{document}

\maketitle

\setlength{\parskip}{10pt}

\newpage

\section{Introduction}\label{sect 1}

The extremely narrow natural linewidth of rovibrational states in the hydrogen molecular ion 
${\rm H}_2^{\,+}$ \cite{SchillerCP}, together with recent theoretical and experimental developments 
\cite{SAS2024,SchillerKarr2024}, opens up the possibility 
of testing fundamental physics principles such as Lorentz and \textsf{CPT} symmetry
with high-precision rovibrational spectroscopy at levels potentially approaching
1 part in $10^{17}$.  Meanwhile, the increasing precision of $1S$\,-\,$2S$ spectroscopy
in atomic antihydrogen \cite{ALPHA:2016siw, Ahmadi:2018eca} further raises the
prospect of future high-precision rovibrational spectroscopy with the 
anti-molecular ion $\overline{\rm H}_2^{\,-}$  \cite{Myers,Zammit,Zammit:2025nky}.

In a recent paper \cite{Shore:2024ked} (hereafter referred to as Paper I), we presented a detailed
investigation of the potential effects of Lorentz and \textsf{CPT} violation on the 
rovibrational spectrum of ${\rm H}_2^{\,+}$ and $\overline{\rm H}_2^{\,-}$
\cite{Muller:2004tc}, \cite{Kostelecky:2015nma}. In particular,
we showed how, in addition to improved experimental precision, rovibrational spectroscopy
of the ${\rm H}_2^{\,+}$ and $\overline{\rm H}_2^{\,-}$ molecular ions
permits an enhancement of $O(m_p/m_e)$ in sensitivity 
to Lorentz and \textsf{CPT} violation in the proton sector compared to the corresponding
${\rm H}$ and $\overline{\rm H}$ atomic spectroscopy 
\cite{Kostelecky:2015nma, Charlton:2020kie, Baker:2025ehs}.

Our analysis was conducted in the framework of the effective theory of Lorentz and \textsf{CPT}
violation known as the Standard Model Extension (SME) \cite{Colladay:1998fq, Kostelecky:2013rta}.
Here, the standard QED Lagrangian is augmented by Lorentz tensor operators with couplings
which, if non-zero, would break Lorentz invariance and in some cases also \textsf{CPT}. 
In the form we use in this paper, the Lagrangian for a single Dirac fermion field $\psi(x)$ is
taken to be,
\begin{align}
{\cal L}_{\rm SME} \,=\,& \frac{1}{2}\int d^4 x~\Bigl[ 
\bar\psi\left(i \c^\m \partial_\m \,-\, m\right)\psi
\,-\, a^{}_\m \,\bar\psi \c^\m \psi \,+\, i c^{}_{\m\n}\,\bar\psi \c^\m \partial^\n\psi \,+\,
a^{}_{\m\n\l}\, \bar\psi \c^\m \partial^\n \partial^\l\psi     \nonumber \\[3pt]
& -\, b^{}_\m\, \bar\psi\c^5 \c^\m \psi 
\,+\, i d^{}_{\m\n}\, \bar\psi \c^5\c^\m \partial^\n \psi  
\,-\,  \tfrac{1}{2} H^{}_{\m\n}\, \bar\psi\s^{\m\n}\psi  \,+\, 
\tfrac{1}{2}i g^{}_{\m\n\l}\, \bar\psi\s^{\m\n} \partial^\l \psi
~+~\ldots~ \Bigr] \nonumber \\[3pt]
&+~~ {\rm h.c.}
\label{a1}
\end{align}
While the SME does not by any means exhaust the possibilities for Lorentz and \textsf{CPT}
violation -- for example, it remains a local Lagrangian quantum field theory built from
causal fields and respecting the equality of masses for particles amd antiparticles -- it
is especially valuable in showing in a systematic way how Lorentz and \textsf{CPT} violating effects 
may manifest themselves in many different ways, {\it e.g.}~appearing in certain spectroscopic
transitions but not others. Further motivation and discussion of our approach is given 
in Paper I. 

In this sequel, we extend the analysis of Paper I in two main directions.
First, we take fully into account the hyperfine and Zeeman structure of the ${\rm H}_2^{\,+}$ 
spectrum, including the mixing of states in an applied magnetic field.
We consider in detail both high and low magnetic field regimes, keeping in mind that
while current measurements with ${\rm H}_2^{\,+}$ are carried out in small fields, future 
$\overline{\rm H}_2^{\,-}$ spectroscopy will most probably be
conducted in a high-field regime \cite{Zammit,Myers,SAS2024}.

Second, whereas in Paper I we considered only the spin-independent couplings in the 
SME Hamiltonian derived from (\ref{a1}) 
\cite{Kostelecky:1999zh,Yoder:2012ks,Kostelecky:2013rta},
(see eqs.~(\ref{bb1}), (\ref{bb2}) below), here we extend our analysis to the full set
of SME couplings including those ($b_\m, \,g_{\m\n\l}, \,d_{\m\n}, \,H_{\m\n}$)
which couple to the electron and proton spins. While these are more tightly constrained 
\cite{Kostelecky:2008ts}
by existing spin-precession experiments than the spin-independent couplings 
($c_{\m\n}, \, a_{\m\n\l}$), the rich structure and high measurement precision possible
with the molecular ion motivates their study in this context.

In a recent paper \cite{Vargas:2025efi}, the issue of testing Lorentz and \textsf{CPT} symmetry 
with ${\rm H}_2^{\,+}$ and $\overline{\rm H}_2^{\,+}$ spectroscopy was also addressed
within the SME framework, though with a rather different focus and application
from the work presented here.
In particular, a main focus of \cite{Vargas:2025efi} is the possibility of constraining SME 
couplings through sidereal variations, emphasising the non-minimal SME, though 
only considering some of the leading-order effects in the molecular dynamics described
here and in Paper I.\footnote{Notably, ref.~\cite{Vargas:2025efi} does not consider the
terms here proportional to $\tr_Y\langle\,p_a\,p_b\,\rangle$, and therefore their associated
SME couplings. This expectation value is non-zero because the molecule only has cylindrical 
and not spherical symmetry, and is smaller but comparable in magnitude to 
$\tr\langle\,p_a\,p_b\,\rangle$.  Another important approximation in \cite{Vargas:2025efi}
is the neglect of terms beyond $\d_{\rm SME}$ in the expansion (\ref{a2}), in particular
the leading $N(N+1)$ dependence. On the other hand, \cite{Vargas:2025efi} includes contributions 
to the energy levels of fourth order in momentum associated with higher dimension SME operators, 
which we do not consider here.}
Together, these papers should provide complementary insights
and help provide theoretical input into the design of an
experimental programme of high-precision rovibrational spectroscopy tailored 
to provide maximum sensitivity to potential Lorentz and \textsf{CPT} violation.

The paper is organised as follows. In section \ref{sect 2}, we review the hyperfine-Zeeman structure
of the spectrum of ${\rm H}_2^{\,+}$, emphasising the dependence of the energy levels on the 
rovibrational quantum numbers $(v,N)$. Then, in section \ref{sect 22}, we summarise the application
of the Born-Oppenheimer approximation to ${\rm H}_2^{\,+}$ in the presence of
both spin-independent and spin-dependent Lorentz and \textsf{CPT} violating operators in the 
Hamiltonian giving the effective Schr\"odinger equations in the electron and proton sectors.

Sections \ref{sect 3} and \ref{sect 4} contain the main physics development in the paper.
The extended Born-Oppenheimer analysis, described in detail in Paper I, shows that 
Lorentz and \textsf{CPT} violation affects the rovibrational spectrum in two ways -- directly
through the SME couplings in the Schr\"odinger equation describing the rovibrational
motion of the protons, and indirectly through the modifications of the inter-nucleon
potential due to the binding electron (but note this is also sensitive to the proton SME couplings).
The analysis of Paper I for the spin-independent couplings is extended to include the 
hyperfine-Zeeman mixing and magnetic field dependence in section \ref{sect 3},
while an extensive study of the effects of the spin-dependent couplings is given 
in section \ref{sect 4}.

This development is taken up again in section \ref{sect 6}, while meanwhile section \ref{sect 5}
and Appendix \ref{Appendix A} present an equivalent analysis using the SME Hamiltonian 
in the widely-used spherical tensor formalism, extended and adapted to the dynamics
of the molecular ion. This would provide the basis for an extension of our results to 
the non-minimal SME.  These sections may be omitted by readers interested primarily 
in the physics results for the rovibrational spectrum.

In section \ref{sect 6}, our results for the SME-modified inter-nucleon potential are translated 
into energy-level shifts in the rovibrational spectrum using the general methods developed
in Paper I. We present our results in the form of an expansion of the rovibrational energy levels
as,
\begin{align}
\D E_{v N J M_J }^{\rm SME}  ~&=~{\cal E}_{\rm SME} \,+\, \d_{\rm SME} \,(v+\thalf) \,\w_0  
\,+\, B_{\rm SME} \,N(N+1) \,\w_0   
\nonumber \\[8pt]
&~~~~~~ -\,  x_{\rm SME} \,(v+\thalf)^2 \,\w_0 
~-\,  \a_{\rm SME}\,(v + \thalf) N(N+1) \,\w_0  \nonumber \\[8pt]
&~~~~~~ -\,  D_{\rm SME}\,(N(N+1))^2 \,\w_0  ~+~  \ldots 
\label{a2}
\end{align}
where $\omega_0$ is the fundamental vibration frequency.  We give explicit expressions for 
${\cal E}_{\rm SME}$ and the 
coefficients $\d_{\rm SME}$, $B_{\rm SME}, \ldots$ for each hyperfine-Zeeman state in terms of the 
rovibrational and angular momentum quantum numbers and corresponding SME couplings.

Two further appendices provide a collection of useful relations involving Clebsch-Gordan
coefficients, and a brief summary of the implications of our results for the potential detection
of Lorentz violation through annual, and sidereal, variations of the transition frequencies.

\newpage

\section{The rovibrational, hyperfine and Zeeman spectrum of $\textbf{H}_2^{\,+}$ and
  ${\overline{\textbf{H}}_2^{\,-}}$}\label{sect 2}
 
We begin with a brief review of the main features of the hyperfine-Zeeman spectrum for 
${\rm H}_2^{\,+}$ \cite{Korobov2006,Karr2008,KKH20081}, 
and equivalently $\overline{\rm H}_2^{\,-}$, which we need later to describe
the Lorentz and \textsf{CPT} violating effects.  We focus here on the simplest case,
Para-${\rm H}_2^{\,+}$, since this is sufficient to illustrate all the main principles and
is the main subject of current experiments, but comment on the more complicated case
of Ortho-${\rm H}_2^{\,+}$ where relevant.

In general, the energy eigenstates of ${\rm H}_2^{\,+}$ at zero applied magnetic field are
described by the quantum numbers $|v\,N\,S\,I\,F\,J\,M_J\rangle$.
Before including spin, the rovibrational energy levels are described by $|v\,N\rangle$,
where $v$ is the vibrational quantum number and ${\bf N}$ is the molecular orbital
angular momentum, with quantum numbers $N, M_N$.  The electron spin is ${\bf S}$,
with quantum numbers $S=1/2$ and $M_S$.  The nucleon spin is ${\bf I}$ and we define
the total molecular spin by ${\bf F} = {\bf I} + {\bf S}$.  Spin-statistics requires that for $I=0$ 
(Para-${\rm H}_2^{\,+}$),  $N$ is even, while $I=1$ (Ortho-${\rm H}_2^{\,+}$) requires $N$ odd.
The total angular momentum is then ${\bf J} = {\bf N} + {\bf F}$.

For Para-${\rm H}_2^{\,+}$ therefore, since $I = F= 0$ and $S=1/2$ always, we characterise 
the molecular rovibrational-spin states simply by $|v\,N\,J\,M_J\rangle$.
It will also be useful to consider the alternative representation with states $|v\,N\,M_N\,M_S\rangle$,
which are the eigenstates at large applied magnetic field ${\bf B}$.

\subsection{Hyperfine-Zeeman Hamiltonian}\label{sect 2.1}

The hyperfine and Zeeman interactions for Para-${\rm H}_2^{\,+}$ are then
\begin{equation}
H_{\rm HFS} ~=~ c_e(v,N) \, {\bf N} \cdot {\bf S} ~~~~=~~ \frac{1}{2} c_e(v,N)\,
\left( {\bf J}^2 - {\bf N}^2 - {\bf S}^2 \right) \ ,
\label{b1}
\end{equation}
and
\begin{equation}
H_{\rm Z} ~=~ g_e \mu_B\,{\bf S} \cdot{\bf B} ~-~ g_m(v,N) \mu_B \, {\bf N} \cdot {\bf B} ~~~~=~~
\mu_B B\,\left( g_e\, S_3 ~-~ g_m(v,N)\, N_3\right) \ ,
\label{b2}
\end{equation}
in the \textsf{EXP} frame with $3$-axis aligned with the applied magnetic field.

$H_{\rm HFS}$ describes the spin-orbit interaction between the electron spin and the 
molecular orbital angular momentum.\footnote{We follow the common nomenclature and
refer to all the spin and angular momentum interactions, perhaps loosely,
as ``hyperfine'', even though (\ref{b1}) is a spin-orbit interaction. 
For Ortho-${\rm H}_2^{\,+}$ there are 5 such couplings including
$b_p(v,N)\, {\bf I}\cdot{\bf S}$ and $c_I(v,N)\, {\bf N}\cdot{\bf I}$ together with two
more complicated interactions amongst ${\bf N}$, ${\bf I}$ and ${\bf S}$.
See ref.\cite{Korobov2006} for details and values for the coefficients $b_F, \, c_e, \,c_I, \ldots$.}
Evidently, $J$ and $M_J$ are good quantum numbers for this interaction, so at zero magnetic 
field the energy eigenstates are naturally labelled as $|v\,N\,J\,M_J\rangle$.
The coupling $c_e(v,N)$ depends on the rovibrational state and can be
calculated at $O(\a^2)$ in non-relativistic QED.  Explicit values are given in Table 1
of \cite{Korobov2006}, which includes a very detailed discussion of the hyperfine
structure of both Para- and Ortho-${\rm H}_2^{\,+}$. For illustration, a typical value
for a low-lying rovibrational state is (in $h=1$ units) $c_e(0,2) = 42.1625\,{\rm MHz}$.

The form of the Zeeman interaction (\ref{b2}) requires some justification.  Although
the electron is in the $1s\s_g$ ground state, its orbital angular momentum ${\bf N}_e$
is not exactly zero because unlike the hydrogen atom the molecular ion only possesses
cylindrical and not spherical symmetry. Compared to the orbital angular momentum 
${\bf N}_p$ of the proton it is of $O(m_e/m_p)$, but this is compensated by the $g$-factor
coefficients such that both ${\bf N}_p$ and ${\bf N}_e$ contribute approximately equally to 
the Zeeman energy. 

To justify (\ref{b2}) therefore, we have to demonstrate that the elementary Zeeman Hamiltonian 
in terms of ${\bf N}_e$ and ${\bf N}_p$ satisfies
\begin{align}
&\mu_B B\,\langle v\,N\,J'\,M_J |\,\Big(\,N_{e3} ~-~  \frac{2 m_e}{m_p}\, N_{p3}\,\Big) 
|v\,N\,J\,M_J\rangle \nonumber \\
&~~~~~~~~~~~~~~~~~~~~~~~~~~~~~~~~~~=~
- g_m(v,N)\,\mu_B B\, \langle v\,N\,J'\,M_J |\, N_3\,|v\,N\,J\,M_J\rangle \ ,
\label{b3}
\end{align}
where the factor 2 is due to the reduced mass $m_p/2$ of the protons.

Now, inspection of (\ref{b1})--(\ref{b3}) shows that $M_J$ remains a good quantum number
in the presence of an applied magnetic field but $J$ is not.  The Zeeman interaction
induces mixing between the states with $J = N\pm \thalf$ for fixed $M_J$ (as indicated
in (\ref{b3})), except for the unique (``stretched'') states where $M_J = \pm (N + \thalf)$ which occur 
only for $J= N+\thalf$.
To justify the identification (\ref{b3}) and find the effective $g$-factor $g_m(v,N)$,
we therefore need to compare the matrix elements on both sides for $J',\, J = N\pm \thalf$
including the off-diagonal elements.

To evaluate these, first write the $|v\,N\,J\,M_J\rangle$ states in terms of the $|v\,N\,M_N\, M_S\rangle$
basis states as follows:
\begin{equation}
|v\,N\,J\,M_J\rangle  ~~=~~ \sum_{M_S} \, C_{N \,M_N,\,\thalf \,M_S}^{J \,M_J} \, |v\,N\,M_N\, M_S\rangle \ ,
\label{b4}
\end{equation}
where $M_N = M_J - M_S$. We will make extensive use of these Clebsch-Gordan coefficients 
in what follows, so it is convenient to record them here, for $J = N\pm \thalf$:
\begin{align}
C_{N\,M_J\mp\thalf,\,\thalf\,\pm\thalf}^{N+\thalf\,\,M_J}  ~~&=~~~~ \frac{1}{\sqrt{2N+1}}\, \sqrt{N+\thalf \pm M_J}\ ,   
\nonumber \\[10pt]
C_{N\,M_J\mp\thalf,\,\thalf\,\pm\thalf}^{N-\thalf\,\,M_J}  ~~&=~~ \mp\,\frac{1}{\sqrt{2N+1}}\, \sqrt{N+\thalf \mp M_J} \ .
\label{b5}
\end{align}
The matrix elements of $S_3$ are readily evaluated in the $|v\,N\,M_N\, M_S\rangle$ basis
(see (\ref{bpp4}), giving
\begin{align}
\langle v\,N\,J'=N\pm\thalf \,M_J|\,S_3\, |v\,N\,J=N\pm\thalf\,M_J\rangle ~~&=~~
\pm\,\frac{1}{2N+1}\, M_J \ ,  \nonumber \\[10pt]
\langle v\,N\,J'=N\mp\thalf \,M_J|\,S_3\, |v\,N\,J=N\pm\thalf\,M_J\rangle ~~&=~~
-\,\frac{1}{2N+1}\, \sqrt{(N+\thalf)^2 \,-\, M_J^2}  \ , \nonumber \\
\label{b6}
\end{align}
and the matrix elements of $N_3$ follow directly using $N_3 = J_3\,-\,S_3$. Explicitly,
\begin{align}
\langle v\,N\,J'\,M_J|\,N_{3}\, |v\,N\,J\,M_J\rangle ~~&=~~
\frac{1}{2N+1}~~\begin{cases}
2N \, M_J  \ , ~~~~~~~~~~&(J'=J= N+\thalf)\\[10pt]
2(N+1)\, M_J \ , ~~~~~~&(J'=J=N-\thalf) \\[10pt]
\sqrt{(N+\thalf)^2 \,-\, M_J^2}  \ ,
\end{cases} \nonumber \\
&~~~~~~~~~~~~~~~~~~~~~~~~~~~~~~~~~~~~~~~~~~(J'=N\mp\thalf,\,J=N\pm\thalf)
\nonumber \\
\label{b7}
\end{align}

Next, consider the matrix elements of $N_{e3}$ in (\ref{b3}).  In this case, remembering
that $N_{e3}$ does not act on the electron spin states, we can use the Wigner-Eckart
theorem to write
\begin{equation}
\langle v\,N\,N\,M_N|\,N_{e3}\, |v\,N\,M_N\,M_S\rangle ~~=~~
C_{N\,M_N,\,1\,0}^{N\,M_N}\,\, \langle v\,N||\,N_e\,||v\,N\rangle \ ,\\[7pt]
\label{b8}
\end{equation}
where $C_{N\,M_N,\,1\,0}^{N\,M_N} = 1/\sqrt{N(N+1)}$. Using the Clebsch-Gordan coefficients in (\ref{b5})
we can then express the required matrix elements for $N_{e3}$ in terms of these reduced matrix
elements involving only the rovibrational quantum numbers. We find,
\begin{align}
\langle v\,N\,J'\,M_J|\,N_{e3}\, |v\,N\,J\,M_J\rangle ~~&=~~
\frac{1}{2N+1}\,\frac{1}{\sqrt{N(N+1)}}\,\langle v\,N||\,N_e\,||v\,N\rangle \nonumber \\[8pt]
& ~~~~~~~\times~\begin{cases}
2N \, M_J  \ , ~~~~~~&(J'=J= N+\thalf)\\[10pt]
2(N+1)\, M_J \ , ~~~~~~&(J'=J=N-\thalf) \\[10pt]
\sqrt{(N+\thalf)^2 \,-\, M_J^2}  \ ,&(J'=N\mp\thalf,\,J=N\pm\thalf)
\end{cases}
\label{b9}
\end{align}
with a similar result for $N_{p3}$.

Comparing (\ref{b7}) and (\ref{b9}), we verify the proposed identification (\ref{b2}) for the
Zeeman Hamiltonian in terms of the total molecular orbital angular momentum ${\bf N}$
with the identification,
\begin{equation}
g_m(v,N) ~=~ \frac{1}{\sqrt{N(N+1)}} \,\, \Big(- \langle v\,N||\,N_e\,||v\,N\rangle  
~+~ \frac{2 m_e}{m_p} \, \langle v\,N||\,N_p\,||v\,N\rangle\Big) \ .
\label{b10}
\end{equation}

The reduced matrix elements have been calculated using precision variational estimates for the 
rovibrational wavefunctions for ${\rm H}_2^{\,+}$ and are given in \cite{KKH20081}
for low values of $(v,\,N)$.\footnote{As evident from (\ref{b12}) below, the off-diagonal matrix 
elements only contribute to the energies at $O(B^2)$ and can be neglected in the weak-field
approximation. Noting from above that the diagonal matrix elements are all proportional to
$M_J$, the Zeeman Hamiltonian may be approximated at weak field, only, by
\begin{equation*}
H_{\rm Z}~\simeq~ g_J(v,N,J) \, \mu_B\, {\bf J}\cdot{\bf B} \ ,
\end{equation*}
with an effective Land\'e $g$-factor,
\begin{equation*}
g_J(v,N,J) ~=~ \begin{cases}
\frac{1}{2N+1}\,\left( g_e \,-\, 2 N\,g_m(v,N)  \right)\ , ~~~~~~~~~~~~&(J = N+\thalf) \\[5pt]
-\frac{1}{2N+1}\,\left( g_e \,+\, 2 (N+1)\,g_m(v,N)  \right) \ , &(J=N-\thalf)
\end{cases}
\end{equation*}
as realised in the diagonal elements of (\ref{b12}). 
This may be compared with \cite{KKH20081}, where these two contributions to $g_J$
are denoted $g_1(N,J)$ and $g_3(v,N,J)$ respectively.  Note however that our derivation
here determines the angular momentum $g$-factor $g_m(v,N)$ independently of being in
the weak or strong magnetic field regime.

A similar analysis may be made for Ortho-${\rm H}_2^+$, starting from the representation
(\ref{b10b}) of the relevant states. In this case there are three contributions to the Land\'e
$g$-factor corresponding to the three terms in the Zeeman Hamiltonian (\ref{b10a}),
and they each depend additionally on the total spin quantum number $F$. Full details 
are given in \cite{KKH20081}.
\label{FNLande}} The identification is $g_m(v,N) = g_{rot} (v,N)\, m_e/m_p$, with $g_{rot}(v,N)$ 
given in Table 1 of \cite{KKH20081}. For illustration, a typical value is $g_m(0,2) = 0.9198\,m_e/m_p$.

For Ortho-${\rm H}_2^{\,+}$, the proton spins are aligned so that $I=1$ and spin-statistics
implies $N$ is odd.  The Zeeman interaction is generalised to 
\begin{equation}
H_{\rm Z} ~=~ g_e \mu_B\,{\bf S} \cdot{\bf B} ~-~ g_p \mu_p \,{\bf I}\cdot {\bf B}
 ~-~ g_m(v,N) \mu_B \, {\bf N} \cdot {\bf B} \ ,
\label{b10a}
\end{equation}
with $\mu_p = (m_e/m_p)\mu_B$. Since $I=1$, $S=1/2$ always, we abbreviate the
notation for the states here to $|v\,N\,F\, J\,M_J\rangle$. Again, $M_J$ remains a good quantum 
number for both the hyperfine and Zeeman interactions, but the energy levels are now split
in both $J$ and $F$. 

The effective orbital angular momentum $g$-factor $g_m(v,N)$ in (\ref{b10a})
is naturally unchanged and is still given in terms of reduced matrix elements by (\ref{b10}).
To see this explicitly, and evaluate the weak-field $g$-factors described in footnote \ref{FNLande},
it is convenient first to re-express the states in the form
\begin{equation}
|v\, N\,F\,J\,M_J\rangle ~=~ \sum_{M_F,M_S}\,C_{N\,M_N,\,F\,m_F}^{J\,M_J}\, 
C_{1\,M_I,\,\thalf\,M_S}^{F\,M_F}\, |v\,N\,M_J\,M_F\,M_S\rangle\ ,
\label{b10b}
\end{equation}
where of course $M_I=M_F -M_S$ and $M_N=M_J-M_F$. The analysis above can then
be carried through using orthonormality relations to simplify the additional Clebsch-Gordan
factors and verify (\ref{b10}) still holds.
A complete set of values for the Land\'e $g$-factors for Ortho-${\rm H}_2^{\,+}$ 
is given in \cite{KKH20081}.

\subsection{Hyperfine-Zeeman energy spectrum}\label{sect 2.2}

Given the hyperfine and Zeeman Hamiltonians in the form (\ref{b1}), (\ref{b2}),
we can now readily evaluate the energy spectrum.  With zero magnetic field, 
$|v\,N\,J\,M_J\rangle$ are eigenstates of $H_{\rm HFS}$ and we find immediately
\begin{equation}
\langle v\,N\,J\,M_J|\,H_{\rm HFS}\, |v\,N\,J\,M_J\rangle ~=~
\begin{cases}
\frac{1}{2}\,N\, c_e(v,N) \ , ~~~~~~~~~~ &(J=N+\thalf) \\[8pt]
- \frac{1}{2}\,(N+1)\, c_e(v,N) \ ,  &(J = N-\thalf)
\end{cases}\\[3pt]
\label{b11}
\end{equation}
with hyperfine energy splitting $(N+\thalf)c_e(v,N)$.

Combining with the results above for the matrix elements of $H_{\rm Z}$, we can express the 
matrix elements of the full hyperfine-Zeeman Hamiltonian in the $|v\,N\,J\,M_J\rangle$
basis in the form
\begin{align}
&\langle v\,N\,J'\,M_J|\,H_{\rm HFS}+ H_{\rm Z}\, |v\,N\,J\,M_J\rangle ~=~ \nonumber \\[10pt]
&\begin{pmatrix}
\thalf N c_e +\frac{1}{2N+1} (g_e - 2N g_m) \mu_B B M_J ~~~~~~~~~~
-\frac{1}{2N+1} \sqrt{(N+\thalf)^2 - M_J^2}\,(g_e + g_m) \mu_B B\\[20pt]
~~~~~~~~~~~~~{\rm Sym} ~~~~~~~~~~~~~ 
-\thalf (N+1) c_e(v,N) - \frac{1}{2N+1} (g_e + 2(N+1)g_m)\mu_B B M_J \\
\end{pmatrix}
\label{b12}
\end{align}
with rows/columns corresponding to $J=N\pm\thalf$. 
Notice that for the states $J=N+\thalf, ~M_J =\pm(N+\thalf)$ where there is no mixing,
the off-diagonal elements here vanish. The energy eigenvalues in this case are simply
read off from the top left element,
\begin{equation}
E(B)\big|_{M_J = \pm\left(N+\thalf\right)} ~=~ \frac{1}{2} N\,c_e(v,N) \,\pm\, \frac{1}{2} 
\big(g_e - 2 N\,g_m(v,N) \big) \mu_B B \ .
\label{b13}
\end{equation}

The energy eigenvalues $E(B)$ for the mixed states are given by the eigenvalues of this matrix, 
and after some algebra we find
\begin{align}
E_{\pm}(B) ~&=~ -\frac{1}{4} c_e(v,N)\,-\, g_m \,\mu_B B\, M_J \nonumber \\[3pt]
& ~~~\pm\frac{1}{2} \Big[ (N+\thalf)^2 c_e^2 \,+\, 2 c_e (g_e + g_m) \mu_B B \,M_J 
\,+\, (g_e + g_m)^2 (\mu_B B)^2 \Big]^{\thalf} \ .
\label{b14}
\end{align}

For small $B$, the energies may be expanded as 
\begin{align}
E_+(B) ~&=~ \frac{1}{2} N c_e(v,N) \,+\, \frac{1}{2N+1} \big(g_e -2N\,g_m(v,N)\big) \mu_B B\, M_J  \ ,
\nonumber\\
E_-(B) ~&=~ -\frac{1}{2} (N+1) c_e(v,N) \,-\,\frac{1}{2N+1} \big(g_e + 2(N+1) g_m(v,N) \big)\mu_B B \,M_J \ ,
\label{b15}
\end{align}
where the expansion parameter is $\mu_B B/c_e(v,N) \simeq B/3\,{\rm mT}$ for small
values of $(v,N)$. This defines the range of validity of the `small' or `large' $B$ approximations.

For large $B$, we find
\begin{align}
E_+(B) ~&=~ \big( \thalf g_e \,-\, g_m(v,N) (M_J-\thalf)\big) \mu_B B \,+\, 
\frac{1}{2} c_e(v,N) (M_J-\thalf)\ ,
\nonumber\\
E_-(B) ~&=~ \big( -\thalf g_e \,-\, g_m(v,N) (M_J+\thalf)\big) \mu_B B \,-\, 
\frac{1}{2} c_e(v,N) (M_J+\thalf)\ ,
\label{b16}
\end{align}
which we recognise as the eigenvalues corresponding to the high-field eigenstates
$|v\,N\,M_N\,M_S\rangle$ with $M_S = \pm \thalf$.
The expansion parameter in this case is clearly $c_e(v,N)/\mu_B B$.

\begin{figure}[h!]
\centering{ {\includegraphics[scale=0.65]{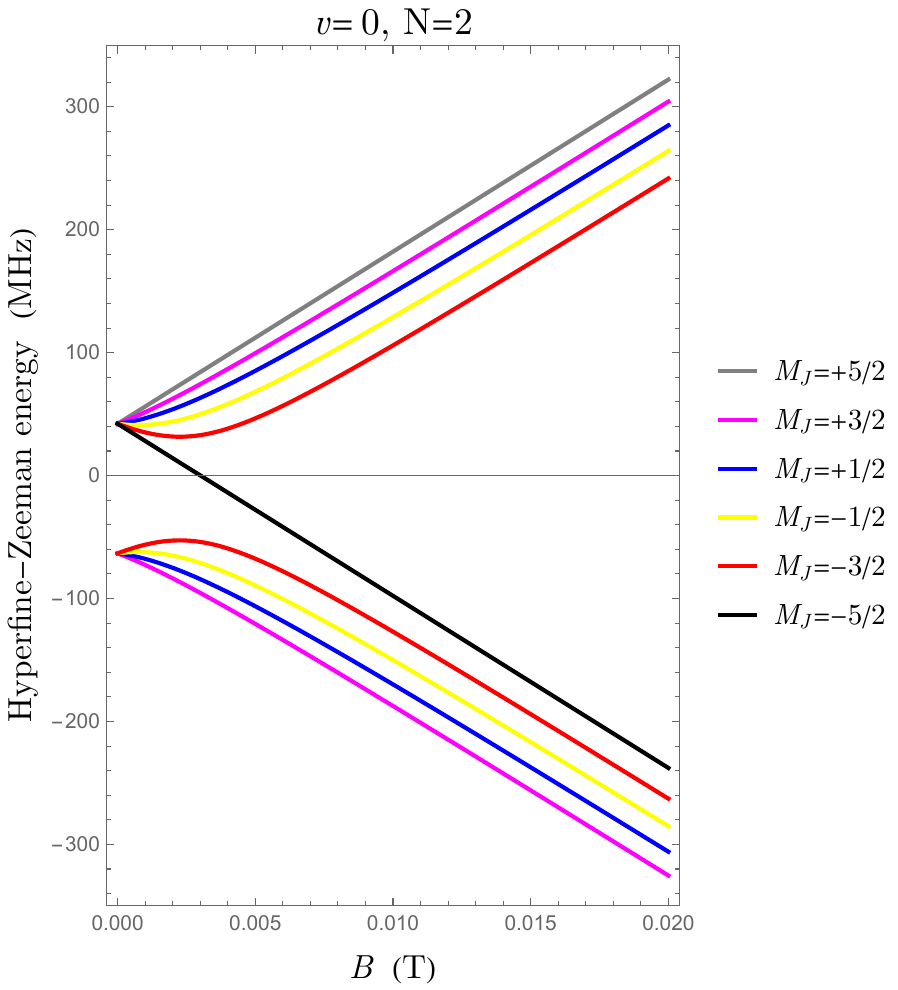} } } 
\caption{Hyperfine-Zeeman energies for the rovibrational level $(v,N) = (0,2)$.
For general $N$, the eigenstates for given $M_J$ are linear combinations of states
with $J= N\pm\thalf$ and are doubled, apart from the unmixed states $J=N+\thalf,\,
M_J = \pm (N+\thalf)$.}
\label{FigHZ}
\end{figure}
These results are illustrated for the particular case of the rovibrational level $v=0$, $N=2$
in Fig.~\ref{FigHZ}, using the input parameters for $c_e(0,2)$ and 
$g_m(0,2)$ quoted above. This reproduces the corresponding plot in the review \cite{SchillerCP}
(Supplementary Information). The hyperfine splitting $(N+\thalf)c_e(v,N)$ at zero magnetic field 
in this case is approx.~105\,MHz.
Of course at the level of resolution of the plots, the effect
of $g_m(v,N) \sim O(m_e/m_p)$ is not visible and the Zeeman energies are dominated by the 
electron spin term. This explains the form of the spectrum at large $B$, which splits into two sets
of states with energies rising (for $M_S=\thalf)$ or falling (for $M_S = -\thalf)$ linearly 
with $B$.\footnote{It is interesting to contrast this with the corresponding hyperfine-Zeeman 
spectrum for the ${\rm H}$ atom with $L\neq 0$, for example the $2P$ states. This shows many 
similarities with the above analysis, except that for the atom the orbital angular momentum 
Zeeman interaction is not suppressed by $O(m_e/m_p)$ as here, resulting in a qualitatively different
energy level diagram (see, for example, refs.~\cite{ALPHA:2018sre} or \cite{Charlton:2020kie}).}

\section{Born-Oppenheimer analysis  with Lorentz and $\textsf{CPT}$ violation}\label{sect 22}

The dynamical analysis of the spectrum of the ${\rm H}_2^{\,+}$ and $\overline{{\rm H}}_2^{\,-}$
molecular ions is carried out in the framework of the Born-Oppenheimer approximation,
extended here to include Lorentz and \textsf{CPT} violation.
This was described in Paper I for the spin-independent SME couplings $c_{\m\n}$ and $a_{\m\n\l}$,
and here we extend this to include also the spin-dependent couplings $b_{\m}$, $g_{\m\n\l}$, 
$d_{\m\n}$ and $H_{\m\n}$. 

First recall the form of the non-relativistic SME Hamiltonian for a single Dirac fermion, derived 
from the original QFT Lagrangian (\ref{a1}):
\begin{equation}
H_{\rm SME} ~=~ \bigl(A\,+2\,B_k S^k\bigr) \,+\, \bigl(C_i \,+\, 2D_{ik} S^k\bigr)\frac{p^i}{m} 
\,+\, \bigl(E_{ij} \,+\, 2F_{ijk} S^k\bigr) \frac{p^i p^j}{m^2} \ ,
\label{bb1}
\end{equation}
We require only the even-parity operators, with coefficients
\cite{Kostelecky:1999zh, Yoder:2012ks, Kostelecky:2013rta}
\begin{align}
A ~&=~ a_0 \,-\, m\, c_{00} \,+\, m^2\, a_{000} \ , 
\nonumber \\[5pt]
B_k ~&=~ - b_k \,+\, m\, d_{k0} \,+\, \tfrac{1}{2}\e_{kmn}
\big(H_{mn} - m\, g_{mn0}\big) \ ,
\nonumber \\[5pt]
E_{ij} ~&=~ -m\big(c_{ij} \,+\, \tfrac{1}{2}\, c_{00}\, \d_{ij}\big) \,+\,
m^2 (3\, a_{0ij} \,+\, a_{000}\, \d_{ij} )  \ ,
\nonumber \\[5pt]
F_{ijk} ~&=~ \frac{1}{2}\bigg[\, \tfrac{1}{2}\big(b_k \,\d_{ij} \,-\, b_j \,\d_{ik}\big) \,+\, 
m\,\big(d_{0j} \,+\, \tfrac{1}{2}\,d_{j0}\big) \d_{ik} \,-\, 
\tfrac{1}{4} \d_{ik} \e_{jmn}\, H_{mn} \nonumber \\[2pt]
&~~~~~~~~~~~~~~~~~~~~~~~~~~~~~~~~
\,-\, m\, \e_{ikm} \big(g_{m0j} + \tfrac{1}{2}\, g_{mj0}\big) \,\bigg] 
~+~ (\,i \leftrightarrow j \,)  \ ,
\label{bb2}
\end{align}
where for later convenience we have explicitly symmetrised the $F_{ijk}$ coupling on $i,j$.
Note also that the overall constant term $A$ is unobservable in spectroscopy.

\begin{figure}[h!]
\centering{{\includegraphics[scale=0.59]{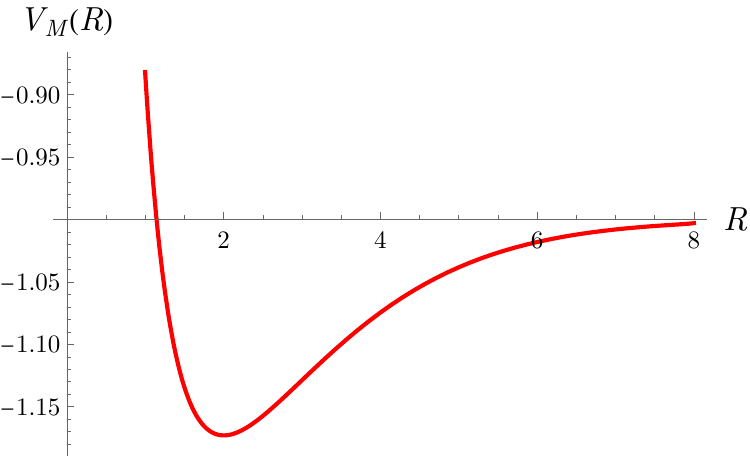} }
\hskip0.2cm{\includegraphics[scale=0.59]{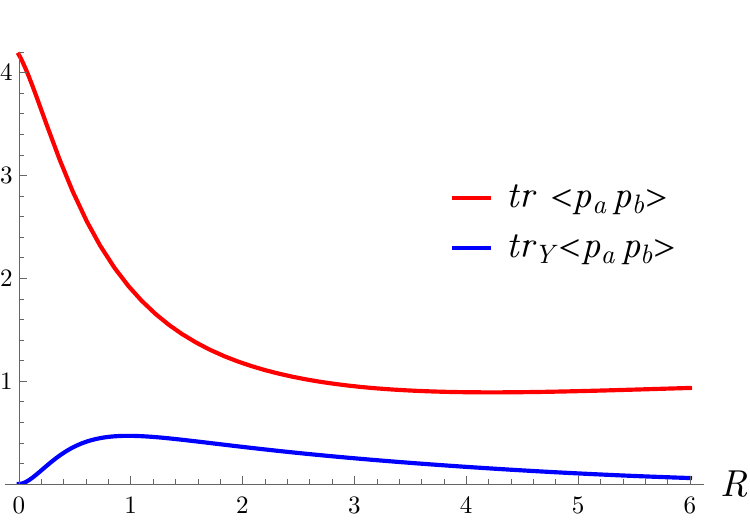} } } 
\caption{({\bf a})~The inter-nucleon potential $V_M(R)$  as a function of the 
bond length $R$. The equilibrium bond length is $R_0=2.003$
(in ``atomic units'', see Paper I).~
({\bf b})~The momentum expectation values in the electron $1s\sigma_g$ ground state
which determine the SME potential $V_{\rm SME}^e(R)$. }
\label{Figpp}
\end{figure}

The fundamental idea of the Born-Oppenheimer analysis is to separate the full Schr\"odinger
equation for the molecular ion into two separate equations, the first describing the electron
motion relative to the nucleon CM and the second describing the rovibrational motion of
the nucleons. The energy eigenvalues of the electron Schr\"odinger equation depend on the
inter-nucleon separation $R$ and feed back into the nucleon Schr\"odinger equation as a
potential $V_M(R)$, which determines the rovibrational motion. $V_M(R)$ has the characteristic
shape of a Morse potential, and a numerical determination is shown in Fig.~\ref{Figpp},
taken from Paper I. 

The key step is to factorise the spatial part of the molecular wavefunction 
into nucleon and electron parts:~
 $\Psi(\boldsymbol{R},\boldsymbol{r}) = \Phi(\boldsymbol{R})\, \psi(\boldsymbol{r};R)$.
The spin operators are spectators at this stage,
but the momentum dependence of the SME interactions is important. The effective
electron Schr\"odinger equation is then,\footnote{The electron position relative to the nucleon CM is 
denoted by ${\bf r}$ with corresponding momentum ${\bf p} = \hat{\m} \,\dot{\bf r}$,
where $\hat{\m} = 2m_e m_p / (m_e + m_p)$ is a reduced mass.
The relative motion of the nucleons (protons/antiprotons in this case) is treated as that 
of a single particle at point ${\bf R}$ with momentum ${\bf P} = \m \,\dot{\bf R}$ 
and reduced mass $\m = m_p/2$.} 
\begin{align}
&\bigg[- \frac{1}{2\hat{\mu}} \nabla_{\boldsymbol{r}}^2 \,+\, V_{mol}(R,r_{1e},r_{2e}) 
\nonumber \\[6pt]
&    +~\frac{1}{m_e^2} \big(E_{ab}^e \,+\, F_{abk}^e\,S_k \big)p_a p_b ~+~
\frac{1}{2 m_p^2} \big(E_{ab}^p \,+\, F_{abk}^p\,I_k \big) p_a p_b
\,\bigg]\,\psi(\boldsymbol{r};R) \,=\, E_e(\boldsymbol{R}) \,\psi(\boldsymbol{r};R) \ ,
\label{bb3}
\end{align}
where we understand $\boldsymbol{p} \rightarrow -i \nabla_{\boldsymbol{r}}$.
$V_{mol}(R,r_{1e},r_{2e})$ is the electrostatic potential binding the molecule, and we assume
the electron is in the $1s\s_g$ ground state. The electron momentum components $p_a$ 
are defined in a frame, denoted MOL, which is fixed with respect to the molecular
axis.\footnote{ MOL frame components are indicated 
by indices $a,b,c, \ldots$.  The indices $i,j,k, \ldots$ are used to specify the EXP frame,
usually fixed with respect to the externally applied magnetic field, and in which the
molecular axis rotates. For convenience we temporarily define the SME couplings $E_{ab}$ {\it etc.}
here with MOL frame components; the rotation between the MOL and EXP frames is
discussed in section \ref{sect 3}. Spins are always referred to the EXP frame, indicated by 
$S_k$, $I_k$.}

As discussed in Paper I, the proton terms in the SME Hamiltonian acquire a dependence
on ${\bf p}$ in this kinematics and so affect the energy levels in two ways -- directly from 
the nucleon Schr\"odinger equation but also indirectly through their effect on the 
inter-nucleon potential via (\ref{bb3}), the latter with the same parametric dependence
on $m_e$ and $m_p$ as in atomic hydrogen spectroscopy.  

The eigenvalues define the inter-nucleon potential,
\begin{equation}
E_e({\bf R}) ~=~  V_M(R) \,+\, V_{\rm SME}^e({\bf R};S_k) \,+\, V_{\rm SME}^p({\bf R};I_k)  \ ,
\label{bb4}
\end{equation}
where the SME contributions depend on the expectation values 
$\langle\,p_a\,p_b\rangle$ of the electron momentum
in the MOL frame. These have been calculated numerically in \cite{Muller:2004tc} and Paper I 
and are shown in Fig.~\ref{Figpp}.  Inserting these back into the nucleon Schr\"odinger
equation we have,
\begin{align}
&\bigg[-\frac{1}{2\mu} \nabla_{\boldsymbol{R}}^2 \,+\, V_M(R) 
\,+\, V_{\textrm{SME}}^e(\boldsymbol{R};S_k) \,+\, V_{\textrm{SME}}^p(\boldsymbol{R};I_k) 
\nonumber \\
&~~~~~~~~~~~~~~\,+\, 2B_k^e\,S_k  \,+\, 2B_k^p \,I_k \,+\, 
\frac{2}{m_p^2}\big( E_{ij}^p \,+\, F_{ijk}^p I_k\,\big)P^i P^j \, \bigg]
\Phi_{NM_N}(\boldsymbol{R})\,|M_I\, M_S\rangle 
\nonumber \\
&\,=\, 
E_{vNM_N M_I M_S}\, \,\Phi_{NM_N}(\boldsymbol{R})\,|M_I\, M_S\rangle  \ .
\label{bb5}
\end{align}

The final step is to separate the contributions and write the total energy eigenvalues as,
\begin{equation}
E_{v N M_N M_I M_S}~=~ \tilde{E}_{v N M_N M_I M_S} ~+~ \D E_{\rm SME}^{e\,B} ~+~ \D E_{\rm SME}^{p\,B} ~+~
\D E_{\rm SME}^n \ ,
\label{bb6}
\end{equation}
where $\tilde{E}_{vN M_N M_I M_S}$ is found by setting
$\Phi_{NM_N}(\boldsymbol{R}) =\dfrac{1}{R} \,\phi(R) \,Y_{NM}(\theta,\phi)$ in (\ref{bb5}) and
solving the rovibrational equation,
\begin{align}
&\bigg[- \frac{1}{2\mu}\, \frac{d^2}{dR^2} \,+\, \frac{1}{2\mu R^2} N(N+1) \,+\, V_M(R) \,+\,
V_{\textrm{SME}}^e(R;S_k) \,+\,V_{\textrm{SME}}^p(R;I_k)\bigg]\,\phi(R)\,  |M_I\,M_S\rangle
\nonumber \\[10pt]
&=~ \tilde{E}_{vNM_N M_I M_S}\,\, \phi(R)\, |M_I\,M_S\rangle\ ,
\label{bb7}
\end{align}
where,
\begin{equation}
V_{\textrm{SME}}^e(R;S_k) \,=\, \langle v\,N\, M_N |\, V_{\textrm{SME}}^e(\boldsymbol{R};S_k)\,|v\,N\, M_N \rangle \ ,
\label{bb8}
\end{equation}
together with,
\begin{equation}
\D E_{\rm SME}^{e\,B} ~=~ \langle v\,N\,M_N\,M_I\,M_S|\,2B_k\,S_k\,|v\,N\,M_N\,M_I\,M_S\rangle \ ,
~~~~~~~~~~~~~~~~~~~~~~
\label{bb9}
\end{equation}
and
\begin{equation}
\D E_{\rm SME}^n ~=~ \frac{2}{m_p^2} \langle v\,N\,M_N\,M_I\,M_S|\,
\big(E_{ij}^p \,+\,F_{ijk}^p I_k\big)\,P_i P_j |v\,N\,M_N\,M_I\,M_S\rangle \ ,
\label{bb10}
\end{equation}
with analogous expressions for $V_{\textrm{SME}}^p(R;I_k)$ and $\D E_{\rm SME}^{p\,B}$.
\vskip0.3cm

In practice, we simplify this by for the most part considering Para-${\rm H}_2^{\,+}$, in which case
the nucleon spin ${\bf I}$ is zero. We also absorb the spin-independent proton coupling into
$V_{\rm SME}^e(R)$ by defining $\tilde{E}_{ij}^e = E_{ij}^e + \tfrac{1}{2}\tfrac{m_e^2}{m_p^2} E_{ij}^p$, leaving
\begin{equation}
V_{\rm SME}^e({\bf R}) ~=~ \frac{1}{m_e^2}\,\big(\tilde{E}_{ab}^e \,+\, F_{abk}^e\,S_k\big) \,\,
\langle\,p_a\,p_b\,\rangle \,
\label{bb11}
\end{equation} 
and
\begin{equation}
V_{\rm SME}^e(R) ~=~  \langle v\,N\,M_N\, M_S|\, V_{\rm SME}^e({\bf R}) \,|v\,N\,M_N\,M_S\rangle \ ,
\label{bb12}
\end{equation}
suppressing the spin labels.

In what follows, we evaluate (\ref{bb9}), (\ref{bb10}) and (\ref{bb11}), (\ref{bb12}) for 
Para-${\rm H}_2^{\,+}$, initially in the basis states $|v\,N\,M_N\,M_S\rangle$, then in the
hyperfine states $|v\,N\,J\,M_J\rangle$ described in section \ref{sect 2}. Including the
Zeeman as well as the hyperfine interaction, we find that the SME couplings give rise to 
spin and magnetic field-dependent corrections to the rovibrational energy levels, due to the
mixing of states induced by the magnetic field.

\vskip0.3cm
A full description of how the SME-modified inter-nucleon potential $V_M(R) + V_{\rm SME}^e(R)$, 
together with the direct nucleon contribution $\D E_{\rm SME}^n$, determines the rovibrational energy
levels $E_{v N M_N}$ (for spin-independent SME couplings) was given in Paper I. 
It was shown there that they admit an expansion of the form,
\begin{align}
E_{v NM_N} ~=~ &V_{\rm SME}^e \,+\,(1 + \d_{\rm SME}^e + \d_{\rm SME}^n)\,(v+\thalf) \,\w_0  \,
-\,  (x_0 + x_{\rm SME}^e + x_{\rm SME}^n)\,(v+\thalf)^2 \,\w_0 \nonumber \\[2pt]
&+\, (B_0 + B_{\rm SME}^e + B_{\rm SME}^n) \,N(N+1) \,\w_0   \,  \nonumber \\[2pt]
&-\,  (\a_0 +\a_{\rm SME}^e + \a_{\rm SME}^n)\,(v + \thalf) N(N+1) \,\w_0 \nonumber \\[2pt]
&-\,  (D_0 + D_{\rm SME}^e + D_{\rm SME}^n) \,(N(N+1))^2 \,\w_0 \,+\,  \ldots 
\label{bb13}
\end{align}
where $\w_0$ is the fundamental vibration frequency.  
Each coefficient of $\w_0$ is an expansion
in powers of the small parameter $\l = 2/ m_p \w_0 R_0^2 \simeq 0.027$, with the leading terms 
in the coefficients $x_0, B_0, \a_0, D_0$ being of order $\l, \l, \l^2, \l^3$ resp.
(Recall from Paper I that parametrically, $\l \sim \sqrt{\frac{m_e}{m_p}}$\,.)

In (\ref{bb13}), we have defined  $V_{\rm SME}^e \equiv V_{\rm SME}^e(R_0)$ where $R_0$ is the mean 
bond length, {\it i.e.}~the minimum of the inter-nucleon potential. 
The coefficients $\d_{\rm SME}^e, B_{\rm SME}^e, \ldots$ are given in terms of its derivatives
at $R_0$, {\it e.g.}
\begin{equation}
\d_{\rm SME}^e \,=\, \frac{1}{2} \,\frac{1}{V_M^{''}}\, \Big[\, V_{\rm SME}^{e\,''}  \,-\, 
\frac{V_M^{'''}}{V_M^{''}}\,\, V_{\rm SME}^{e\,'}\Big] \ ,
~~~~~~~~~~~~~
B_{\rm SME}^e \,=\,  \l\, \,\frac{1}{V_M^{''}}\, \Big[\frac{1}{R_0}\,V_{\rm SME}^{e\,'} \,\Big] \ ,
\label{bb14}
\end{equation} 
while the $\d_{\rm SME}^n, B_{\rm SME}^n, \ldots$ are determined by reducing (\ref{bb10})
to a product of proton SME couplings and the rovibrational kinetic energy, then expanding 
the latter in powers of $(v + \thalf)$ and $N(N+1)$ as above.

Everything in this analysis now goes through exactly as before, where the expansion 
applies to each hyperfine-Zeeman eigenstate individually. The required $V_{\rm SME}^e(R)$,
$\D E_{\rm SME}^n$ and $\D E_{\rm SME}^{e\,B}$ are evaluated in these hyperfine-Zeeman 
eigenstates in the following two sections.

\section{Lorentz and CPT violation in the $\textbf{H}_2^{\,+}$ and
${\overline{\textbf{H}}_2^{\,-}}$ spectrum -- spin-independent couplings}\label{sect 3}

We begin in this section with the spin-independent SME couplings $E_{ij}$ in (\ref{bb1}). 
We restrict to Para-$\rm{H}_2^{\,+}$ for simplicity, the extension to Ortho-$\rm{H}_2^{\,+}$ 
being straightforward in principle.

Following the Born-Oppenheimer analysis described in Paper I and section \ref{sect 22}, the first step
is to determine the potential $V_{\rm SME}^e({\bf R})$ by evaluating the expectation value of
the SME Hamiltonian term,
\begin{equation}
H_{\rm SME}^{e\,E} ~=~ \frac{1}{m_e^2}\, \tilde{E}_{ab}\,\, p_a p_b \ ,
\label{c1}
\end{equation}
in the electron $1s\sigma_g$ ground state, where the couplings and momenta are in the 
\MOL frame.  Recall that the coupling here is 
$\tilde{E}_{ab}^e \,=\, E_{ab}^e + \thalf \tfrac{m_e^2}{m_p^2} E_{ab}^p$
since the proton SME couplings also contribute to the electron Schr\"odinger equation.
Using the cylindrical symmetry of the $1s\sigma_g$ wavefunction, the expectation values
satisfy $\langle\,p_a\,p_b\,\rangle = 0 $ for $a\neq b$, and $\langle\,p_x^2\,\rangle 
= \langle\,p_y^2\,\rangle$. It is then convenient to write $\langle\,p_a\,p_b\,\rangle$
in terms of the two independent expectation values as,
\begin{equation}
\langle\,p_a\,p_b\,\rangle ~=~   \frac{1}{3}\, \tr\langle\,p_a\,p_b\,\rangle\,\d_{ab}
~+~ \frac{1}{6}\,  \tr_Y\langle\,p_a\,p_b\,\rangle\, \textsf{Y}_{ab} \ ,
\label{c2}
\end{equation}
where $\textsf{Y}_{ab} = \begin{pmatrix} 1~\,&0~\,&0 \\0~\,&1~\,&0 \\0~\,&0~\,&-2 \\
\end{pmatrix}$ and $\tr_Y\,p_a\,p_b\, = p_x^2 + p_y^2 -2p_z^2$. 
These expectation values have been evaluated numerically in \cite{Muller:2004tc}
and Paper 1 and are shown here in
Fig.~\ref{Figpp}.

We then have
\begin{equation}
V_{\rm SME}^{e\,E}({\bf R}) ~=~ \frac{1}{m_e^2}\,\Big(
\frac{1}{3}\, \tr\langle\,p_a\,p_b\,\rangle\, \tr\tilde{E}_{ab}^e ~+~ 
\frac{1}{6}\, \tr_Y\langle\,p_a\,p_b\,\rangle\,  \tr_Y \tilde{E}_{ab}^e\,\Big) \ .
\label{c3}
\end{equation}

The next step is to re-express the SME couplings in terms of their components
$\tilde{E}_{ij}^e$ the \EXP frame. With the rotation matrix $R_{ai}$ introduced in
Paper I,
\begin{equation}
R_{ai} \,=\, \begin{pmatrix}
\cos\theta \,\cos\phi &~~~ \cos\theta\,\sin\phi &~~~-\sin\theta  \,\,\\
-\sin\phi &~~~ \cos\phi &~~~0 \\
\sin\theta\,\cos\phi &~~~\sin\theta\,\sin\phi &~~~\cos\theta 
\end{pmatrix} \ ,
\label{c4}
\end{equation}
where $(\theta,\phi)$ specify the orientation of the molecular axis in the \EXP frame,
\begin{equation}
\tilde{E}_{ab}^e ~=~ R_{ai}\, \tilde{E}_{ij}^e\, R_{jb}^{\textsf{T}} \ .
\label{c5}
\end{equation}
It then follows immediately that $\tr\,\tilde{E}_{ab}^e \,=\, \tr\,\tilde{E}_{ij}^e$,
while
\begin{align}
\tr_Y \tilde{E}_{ab}^e ~&=~ \tr \,\tilde{E}_{ab}^e ~-~ 3 \tilde{E}_{zz}^e  \nonumber \\[6pt]
&=~ \tr \,\tilde{E}_{ij}^e ~-~3\, \tilde{E}_{ij} \,R_{iz}^{\textsf{T}}\,R_{zj} \ .
\label{c6}
\end{align}
The product of rotation matrices can be expanded in terms of spherical harmonics as
\begin{equation}
R_{iz}^{\textsf{T}}\,R_{zj} ~=~ \frac{1}{3} \,\d_{ij} ~+~ \sum_M \,C_{ij}^M\, Y_{2M}(\theta,\phi) \ ,
\label{c7}
\end{equation}
with known coefficients $C_{ij}^M$ \cite{Yoder:2012ks,Shore:2024ked},
so 
\begin{equation}
\tr_Y\tilde{E}_{ab}^e ~=~ -3\,\sum_M\, C_{ij}^M\, Y_{2M}(\theta,\phi) \, \tilde{E}_{ij}^e \ .
\label{c8}
\end{equation}
Substituting back into (\ref{c3}) then gives the required expression for $V_{\rm SME}^{e\,E}({\bf R})$
in terms of the \EXP frame SME couplings:\footnote{Looking ahead to section \ref{sect 5}
where we express our results in terms of SME couplings in a spherical tensor formalism, we can compare
(\ref{c9}) and (\ref{c14}) directly with (\ref{e17}),(\ref{e18}). The dictionary is
\begin{equation*}
\sqrt{\tfrac{1}{4\pi}}\,\VV_{200}^{\NR}\,=\, \tfrac{1}{3}\, \tfrac{1}{m^2} \,\tr\,E_{ij} \ ,
~~~~~~~~~~~~~~~~
\sqrt{\tfrac{5}{4\pi}}\,\VV_{220}^{\NR}\,=\, -\tfrac{1}{3}\,\tfrac{1}{m^2}\, \tr\,E_{ij} \ ,
\end{equation*}
the latter being a special case of $\VV_{22m}\,=\, \tfrac{1}{m^2}\, C_{ij}^m\,E_{ij}$.  See section \ref{sect 5} for
further discussion.}
\begin{equation}
V_{\rm SME}^{e\,E}({\bf R}) ~=~ \frac{1}{m_e^2}\,\Big(
\frac{1}{3}\, \tr\langle\,p_a\,p_b\,\rangle\, \tr\,\tilde{E}_{ij}^e ~-~ 
\frac{1}{2}\, \tr_Y\langle\,p_a\,p_b\,\rangle\, \sum_M\, C_{ij}^M\, Y_{2M}(\theta,\phi) \,
\tilde{E}_{ij}^e \, \Big) \ .
\label{c9}
\end{equation}

To evaluate the matrix elements of $V_{\rm SME}^{e\,E}({\bf R})$ in the hyperfine states
$|v\,N\,J\,M_J\rangle$ we use (\ref{b4}) to write 
\begin{align}
{}&\langle v\,N\,J'\,{M_J}'|\,Y_{2M}(\theta,\phi)\, |v\,N\,J\,M_J\rangle \nonumber \\[8pt]
&~~~~=~\sum_{M_S}\, C_{N\,{M_N}',\,\thalf\,M_S}^{J'\,{M_J}'}\,\, C_{N\,M_N,\,\thalf\,M_S}^{J\,M_J}\, 
\langle v\,N\,{M_N}'\,M_S|\,Y_{2M}(\theta,\phi)\, |v\,N\,M_N\,M_S\rangle \ ,
\label{c10}
\end{align}
where the matrix element is given by the Gaunt integral,
\begin{align}
\langle v\,N\,{M_N}'\,M_S|\,Y_{2M}(\theta,\phi)\, |v\,N\,M_N\,M_S\rangle ~&=~
\int d\Omega\, Y_{N{M_N}'}^*\, Y_{2M}\, Y_{NM_N} \nonumber \\[8pt]
&=~ \sqrt{\frac{5}{4\pi}}\,\, C_{N\,M_N,\,2\,M}^{N\,{M_N}'}\, \,C_{N\,0,\,2\,0}^{N\,0}  \ .
\label{c11}
\end{align}

We now come to an important simplification. The hyperfine-Zeeman interactions in 
(\ref{b1}) and (\ref{b2}) commute with $J_z$, so $M_J$ is a good quantum number
for their eigenstates.  This is no longer the case when the SME Hamiltonian is included
and in general (\ref{c9}) has matrix elements which are off-diagonal in $M_J$. 
However, since there are no corresponding matrix elements for 
$\langle\,H_{\rm HFS} + H_{\rm Z}\,\rangle$,
when we extract the energy eigenvalues including the SME Hamiltonian, these off-diagonal 
matrix elements only contribute at second order in the SME couplings.
Such $O({\rm SME})^2$ contributions to energy levels are universally neglected in the 
SME formalism.  

It follows that we should only retain the matrix elements in (\ref{c9}) with ${M_N}' = M_N$, 
that is, with $M=0$. In this case,
\begin{equation}
C_{ij}^0 ~=~ - \frac{1}{3} \sqrt{\frac{4\pi}{5}} \,\,\textsf{Y}_{ij} \ ,
\label{c12}
\end{equation}
and the Clebsch-Gordan coefficients are
\begin{equation}
C_{N\,M_N,\,2\,0}^{N\,{M_N}}\, \,C_{N\,0,\,2\,0}^{N\,0} ~=~ \frac{N(N+1) - 3 M_N^2}{(2N-1)(2N+3)}
~=~ c_{NM_N} \ ,
\label{c13}
\end{equation}
in the notation of Paper I. So keeping only the terms in (\ref{c9}) which will contribute to energy levels
at $O({\rm SME})$, we have
\begin{equation}
V_{\rm SME}^{e\,E}({\bf R}) ~=~ \frac{1}{m_e^2}\,\Big(
\frac{1}{3}\, \tr\langle\,p_a\,p_b\,\rangle\, \tr\,\tilde{E}_{ij}^e ~+~ 
\frac{1}{6}\, \sqrt{\frac{4\pi}{5}} \,\,
\tr_Y\langle\,p_a\,p_b\,\rangle\, \tr_Y\tilde{E}_{ij}^e \,\,Y_{20}(\theta,\phi) \, \Big) \ ,
\label{c14}
\end{equation}
and from (\ref{c10}) the matrix elements in the hyperfine states giving rise to the inter-nucleon
potential are therefore,\footnote{We drop the implicit matrix element labels $J', J, M_J$
on $V_{\rm SME}^{e\,E}(R)$ here simply to avoid cluttering the notation.}
\begin{align}
&V_{\rm SME}^{e\,E}(R) ~=~\langle v\,N\,J'\,{M_J}|\,V_{\rm SME}^{e\,E}({\bf R})\, |v\,N\,J\,M_J\rangle
\nonumber \\[8pt]
&=~\sum_{M_S}\, C_{N\,{M_N},\,\thalf\,M_S}^{J'\,{M_J}}\,\, C_{N\,M_N,\,\thalf\,M_S}^{J\,M_J}\, \,
\frac{1}{m_e^2}\,\Big( \frac{1}{3}\, \tr\langle\,p_a\,p_b\,\rangle\, \tr\,\tilde{E}_{ij}^e ~+~ 
\frac{1}{6} \,
\tr_Y\langle\,p_a\,p_b\,\rangle\, \tr_Y\tilde{E}_{ij}^e \,\, c_{N M_N} \, \Big) \ .
\label{c15}
\end{align}

For the coefficient of $\tr\,\tilde{E}_{ij}$, orthonormality of the Clebsch-Gordan coefficients
implies these matrix elements are diagonal in $J',J$. However, the coefficient of 
$\tr_Y\tilde{E}_{ij}$ has off-diagonal elements mixing $J= N\pm\thalf$ states in the same
way as $\langle\,H_{\rm HFS} + H_{\rm Z}\,\rangle$ in (\ref{b12}).  The required sums over 
Clebsch-Gordan factors, weighted with $c_{NM_N}$, are given explicitly in 
Appendix \ref{Appendix B}, eq.(\ref{bpp5}).

Adding (\ref{c15}) to $\langle v\,N\,J'\,M_J|\,H_{\rm HFS}+ H_{\rm Z}\, |v\,N\,J\,M_J\rangle$ 
from (\ref{b12}) now gives the full SME modification to the hyperfine-Zeeman energy levels.
Calculating the eigenvalues of this combined matrix
and, for simplicity, just quoting the results in the weak magnetic field limit,
we find\footnote{An elementary but useful algebraic result to quickly read off the 
$O({\rm SME})$ corrections to the hyperfine-Zeeman energy levels here and in 
section \ref{sect 4} is found by first denoting
\begin{equation*}
\langle\,H_{\rm HFS} + H_{\rm Z}\,\rangle ~=~
\begin{pmatrix}
\textsf{A}~~&\textsf{B} \\ \textsf{B}~~&\textsf{D} 
\end{pmatrix} \ ,
~~~~~~~~~~~~~~~~~~
V_{\rm SME}^{e}(R) ~=~ \begin{pmatrix}
\a~~&\b \\
\c~~&\d 
\end{pmatrix}  \ .
\end{equation*} 
Then adding these matrices and calculating the SME corrections $\D\l_{\rm SME}^\pm$ to the 
eigenvalues gives, at $O(SME)$,
\begin{equation*}
\D\l_{\rm SME}^\pm ~=~ \thalf (\a + \d) ~\pm~ \thalf \, \frac{1}{\sqrt{(A-D)^2 + 4B^2}}
\,\big((A-D)(\a-\d) + 2B(\b+\c)\,\big) \ .
\end{equation*}
In the case considered here, the off-diagonal element \textsf{B} is proportional to the magnetic
field, so in the small-field regime where we can neglect terms of $O(\textsf{B}^2)$, we have the
simple forms:
\begin{align*}
\D \l_{\rm SME}^+ ~&=~ \a \,+\, \frac{\textsf{B}}{\textsf{A} - \textsf{D}}\,\, (\b +\c)   \ ,
\nonumber\\
\D \l_{\rm SME}^- ~&=~ \d \,\,-\, \frac{\textsf{B}}{\textsf{A} - \textsf{D}}\,\, (\b + \c)  \ .
\end{align*}
which are used to deduce (\ref{c18}) and (\ref{c19}) above.   Usually $V_{\rm SME}^e(R)$ here is symmetric
so $\c = \b$ but in section \ref{sect 4} we encounter a case where it is hermitian, $\c = \b^*$.
Also note that this method of calculating
the eigenvalues gives the same result at $O({\rm SME})$ as evaluating the expectation values of the 
SME perturbation in the hyperfine-Zeeman eigenstates, without the need to explicitly calculate the 
corresponding mixing angles.\label{FNalgebra} }
\begin{equation}
V_{{\rm SME}\,\pm}^{e\,E}(R)~=~ \frac{1}{m_e^2}\,\Big(
\frac{1}{3} \,\tr\,\langle\,p_a\,p_b\,\rangle\, \tr\,\tilde{E}_{ij}^e ~+~ 
\frac{1}{6} \,\tr_Y\langle\,p_a\,p_b\,\rangle\, \tr_Y\tilde{E}_{ij}^e \,\, \wh{c}_{N M_J}^{\,\pm}(B) \, \Big) \ ,
\label{c17}
\end{equation}
with
\begin{align}
\wh{c}_{NM_J}^{\,+}(B) ~&=~ \frac{1}{(2N+1)(2N+3)}\,\big[(N+\thalf)(N+\tfrac{3}{2}) - 3 M_J^2\big] \ ,  
\nonumber \\[8pt]
&+~24 \,\frac{\big[(N+\thalf)^2 - M_J^2\big]}{(2N+1)^3 (2N-1) (2N+3)}\,\, 
\frac{1}{c_e(v,N)}\, \big(g_e + g_m(v,N)\big) \mu_B B\, M_J  \ ,
\label{c18}
\end{align}
and 
\begin{align}
\wh{c}_{NM_J}^{\,-}(B) ~&=~ \frac{1}{(2N-1)(2N+1)}\,\big[(N-\thalf)(N+\thalf) - 3 M_J^2\big] 
\nonumber \\[8pt]
&-~24 \,\frac{\big[(N+\thalf)^2 - M_J^2\big]}{(2N+1)^3 (2N-1) (2N+3)}\,\, 
\frac{1}{c_e(v,N)}\, \big(g_e + g_m(v,N)\big) \mu_B B\, M_J  \ ,
\label{c19}
\end{align}
while for the unmixed states, the $\wh{c}_{NM_J}$ factor is independent of the
magnetic field:
\begin{equation}
\wh{c}_{NM_J} ~=~ - \frac{N}{(2N+3)} \ .
\label{c20}
\end{equation}
Notice that the leading terms in $\wh{c}_{NM_J}^{\,\pm}$ can be written as $\frac{1}{4} \,\frac{1}{J(J+1}\,
\big[J(J+1) - 3M_J^2\big]$ with $J= N\pm \thalf$ respectively. 

This magnetic field dependence of the SME potential $V_{{\rm SME}\,\pm}^{e\,E}(R)$ 
of course arises purely from the mixing of the states due to the hyperfine-Zeeman
interactions. These terms are proportional to $\tr_Y\tilde{E}_{ij}^e\, \mu_B B\, M_J$
but arise with a distinctive $N$-dependence.  
While high-precision spectroscopy will use combinations of transitions with different $\D M_J$ 
to attempt to cancel the Zeeman effect \cite{SchillerKorobov2018, SAS2024}, 
this SME-specific $N$-dependence means these terms may still contribute even to 
otherwise Zeeman-free combinations of transitions. This important issue will
be discussed in detail elsewhere.

Anticipating the discussion in section \ref{sect 6},
it is immediately clear that while high-precision spectroscopy will use combinations of
transitions with different $\D M_J$ to attempt to cancel the Zeeman effect, this
SME-specific $N$-dependence means that these terms will still contribute even to 
otherwise Zeeman-free combinations of transitions.

\vskip0.2cm

For high magnetic fields, the hyperfine-Zeeman eigenstates just reduce to the
$|v\,N\,M_N\,M_S\rangle$ states, for which the SME potential is simply
\begin{equation}
V_{{\rm SME}\,\pm}^{e\,E}(R)~=~ \frac{1}{m_e^2}\,\Big(
\frac{1}{3} \tr\,\langle\,p_a\,p_b\,\rangle\, \tr\,\tilde{E}_{ij}^e ~+~ 
\frac{1}{6} \,\tr_Y\langle\,p_a\,p_b\,\rangle\, \tr_Y\tilde{E}_{ij}^e \,\, c_{N M_N} \, \Big) \ ,
\label{c21}
\end{equation}
with $c_{NM_N}$ given in (\ref{c13}). This was the case analysed in detail in Paper I.

\vskip0.3cm

The direct contribution of the proton SME couplings in the Born-Oppenheimer analysis
is described in Paper I and, while conceptually different, involves a similar calculation 
encompassing matrix elements in the mixed hyperfine-Zeeman states. 
In the notation introduced there, we require
\begin{equation}
\D E_{\rm SME}^n ~=~ \frac{2}{m_p^2}\, E_{ij}^p\,\langle v\,N\,J'\,M_J|\,P_i P_j\,
|v\,N\,J\,M_J\rangle \ ,
\label{c22a}
\end{equation}
where we resolve the relative motion of the protons to that of a single particle with
momentum ${\bf P}$ and reduced mass $\m = m_p/2$. 

Here, we can expand the product of momenta in an analogous way to (\ref{c7}),
that is \cite{Yoder:2012ks,Shore:2024ked}
\begin{equation}
P_i\,P_j ~=~ |{\bf P}|^2 \,\big( \frac{1}{3} \d_{ij} ~+~ C_{ij}^M\, Y_{2M}(\theta,\phi)\,\big) \ .
\label{c22}
\end{equation}
Then, since $K=|{\bf P}|^2/2\mu$ is the kinetic energy of the protons, and is independent 
of the spin state, we have
\begin{equation}
\D E_{\rm SME}^n ~=~ \frac{2}{m_p}\, \langle v\,N|\,K\,|v\,N\rangle\,\,
\Big(\frac{1}{3} \tr\,E_{ij}^p ~+~ C_{ij}^M\, E_{ij}^p\, \langle v\,N\,J'\,M_J|\,Y_{2M}(\theta,\phi)\,
|v\,N\,J\,M_J\rangle \, \Big) \ .
\label{c23}
\end{equation}

The evaluation of the expectation value of the spherical harmonic, which physically is describing 
the dynamics of the orientation of the molecular axis in the relevant angular momentum
state, now repeats the calculation following (\ref{c10}) above.  We may again specialise
to the case $M=0$, and the $c_{NM_J}$ factors are found in the same way. 
We therefore find, in the mixed hyperfine-Zeeman eigenstates,
\begin{equation}
\D E_{\rm SME \,\pm}^n ~=~ \langle v\,N|\,K\,|v\,N\rangle \,\, \tilde{V}_{\rm SME\,\pm}^n \ ,
\label{c24}
\end{equation}
with 
\begin{equation}
\tilde{V}_{\rm SME\,\pm}^n ~=~ \frac{2}{3}\,\frac{1}{m_p}\,\Big[ \,\tr\,E_{ij}^p ~-~ 
\tr_Y E_{ij}^p\, \, \wh{c}_{NM_J}^{\,\pm}(B) \Big]\ ,
\label{c25}
\end{equation}
with $\wh{c}_{NM_J}$ as in (\ref{c20}) for the unmixed states.  
For large applied magnetic fields, $\wh{c}_{NM_J}^{\,\pm}(B) \rta c_{NM_N}$ as described above. 

\vskip0.2cm
Finally, we can re-express these results expressed in terms of $E_{ij}$ with their
equivalents with the spin-independent Lagrangian couplings $c_{\m\n}$ and $a_{\m\n\l}$.
The dictionary, first in terms of the spherical tensor description, with $\VV_{njm}^{\NR} \,=\, c_{njm}^{\NR}
- a_{njm}^{\NR}$, is
\begin{equation}
\tr\, {E}_{ij} \,=\, -3m^2\, \frac{1}{\sqrt{4\pi}}\,\big(c_{200}^{\rm NR} \,-\, a_{200}^{\rm NR}\big) \ , 
~~~~~~~~~~~~
\tr_Y {E}_{ij} \,=\, 3m^2\, \sqrt{\frac{5}{4\pi}}\,\big(c_{220}^{\rm NR} \,-\, a_{220}^{\rm NR}\big)  \ .
~~~~\\[3pt]
\label{c26}
\end{equation}
then with the original SME couplings in (\ref{a1}),  
\begin{align}
\frac{1}{\sqrt{4\pi}}\,c_{200}^{\rm NR} \,&=\,\,\frac{1}{3m}\big(c_{ii} \,+\, \tfrac{3}{2}c_{00}\big) 
~=~ \frac{5}{6m} c_{00}\ ,
~~~~~~~~~~~
\frac{1}{\sqrt{4\pi}}\,a_{200}^{\rm NR} \,=\,\, a_{0ii}  \,+\, a_{000} \ , \nonumber \\[5pt]
\sqrt{\frac{5}{4\pi}}\,c_{220}^{\rm NR} \,&=\, -\frac{1}{3m}\, \tr_Y c_{ij} \ ,
~~~~~~~~~~~~~~~~~~~~~~~~~~~~~~~
\sqrt{\frac{5}{4\pi}}\,a_{220}^{\rm NR} \,=\, - \tr_Ya_{0ij} \ ,
\label{c27}
\end{align}
since we may assume the spacetime trace of $c_{\m\n}$ vanishes (see \cite{Colladay:1998fq}).

\section{Lorentz and CPT violation in the $\textbf{H}_2^{\,+}$ and
${\overline{\textbf{H}}_2^{\,-}}$ spectrum -- spin-dependent couplings}\label{sect 4}

In this section, we extend the analysis of Lorentz and \textsf{CPT} violation in the ${\rm H}_2^{\,+}$
spectrum to include the spin-dependent couplings $B_k$ and $F_{ijk}$ in (\ref{bb1}),
corresponding to the couplings $b_\m, \,g_{\m\n\l}, \,d_{\m\n}, \,H_{\m\n}$ in the fundamental
SME Lagrangian (\ref{a1}). 

We again focus on Para-${\rm H}_2^{\,+}$ as the simplest and most experimentally favoured case,
leaving comments on the extension to Ortho-${\rm H}_2^{\,+}$ until later. 
This means that the only spin-dependence arises from the electron SME couplings
$B_k^e$ and $F_{ijk}^e$.  The main technical task of this section is therefore to evaluate 
the contribution to the inter-nucleon potential $V_{\rm SME}^e(R)$ from the momentum-dependent
$F_{ijk}^e$ terms in (\ref{bb1}).

\subsection{SME $B_k^e$ couplings}\label{sect 4.1}

First though, we consider the momentum-independent term,
\begin{equation}
H_{\rm SME}^{e\,B} ~=~~ 2 B_k^e\, S_k~~ =~~ 2 B_3^e\, S_3 \,+\, 
\big(B_+^e\, S_- \,+\, B_-^e \,S_+\big) \ ,
\label{d1}
\end{equation}
in the SME Hamiltonian. Clearly this just gives a direct addition $\D E_{\rm SME}^{e\,B}$
to the energy in the hyperfine states, where
\begin{equation}
\D E_{\rm SME}^{e\,B} ~=~ 2 B_k^e\, \langle v\,N\,J'\,M_J|\,S_k\,|v\,N\,J\,M_J\rangle \ .
\label{d2}
\end{equation}

As always, to evaluate these matrix elements we first expand the hyperfine states in terms of the 
basis states $|v\,N\,M_N\,M_S\rangle$ as in (\ref{b4}).  It is then clear that the operators $S_{\pm}^e$ 
only contribute to matrix elements with $\D M_J \neq 0$ and so, by the argument following (\ref{c11}),
only give corrections to the energy levels at $O({\rm SME})^2$.  The SME couplings $B_{\pm}^e$ can
therefore be neglected. This leaves only the matrix elements of $S_3$ to be evaluated, which we have
already found in (\ref{b6}) (see also (\ref{bpp4}). Indeed, no new calculations are required here. 
The SME term $B_k^e\,S_3$ 
acts entirely analogously to a background magnetic field and the required results may be read off
immediately from section \ref{sect 2}.

We therefore find,
\begin{equation}
\D E_{\rm SME}^{e\,B} ~=~ 2\,B_3^e\,\frac{1}{2N+1}\,
\begin{pmatrix}
~~M_J ~~~~~~&~~~- \sqrt{(N+\thalf)^2 - M_J^2}~~\\[10pt]
- \sqrt{(N+\thalf)^2 - M_J^2}  & -M_J \\
\end{pmatrix}
\label{d3}
\end{equation}
for the mixed $J',J=N\pm \thalf$ sector. 
Determining the corresponding eigenvalues, we find in the low magnetic field regime,
\begin{align}
\D E_{{\rm SME}\,\pm}^{e\,B} ~=~ &\pm 2\,B_3^e\, \frac{1}{2N+1}\,M_J  \nonumber \\[8pt]
&\pm 8\,B_3^e\, \frac{1}{(2N+1)^3}\,\sqrt{(N+\thalf)^2 - M_J^2}\,\,\,\frac{1}{c_e(v,N)}\,
\big(g_e + g_m(v,N)\big) \mu_B B \ ,
\label{d4}
\end{align}
to $O(\m_B B/c_e)$.  For the unmixed states with $J=N+\thalf$,
$\,M_J=\pm (N+\thalf)$ we have simply $\D E_{\rm SME}^{e\,~B} \,=\,\pm\,B_3^e$
with no $O(\m_B B/c_e)$ correction.

In the large magnetic field limit, the eigenstates are $|v\,N\,M_N\,M_S\rangle$ and we have,
\begin{equation}
\D E_{{\rm SME}}^{e\,B} ~=~ 2\,B_3^e\, M_S  \ .
\label{d5}
\end{equation}
In this case there is no $O(c_e/\m_B B)$ correction since, restricting to states with fixed $M_J$,
the matrix elements of $H_{\rm SME}^{e\,B}$ are diagonal in the $|v\,N\,M_N\,M_S\rangle$
basis (see footnote \ref{FNalgebra}).

\subsection{SME $F_{ijk}^e$ couplings}\label{sect 4.2}

Now consider the spin and momentum-dependent terms in the SME Hamiltonian,
\begin{equation}
H_{\rm SME}^{e\,F} ~=~ 2\,\frac{1}{m_e^2}\,F_{ijk}^e\, p_i\,p_j\,S_k  ~~~
=~~  2\,\frac{1}{m_e^2}\,F_{abk}^e\, p_a\,p_b\,S_k  \ ,
\label{d6}
\end{equation}
where we re-express the electron momenta in the \textsf{MOL} frame in which their
expectation values are evaluated (see Fig.~\ref{Figpp}).

In this case, we first need to calculate $V_{\rm SME}^{e\,F}({\bf R})$, extending the corresponding 
expression (\ref{c3}) for the spin-independent couplings. Here,
\begin{equation}
V_{\rm SME}^{e\,F}({\bf R}) ~=~ 2\,\frac{1}{m_e^2}\,\Big(
\frac{1}{3}\, \tr\langle\,p_a\,p_b\,\rangle\, \tr F_{abk}^e ~+~ 
\frac{1}{6}\, \tr_Y\langle\,p_a\,p_b\,\rangle\,  \tr_Y F_{abk}^e\,\Big) S_k \ 
\label{d7}
\end{equation}
where the traces act only on the first two indices of $F_{abk}^e$.  Following the analysis 
in section \ref{sect 3}, we relate these \MOL frame components to those in the \EXP frame
by,
\begin{equation}
\tr_YF_{abk}^e ~=~ - 3 \sum_M\, C_{ij}^M \, Y_{2M}(\theta,\phi)\,\, F_{ijk}^e \ ,
\label{d8}
\end{equation}
while of course $\tr\,F_{abk}^e \,=\, \tr\,F_{ijk}^e$, giving
\begin{equation}
V_{\rm SME}^{e\,F}({\bf R}) ~=~ 2\,\frac{1}{m_e^2}\,\Big(
\frac{1}{3}\, \tr\,\langle\,p_a\,p_b\,\rangle\, \tr\,F_{ijk}^e ~-~ 
\frac{1}{2}\, \tr_Y\langle\,p_a\,p_b\,\rangle\, \sum_M\, C_{ij}^M\, Y_{2M}(\theta,\phi) \,
F_{ijk}^e \, \Big) \, S_k\ .
\label{d9}
\end{equation}
Then, with $C_{ij}^0$ from (\ref{c12}) and
\begin{equation}
C_{ij}^{\pm 1} ~=~ \mp \sqrt{\frac{2\pi}{15}}\,\, \big( \d_{i\,\mp}\,\d_{j\,3} 
\,+\, \d_{i\,3}\,\d_{j\,\mp}\big) \ ,
\label{d13}
\end{equation}
and keeping only the $M=0, \pm 1$ contributions as explained below, we have
\begin{align}
V_{\rm SME}^{e\,F}({\bf R}) ~&=~ \frac{1}{m_e^2}\,\bigg[
\frac{2}{3}\, \tr\,\langle\,p_a\,p_b\,\rangle\, \tr\,F_{ij3}^e\,\,S_3 
~+~\frac{1}{3}\,\sqrt{\frac{4\pi}{5}} \,\, \tr_Y\langle\,p_a\,p_b\,\rangle\, 
\tr_Y F_{ij3}^e\, Y_{20}(\theta,\phi)\,\, S_3 ~
\nonumber \\[8pt]
&~~~~~~ +~\sqrt{\frac{1}{6}}\,\sqrt{\frac{4\pi}{5}} \,\, \tr_Y\langle\,p_a\,p_b\,\rangle\,
\Big(F_{-3+}^e \, Y_{21}(\theta,\phi) \,S_-
~+~ F_{+3-}^e \,Y_{2\,-1}(\theta,\phi) \,\, S_+ \Big)\,\,\bigg]  \ .
\label{d9a}
\end{align}

The next step  is to determine the contribution to the inter-nucleon potential 
$V_{\rm SME}^{e\,F}(R)$ by taking the matrix elements of (\ref{d9a}). We evaluate first 
in the $|v\,N\,M_N\,M_S\rangle$ basis states, then extend to the hyperfine states
$|v\,N\,J\,M_J\rangle$ using Clebsch-Gordan coefficients as usual.

The coefficient of $\tr\langle\,p_a\,p_b\,\rangle$ is straightforward since we only 
need the matrix elements of $S_k$, so this is carried out as above for $B_k^e$, with only
$\tr\,F_{ij3}^e$ contributing at $O({\rm SME})$ because of the $\D M_J=0$ criterion.

For the coefficient of $\tr_Y\langle\,p_a\,p_b\,\rangle$  we need the matrix elements of 
$Y_{2M}(\theta,\phi)$, which are given by the Gaunt integral in (\ref{c11}). Here, however, 
due to the presence of the spin operator in (\ref{d9a}), we must also take 
into account contributions with $M\neq 0$. Consider these in turn:

\noindent{(i)~~~$M=0$:~~~~In this case,
\begin{equation}
\langle v\,N\,M_N'\,M_S'|\, Y_{20}(\theta,\phi)\,S_k\,|v\,N\,M_N\,M_S\rangle ~=~
\sqrt{\frac{5}{4\pi}} \,\, c_{NM_N}\, M_S \ ,
\label{d10}
\end{equation}
since $M_N'=M_N$ and then $\D M_J=0$ requires $M_S'=M_S$, so only $S_3$ contributes.
So here, we need only
\begin{equation}
\langle v\,N\,M_N\,M_S|\, Y_{20}(\theta,\phi)\,S_3\,|v\,N\,M_N\,M_S\rangle\, C_{ij}^0\, F_{ij3}^e ~=~
-\frac{1}{3} \, c_{NM_N}\, M_S \,\tr_Y F_{ij3}^e \ .
\label{d11}
\end{equation}

\noindent{(ii)~~~$M=\pm 1$:~~~~Here, $M_N'=M_N\pm1$, so we require $M_S' = M_S\mp 1$
to maintain $\D M_J=0$. We therefore have contributions from the raising and lowering
spin operators $S_{\pm}$. For the matrix elements, we need
\begin{equation}
\langle v\,N\,M_N\pm 1\,\,M_S\mp1\,|\,Y_{2\pm 1}(\theta,\phi)\,|v\, N\,M_N\,M_S\rangle
~=~ \sqrt{\frac{5}{4\pi}} \, \, C_{N\,M_N,\,2\,\pm1}^{N\,(M_N\pm1)}\,\, C_{N\,0,\,2\,0}^{N\,0} \ .
\label{d12}
\end{equation}
Evaluating the Clebsch-Gordan coefficients using (\ref{bpp7}), we find
\begin{align}
&\langle v\,N\,M_N\pm 1\,\,M_S\mp1\,|\,Y_{2\pm 1}(\theta,\phi)\,|v\, N\,M_N\,M_S\rangle\,\,
C_{ij}^{\pm 1}\, F_{ij\pm}^e \nonumber \\[8pt]
&~~~~~~~~~~~~~~~~~~~~~~~~~~~~~~~~~~~~~ ~=~ 
- \, \frac{1}{(2N-1)(2N+3)}\,\big[(N+\thalf)^2 - M_J^2\big]\,\, M_J\,\, F_{\mp 3 \pm}   \ .
\label{d14}
\end{align}
Evidently there is no contribution from $M=2$ given the constraint $\D M_J=0$. 

Putting all this 
together, the matrix elements $\hat{V}_{\rm SME}^{e\,F}(R)$ expressed in an $|v\,N\,M_J\,M_S\rangle$
basis with fixed $M_J$, with $M_S = \pm \thalf$ rows and columns, are
\begin{align}
{\hat V}_{\rm SME}^{e\,F}(R) ~&= ~ \frac{1}{3}\frac{1}{m_e^2} \, \tr\,\langle\,p_a\,p_b\,\rangle\,
\tr\, F_{ij3}\,\,\begin{pmatrix}
~1~~&0\,\,\\[5pt]
~0~~&-1 \\
\end{pmatrix}  \nonumber \\[10pt]
& +\frac{1}{m_e^2}\, \tr_Y\langle\,p_a\,p_b\,\rangle\,\,
\begin{pmatrix}
\frac{1}{6}\,c_{N\,(M_J-\thalf)} \,\tr_Y F_{ij3}^e ~~~~~~~~~~
&\frac{\sqrt{(N+\thalf)^2 - M_J^2}}{(2N-1)(2N+3)}\,\, M_J\,\,F_{+3-} ~~\\[20pt]
\frac{\sqrt{(N+\thalf)^2 - M_J^2}}{(2N-1)(2N+3)}\,\, M_J\,\, F_{-3+} 
&-\frac{1}{6}\,c_{N\,(M_J+\thalf)} \,\tr_Y F_{ij3}^e \\
\end{pmatrix}
\label{d15}
\end{align}
For the unmixed states with $M_J=\pm (N+\thalf)$, the off-diagonal elements
vanish and both $c_{N (M_J\mp\thalf)}$ coefficients reduce to $c_{N M_N}$. 

The next step is to evaluate $V_{\rm SME}^{e\,F}(R)$ in the hyperfine basis $|v\,N\,J\,M_J\rangle$,
as required for the weak magnetic field regime. Here, analogously to (\ref{c15}) for the 
spin-independent couplings, we need
\begin{equation}
V_{\rm SME}^{e\,F}(R) ~=~ \sum_{M_S',\,M_S}\,\, C_{N\,(M_J-M_S'),\,\thalf\, M_S'}^{J'\,M_J} \,\,
C_{N\,(M_J-M_S),\,\thalf\, M_S}^{J\,M_J} \,\, \hat{V}_{\rm SME}^{e\,F}(R) \ ,
\label{d16}
\end{equation}
with $\hat{V}_{\rm SME}^{e\,F}(R)$ in (\ref{d15}). Notice that the presence of off-diagonal terms
in (\ref{d15}) means that for these terms we have to sum over $M_S$ and $M_S'$. 

We now need the sums over Clebsch-Gordan coefficients weighted by the appropriate
factors in (\ref{d15}). These are given in Appendix \ref{Appendix B}.
For the $\tr\,F_{ij3}^e$ term, we just need the weight factor $M_S$ as in (\ref{bpp4}),
while for $\tr_YF_{ij3}^3$ we need the combined weight factor $c_{N (M_J-M_S)}\, M_S$
which is given in (\ref{bpp6}). 
This leaves the coefficients of $F_{+3-}^e$ and $F_{-3+}^e$. In this case we simply need 
a single $M_S',\, M_S$ term, and using the Clebsch-Gordan coefficients in (\ref{b5}) and 
Appendix \ref{Appendix B}, we readily find,  
\begin{equation}
C_{N\,(M_J+\thalf),\,\thalf\,-\thalf}^{J'\,M_J} \,\, C_{N\,(M_J-\thalf),\,\thalf\,\thalf}^{J\,M_J}  ~=~
\frac{1}{(2N+1)} 
\begin{pmatrix}
\sqrt{(N+\thalf)^2 - M_J^2} ~~~~~&-(N+\thalf - M_J)  \\[12pt]
(N+\thalf + M_J) & \sqrt{(N+\thalf)^2 - M_J^2} \\
\end{pmatrix} 
\label{d18}
\end{equation}
for $J', J = N\pm \thalf$. This determines the coefficient of $F_{-3+}^e$, while the transpose
matrix gives the coefficient for $F_{+3-}^e$.

The simplest presentation of the full result is to express $V_{\rm SME}^{e\,F}(R)$ as a matrix 
for fixed $M_J$ with $J',J = N\pm \thalf$ as
\begin{equation}
V_{\rm SME}^{e\,F}(R) ~=~  \begin{pmatrix}
~\a~~&\b \,\,\\[8pt]
~\c~~ & \d  \\ 
\end{pmatrix}\ ,
\label{d19}
\end{equation}
with $\c = \b^*$, where
\begin{align}
\a ~&=~ \frac{2}{3} \,\frac{1}{m_e^2} \, \tr\,\langle\,p_a\,p_b\,\rangle\,
\,\tr\,F_{ij3}^e\,\,\frac{1}{2N+1}\,M_J  \nonumber \\[8pt]
&~~ + \frac{1}{m_e^2} \, \tr_Y\langle\,p_a\,p_b\,\rangle\, \, \bigg[
\, \frac{1}{3}\,\tr_Y F_{ij3}^e\,\,\frac{1}{(2N+1)(2N-1)(2N+3)} \Big(N^2 + 4N + \tfrac{3}{4} - 3M_J^2 \Big) \,M_j
\nonumber \\[8pt]
&~~ + \big(F_{+3-}^e\,+\, F_{-3+}^e\big)\,\,\frac{1}{(2N+1)(2N-1)(2N+3)} \big[(N+\thalf)^2 - M_J^2\big]\,M_J \,\bigg]\ ,
\label{d20}
\end{align}
and 
\begin{align}
\d ~&=~ - \frac{2}{3}\, \frac{1}{m_e^2} \, \tr\,\langle\,p_a\,p_b\,\rangle\, \,
\tr\,F_{ij3}^e \,\,\frac{1}{2N+1}\,M_J  \nonumber \\[8pt]
&~~ - \frac{1}{m_e^2} \, \tr_Y\langle\,p_a\,p_b\,\rangle\, \, \bigg[
\frac{1}{3} \,\tr_Y F_{ij3}^e\,\,\frac{1}{(2N+1)(2N-1)(2N+3)} \Big(N^2 - 2N - \tfrac{9}{4} - 3M_J^2 \Big) \,M_j
\nonumber \\[8pt]
&~~ + \big(F_{+3-}^e\,+\, F_{-3+}^e\big)\,\,\frac{1}{(2N+1)(2N-1)(2N+3)} \big[(N+\thalf)^2 - M_J^2\big]\,M_J \, \bigg] \ ,
\label{d21}
\end{align}
while,
\begin{align}
\Re\,\b ~&=~ \sqrt{(N+\thalf)^2 -M_J^2} \,\,\bigg(- \frac{2}{3} \,
\frac{1}{m_e^2} \, \tr\,\langle\,p_a\,p_b\,\rangle\, \, \tr\,F_{ij3}^e\,\,\frac{1}{2N+1}  
\nonumber \\[8pt]
&~ + \frac{1}{m_e^2} \, \tr_Y\langle\,p_a\,p_b\,\rangle\, \bigg[
-\frac{1}{3} \,\tr_Y F_{ij3}^e \,\, \frac{1}{(2N+1)(2N-1)(2N+3)} \Big(N^2 +N - \tfrac{3}{4} - 3M_J^2 \Big) 
\nonumber \\[8pt]
&~ + \big(F_{+3-}^e\,+\, F_{-3+}^e\big)\,\, \frac{1}{(2N+1)(2N-1)(2N+3)} \,M_J^2 \,\,\bigg]\,\bigg)\ .
\label{d21b}
\end{align}

 The expectation values in the hyperfine-Zeeman energy eigenstates are read off by 
applying footnote \ref{FNalgebra} and we find,
\begin{equation}
V_{\rm SME \,+}^{e\,F}(R) ~=~ \a ~-~ 4\,\Re\,\b\,\,\frac{1}{(2N+1)^2}\,\sqrt{(N+\thalf)^2 -M_J^2}\, \,
\frac{1}{c_e(v,N)}\,\big(g_e + g_M(v,N)\big) \m_B B \ ,
\label{d22}
\end{equation}
and
\begin{equation}
V_{\rm SME \,-}^{e\,F}(R) ~=~ \d ~+~ 4\, \Re\,\b \,\frac{1}{(2N+1)^2}\,\sqrt{(N+\thalf)^2 -M_J^2}\, \,
\frac{1}{c_e(v,N)}\,\big(g_e + g_M(v,N)\big) \m_B B   \ .
\label{d23}
\end{equation}
For the unmixed states, there is no dependence on $F_{+3-}^e$ and $F_{-3+}^e$, and no sub-leading
term of $O(B/c_e)$,  and we simply have,
\begin{equation}
V_{\rm SME }^{e\,F}(R) ~=~ \pm \frac{1}{3}\, \frac{1}{m_e^2} \, \tr\,\langle\,p_a\,p_b\,\rangle\,  \, \tr\,F_{ij3}^e
~~\mp~~
\frac{1}{6}\,\frac{1}{m_e^2} \, \tr_Y\langle\,p_a\,p_b\,\rangle\, \, \tr_YF_{ij3}^e  \,\, \frac{N}{2N+3} \ ,
\label{d24}
\end{equation}
for $M_J = \pm (N+\thalf)$ respectively.  In fact, the final factor has a direct interpretation as $c_{NN}$
since, from (\ref{c13}), $c_{N\,M_N=N} = N/(2N+3)$. The reason for this becomes clear in the large
field limit, (\ref{d28}).

These expressions certainly appear complicated in this level of generality. However, in the 
following sub-section, we show how they simplify remarkably in the case of the minimal SME,
due to the relations (\ref{d33}), (\ref{d34}) amongst the $F_{ijk}^e$ couplings.

\vskip0.3cm
In the large magnetic field limit, we may again exploit the fact that the eigenstates of the
hyperfine-Zeeman Hamiltonian are $|v\,N\,M_J\,M_S\rangle$ to calculate the leading and 
sub-leading (in $c_e/\m_B B$) corrections to $V_{\rm SME }^{e\,F}(R)$. Here, we have already evaluated 
$\hat{V}_{\rm SME }^{e\,F}(R)$ in these states, so it is straightforward to read off the large-$B$ limit.

First we evaluate the matrix elements $\langle\, H_{\rm HFS} + H_{\rm Z}\,\rangle$ in these states, which we
use as $\begin{pmatrix} \textsf{A} ~\,&\textsf{B} \\\textsf{B}~\,&\textsf{D}\end{pmatrix}$ in the 
method of footnote \ref{FNalgebra}:
\begin{align}
&\langle v\,N\,M_J\,M_S'|\,H_{\rm HFS}+H_{\rm Z}\,|v\,N\,M_J\,M_S\rangle ~=~\nonumber \\[8pt]
&\begin{pmatrix}
\thalf c_e (M_J-\thalf) +\big(\thalf g_e - g_m (M_J-\thalf)\big) \mu_B B 
& \thalf c_e \sqrt{(N+\thalf)^2 - M_J^2}  \\[15pt]
\thalf c_e \sqrt{(N+\thalf)^2 - M_J^2}   & -\thalf c_e (M_J+\thalf) -
\big(\thalf g_e +g_m (M_J+\thalf)\big) \mu_B B 
\end{pmatrix}  \nonumber \\[5pt]
\label{d25}
\end{align}

The expectation values of the inter-nucleon potential in the hyperfine-Zeeman eigenstates
are then found from footnote \ref{FNalgebra} as
\begin{align}
V_{\rm SME \,+}^{e\,F}(R) ~&=~ \frac{1}{3} \frac{1}{m_e^2}\,\tr\,\langle\,p_a\,p_b\,\rangle\,\, \tr\,F_{ij3}^e 
\nonumber \\[8pt]
&~+ \frac{1}{m_e^2} \, \tr_Y\langle\,p_a\,p_b\,\rangle\,\,\bigg[
\frac{1}{6} \, c_{N(M_J-\thalf)} \,\, \tr_Y F_{ij3}^e  \nonumber \\[8pt]
&~~~~~~~~~~~~~~~~~~~ +\frac{1}{2} \frac{c_e}{(g_e + g_m) \m_B B}\,\,
\big(F_{+3-}^e + F_{-3+}^e\big)\,\, 
\frac{\big[(N+\thalf)^2 - M_J^2\big]}{(2N-1)(2N+3)}\,\, M_J \,\bigg] 
\label{d26}
\end{align}
and
\begin{align}
V_{\rm SME \,-}^{e\,F}(R) ~&=~ -\,\frac{1}{3} \frac{1}{m_e^2}\,\tr\,\langle\,p_a\,p_b\,\rangle\,\, \tr\,F_{ij3}^e 
\nonumber \\[8pt]
&~- \frac{1}{m_e^2} \, \tr_Y\langle\,p_a\,p_b\,\rangle\,\,\bigg[
\frac{1}{6} \, c_{N(M_J+\thalf)} \,\, \tr_Y F_{ij3}^e  \nonumber \\[8pt]
&~~~~~~~~~~~~~~~~~~~ +\frac{1}{2} \frac{c_e}{(g_e + g_m) \m_B B}\,\,
\big(F_{+3-}^e + F_{-3+}^e\big)\,\, 
\frac{\big[(N+\thalf)^2 - M_J^2\big]}{(2N-1)(2N+3)}\,\, M_J \,\bigg] 
\label{d27}
\end{align}
The $O(c_e/\m_B B)$ term depends solely on the mixing of states and is proportional
to the off-diagonal term in $\hat{V}_{\rm SME}(R)$ in the $|v\,N\,M_J\,M_S\rangle$ basis. 

The unmixed states in this basis have $M_J=\pm (N+\thalf),~M_S = \pm \thalf$,
and as usual the mixing factor $\big[(N+\thalf)^2 - M_J^2\big]$ vanishes. This
leaves simply,
\begin{equation}
V_{\rm SME}^{e\,F}(R) ~=~ \pm\,\frac{1}{3} \frac{1}{m_e^2}\,\tr\,\langle\,p_a\,p_b\,\rangle\,\, \tr\,F_{ij3}^e
~\mp~ \frac{1}{6}\frac{1}{m_e^2} \, \tr_Y\langle\,p_a\,p_b\,\rangle\,
\, c_{N\,N} \,\, \tr_Y F_{ij3}^e  \ ,
\label{d28}
\end{equation}
with no $O(c_e/\m_B B)$ correction. Notice that the factor $c_{N\,N} = N/(2N+3)$, and so
$V_{\rm SME}^{e\,F}(R)$ here and in (\ref{d24}) are the same for the unmixed states, as they must be
since the states are identical, irrespective of whether they are expressed in the $|v\,N\,J\,M_J\rangle$
or $|v,N\,M_J\,M_S\rangle$ basis.

\vskip0.3cm
This completes the derivation of the contributions of the spin-dependent SME couplings to the
inter-nucleon potential $V_{\rm SME}^e(R)$, in both the large and small magnetic field regimes
including sub-leading corrections of $O(\m_B B/c_e)$ and $O(c_e/\m_B B)$ respectively.

\subsection{Minimal SME}\label{sect 4.3}

We now re-express these combinations $B_3$, $\,\tr\,F_{ij3}$, $\,\tr_YF_{ij3}$ and 
$(F_{+3-} + F_{-3+})$ of the spin-dependent SME couplings more directly in terms of the
$b_\m$, $\,g_{\m\n\l}$, $\,d_{\m\n}$ and $H_{\m\n}$ appearing in the QFT Lagrangian (\ref{a1}).

It is convenient to define the following frequently-occurring combinations of the \textsf{CPT}
odd and \textsf{CPT} even couplings for which experimental bounds are known \cite{Kostelecky:2008ts}.
In fact this also reveals some interesting further simplification.

We now restrict  to the minimal SME, with $B_k$ and $F_{ijk}$ defined in (\ref{bb1}). First define,
\begin{align}
\tilde{b}_i ~&=~ \big(b_i \,+\, \tfrac{1}{2} m\, \e_{ijk}\,g_{jk0} \big)\,-\, 
\big(m\,d_{i0} \,+\,\tfrac{1}{2}\e_{ijk}\, H_{jk} \big)\ ,
\nonumber \\[8pt]
\tilde{b}_i^* ~&=~ \big(b_i \,+\, \tfrac{1}{2} m\, \e_{ijk}\,g_{jk0} \big)\,+\,
\big(\, m\,d_{i0} \,+\,\tfrac{1}{2}\e_{ijk}\, H_{jk} \big) \ ,
\label{d29}
\end{align}
and
\begin{align}
\label{d30}
\tilde{g}_{D i} ~&=~ - b_i \,+\, m\,\e_{ijk}\big(g_{j0k} \,+\,\tfrac{1}{2}\,g_{jk0} \big)  \ ,
\nonumber \\[8pt]
\tilde{d}_i ~&=~ m\big(d_{0i} \,+\, \tfrac{1}{2}\,d_{i0}\big) \,-\, \tfrac{1}{4}\e_{ijk}\,H_{jk} \ .
\end{align}
In particular, the couplings arising here in the calculation of $V_{\rm SME}^e(R)$ are:
\begin{align}
\tilde{b}_3 ~&=~ \big( b_3 \,+\, m\,g_{120}\big) \,-\, \big(m\,d_{30} \,+\,H_{12}\big) \ ,
\nonumber \\[8pt]
\tilde{g}_{D3} ~&=~ -b_3 \,+\, m\big( g_{102} \,-\, g_{201} \,+\, g_{120}\big) \ ,
\nonumber \\[8pt]
\tilde{d}_3 ~&=~ m\big(d_{03} \,+\, \tfrac{1}{2}\,d_{30}\big) \,-\, \tfrac{1}{2} \, H_{12} \ .
\label{d31}
\end{align}

A straightforward calculation from the SME Hamiltonian (\ref{bb1}),(\ref{bb2}) 
now shows that the following 
very simple relations hold in the minimal SME:
\begin{equation}
B_3 ~=~ -\, \tilde{b}_3  \ ,
\label{d32}
\end{equation}
and
\begin{align}
\tr\,F_{ij3} ~&=~ - \tilde{g}_{D3} \,+\, \tilde{d}_3  \ , \nonumber \\[8pt]
\tr_YF_{ij3} ~&=~ - \tilde{g}_{D3} \,-\, 2\, \tilde{d}_3 \ ,
\label{d33}
\end{align}
while
\begin{equation}
\big( F_{+3-} \,+\, F_{-3+}\big) ~=~ \tilde{g}_{D3} \,+\, 2\, \tilde{d}_3 \ .
\label{d34}
\end{equation}

This shows that in the minimal SME, the momentum and spin-dependent contributions 
to $V_{\rm SME}^e(R)$ depend on just {\it two} independent couplings, the \textsf{CPT} odd
$\tilde{g}_{D3}$ and \textsf{CPT} even $\tilde{d}_3$, in two distinct combinations. 

This raises an interesting point. In principle, we can use the different $N$-\,dependence of
the coefficients of the terms in $V_{\rm SME}^{e\,F}(R)$ (see $f_{NM_J}^{\pm}$ and $f_{NM_J}^{Y\,\pm} $ 
below) to extract $\tr\,F_{ij3}^e$ and $\tr_Y F_{ij3}^e$ from combinations of transition energies,
as described in Paper I and here in section \ref{sect 6}. But then according to (\ref{d33}),
we would determine $\tilde{g}_{D3}$ and $\tilde{d}_3$ {\it independently}. A non-zero value
of $\tilde{g}_{D3}$ would indicate \textsf{CPT} violation, deduced from the spectrum of the
pure matter ion ${\rm H}_2^{\,+}$.   

However, this conclusion may simply be an artifact of restricting to the minimal SME. 
In section \ref{sect 5}, the equivalent analysis would lead us to an independent determination
of the two spherical tensor couplings, $\TT_{210}^{\NR\,(0B)} = g_{210}^{\NR\,(0B)} - H_{210}^{\NR\,(0B)}$ 
and $\TT_{210}^{\NR\,(1B)} = g_{210}^{\NR\,(1B)} - H_{210}^{\NR\,(1B)}$,
as identified in (\ref{e20}). So then we would not be able to conclude whether or not one
of the $g_{210}^{\NR}$ couplings was non-zero and so whether or not \textsf{CPT} was violated.
The issue arises because for this particular measurement, the minimal SME has only half the relevant
number of couplings as are permitted in general. This special case arises because of the two zeroes
in the dictionary in (\ref{e24}) and (\ref{e25}), which we have no obvious reason to expect to persist
in higher order. In such cases, we must apply the minimal SME with caution.

In fact, this is a common feature of effective field theories in general, where unless a
sufficiently complete number of operators are included in the low-energy theory,
spurious predictions will follow from its use. It is the logic, for example,  behind the 
inclusion here of the non-minimal \textsf{CPT} odd operator with coupling $a_{\m\n\l}$.

The individual terms $F_{+3-}$ and $F_{-3+}$ are given by
\begin{align}
F_{+3-} ~&=~ \tfrac{1}{2}\, \tilde{g}_{D3} \,-\, \tfrac{i}{2} m \,\tr_Yg_{i0j} \,+\, \tilde{d}_3 \ ,
\nonumber \\[8pt]
F_{-3+} ~&=~ \tfrac{1}{2}\, \tilde{g}_{D3} \,+\, \tfrac{i}{2} m \,\tr_Yg_{i0j} \,+\, \tilde{d}_3 \ ,
\label{d35}
\end{align}
where of course the imaginary part proportional to $\tr_Y g_{i0j}$ cancels in the sum, which 
controls the physical energy levels. 

The couplings $|\tilde{g}_{Di}|$ and $|\tilde{d}_{i}|$ are constrained by several experiments, 
notably spin-precession and clock comparison experiments, with already very stringent bounds 
of $O(10^{-22} \,{\rm GeV})$ \cite{Kostelecky:2008ts}.
For $|\tilde{b}_i|$, even tighter bounds of up to $10^{-24}\,{\rm GeV}$ are quoted in \cite{Kostelecky:2008ts}.
Weaker bounds may be deduced from atomic hydrogen and antihydrogen spectroscopy,
from ${\rm n}S$--${\rm n}' P$ or $D$ and ground-state hyperfine transitions respectively.

\vskip0.3cm
For completeness, we also record here some further results which are useful in the comparison
of the SME Hamiltonian written in terms of these Cartesian tensor couplings and their equivalent
expression as spherical tensors.  The relations with spherical tensors are discussed in section 
\ref{sect 5.2}.

In particular, we find
\begin{align}
B_{\pm} ~&=~  -(b_{\pm} \,\mp\, im\,g_{\pm 30}) \,+\, (m d_{\pm 0} \,\mp\, i\,H_{\pm 3}) \ ,
\nonumber \\[8pt]
\tr\, F_{ij\pm} ~&=~ -\, \tilde{g}_{D\pm} \,+\, \tilde{d}_{\pm}  \ ,
\nonumber \\[8pt]
\tr_YF_{ij\pm} ~&=~ \tfrac{1}{2}\, \tilde{g}_{D\pm} \,+\, \tilde{d}_{\pm}  
\,\mp\, 3i m\big(g_{\pm 03} \,+\, g_{30\pm} \big)  \ .
\label{d37}
\end{align}

\vskip0.3cm
As indicated earlier, we can now use the minimal SME relation
$(F_{+3-}^e + F_{-3+}^e) = - \tr_Y F_{ijk}^e$
from (\ref{d33}), (\ref{d34}) to simplify our earlier expressions 
(\ref{d19}) -- (\ref{d23}) for $V_{\rm SME}^{e\,F}(R)$.

In this case, remarkable cancellations occur between the two terms in the coefficient
of $\tr_Y\langle\,p_a\,p_b\,\rangle$.  Re-expressing $\a$, $\d$ and $\b$
in terms of the couplings $\tilde{g}_{D3}$ and $\tilde{d}_3$, we now find simply,
\begin{align}
\a ~&=~\frac{1}{3} \,\frac{1}{m_e^2} \, \tr\,\langle\,p_a\,p_b\,\rangle \,
\big(-\tilde{g}_{D3} \,+\, \tilde{d}_3\big) \, \,\frac{2}{2N+1} \,\, M_J  
\nonumber \\[8pt]
&~~+~ \frac{1}{6} \,\frac{1}{m_e^2} \, \tr_Y\langle\,p_a\,p_b\,\rangle\,
\big(\tilde{g}_{D3} \,+\, 2 \tilde{d}_3\big) \,\, \frac{2N}{(2N+1)(2N+3)}\,\, M_J \ ,
\label{d38}
\end{align}
and
\begin{align}
\d ~&=~-\frac{1}{3} \,\frac{1}{m_e^2} \, \tr\,\langle\,p_a\,p_b\,\rangle \,
\big(-\tilde{g}_{D3} \,+\, \tilde{d}_3\big) \, \,\frac{2}{2N+1} \,\, M_J  
\nonumber \\[8pt]
&~~-~ \frac{1}{6} \,\frac{1}{m_e^2} \, \tr_Y\langle\,p_a\,p_b\,\rangle\,
\big(\tilde{g}_{D3} \,+\, 2 \tilde{d}_3\big) \,\, \frac{2(N+1)}{(2N-1)(2N+1)}\,\, M_J \ ,
\label{d39}
\end{align}
while
\begin{align}
\Re\,\b ~&=~\frac{1}{3} \,\frac{1}{m_e^2} \, \tr\,\langle\,p_a\,p_b\,\rangle \,
\big(-\tilde{g}_{D3} \,+\, \tilde{d}_3\big) \, \,\frac{(- 2)}{2N+1} \,\, \sqrt{(N+\thalf)^2 -M_J^2} 
\nonumber \\[8pt]
&~~-~ \frac{1}{6} \,\frac{1}{m_e^2} \, \tr_Y\langle\,p_a\,p_b\,\rangle\,
\big(\tilde{g}_{D3} \,+\, 2 \tilde{d}_3\big) \,\, \frac{1}{2(2N+1)}\,\, \sqrt{(N+\thalf)^2 -M_J^2} \ .
\label{d40}
\end{align}
Notice that the diagonal terms $\a$ and $\d$ are proportional to $M_J$, whereas the
off-diagonal term instead has the common factor $\sqrt{(N+\thalf)^2 -M_J^2}$
\,\,(compare (\ref{b6})), which vanishes for the unmixed states where
$M_J = \pm (N+\thalf)$.

We can now write ({\ref{d22})--(\ref{d24}) in a particularly compact form, analogous to 
(\ref{c17})--(\ref{c20}):
\begin{align}
V_{\rm SME\,\pm}^{e\,F}(R) ~&=~ \frac{1}{3} \,\frac{1}{m_e^2} \, \tr\,\langle\,p_a\,p_b\,\rangle \,
\big(-\tilde{g}_{D3} \,+\, \tilde{d}_3\big) \,\, f_{NM_J}^\pm(B) 
\nonumber \\[8pt]
&~~~~~~~~~~~
+~\frac{1}{6} \,\frac{1}{m_e^2} \, \tr_Y\langle\,p_a\,p_b\,\rangle\,
\big(\tilde{g}_{D3} \,+\, 2 \tilde{d}_3\big) \,\,f_{NM_J}^{Y\,\pm}(B) \ ,
\label{d41}
\end{align}
where
\begin{align}
f_{NM_J}^+ ~&=~  \frac{2}{2N+1} \,M_J ~+~ \frac{8}{(2N+1)^3} \,\big[(N+\thalf)^2 - M_J^2\big]
\,\frac{1}{c_e(v,N)}\, \big(g_e + g_m(v,N)\big) \m_B B  \ , \nonumber \\[10pt]
f_{NM_J}^{Y\,+} ~&= \frac{2N}{(2N+1)(2N+3)} \,M_J  \nonumber\\[10pt]
&~~~~~~~~~~~~ ~-~ \frac{2}{(2N+1)^3} \,\big[(N+\thalf)^2 - M_J^2\big]
\,\frac{1}{c_e(v,N)}\, \big(g_e + g_m(v,N)\big) \m_B B  \ , 
\label{d42}
\end{align}
and
\begin{align}
f_{NM_J}^- ~&= - \frac{2}{2N+1} \,M_J ~-~ \frac{8}{(2N+1)^3} \,\big[(N+\thalf)^2 - M_J^2\big]
\,\frac{1}{c_e(v,N)}\, \big(g_e + g_m(v,N)\big) \m_B B  \ , \nonumber \\[10pt]
f_{NM_J}^{Y\,-} ~&= - \frac{2(N+1)}{(2N-1)(2N+1)} \,M_J \nonumber \\[10pt]
&~~~~~~~~~~~~ ~+~ \frac{2}{(2N+1)^3} \,\big[(N+\thalf)^2 - M_J^2\big]
\,\frac{1}{c_e(v,N)}\, \big(g_e + g_m(v,N)\big) \m_B B  \ , 
\label{d43}
\end{align}
while for the unmixed states, we simply write $V_{\rm SME}^{e\,F}(R)$ as in (\ref{d41})
with 
\begin{equation}
f_{NM_J}~=~\pm 1 \ , ~~~~~~~~~~~~~f_{NM_J}^Y ~=~ \pm \frac{N}{2N+3} \ .
\label{d44}
\end{equation}

We return to these expressions in section \ref{sect 6} where we show how, together
with the analysis of Paper I, they immediately give the SME contributions to the 
hyperfine-Zeeman energy levels $E_{vNJM_J}$, including mixing.

\vskip1.5cm

\section{The SME spherical tensor coupling formalism}\label{sect 5}

The SME Hamiltonian and associated parameter constraints may alternatively be described in
terms of the spherical tensor representation of the Lorentz and \textsf{CPT} couplings most widely
used in the analysis of spectroscopy in this theory. This has a number of virtues, notably
allowing a compact description of higher dimensional operators in the extension beyond the
minimal SME. However, it also necessarily obscures the immediate relation with the couplings
as they appear in the QFT Lagrangian.  In order to make contact with the extensive body of 
literature on the SME using this formalism, in this section we show how our results for the
molecular ion spectrum may be rewriten in the spherical tensor form and give a detailed 
dictionary between the Cartesian and spherical tensor couplings.

The non-relativistic SME Hamiltonian in the spherical tensor formalism is presented in
a very general form including higher-dimensional operators in \cite{Kostelecky:2013rta}
and we use this extensively in what follows. In that work, however, the initial Hamiltonian
is written with the couplings defined in what in our context we are calling the \textsf{MOL}
frame, but with the spin operators in a `helicity frame' \textsf{HEL} with $z$-axis aligned 
with the electron momentum. The first task is therefore to find the SME Hamiltonian for 
the electron with both the spin operators and couplings expressed in the \textsf{MOL} frame. 
This construction is described in Appendix \ref{Appendix A}. We distinguish the couplings in the
\textsf{MOL} frame with circumflex accents, {\it e.g.}~$\widehat{\cal V}_{njm}^{\NR}$.

We find the following result, (\ref{app18}), for the SME Hamiltonian $H_{\rm SME}$  in the \textsf{MOL} frame, 
which in the Born-Oppenheimer analysis is incorporated in the electron Schr\"odinger equation 
leading to the potential $V^e_{\rm SME}(\bf R)$:
\begin{align}
H_{\rm SME} ~=~ &-\sqrt{\frac{1}{4\pi}}\, \bigg[\, \wh{\VV}_{000}^{\NR} ~+~ 
\tr\,p_a\,p_b\,\wh{\VV}_{200}^{\NR}
~-~ \sqrt{5}\,\frac{1}{2}\,\tr_Y p_a\,p_b\,\wh{\VV}_{220}^{\NR} \nonumber \\[8pt]
&+ \sqrt{3}\, \wh{\s}_s \d_{sm}\, \wh{\TT}_{01m}^{\NR\,(0B)} 
~+~ \sqrt{3}\, \wh{\s}_s \big( \,\frac{1}{3} \,\tr\,p_a\,p_b\, \d_{sm} ~+~ 
\frac{1}{6}\, \tr_Y p_a\,p_b\, \textsf{Y}_{sm} \big)\,\wh{\TT}_{21m}^{\NR\, (0B)} \nonumber \\[8pt]
&+~ \sqrt{3}\, \wh{\s}_s \big( \,\frac{2}{3} \,\tr\,p_a\,p_b\, \d_{sm} ~-~ 
\frac{1}{6}\, \tr_Y p_a\,p_b\, \textsf{Y}_{sm} \big)\,\wh{\TT}_{21m}^{\NR\, (1B)} \nonumber \\[8pt]
&+~ \sqrt{5}\, \frac{1}{2}\, \tr_Y p_a\,p_b\, \,\wh{\s}_s\,\textsf{T}_{sm}\, 
i \wh{\TT}_{22m}^{\NR\,(1E)} \, \bigg]\ .   
\label{e1}
\end{align}
Recall that here we have omitted the $\wh{{\cal T}}^{\rm NR(0B)}_{23m}$ and
$\widehat{{\cal T}}^{\rm NR(1B)}_{23m}$ terms since these couplings correspond to higher-dimensional
operators not present in the minimal SME, which is our primary focus here. We have also kept only
the electron momentum factors $\tr\,p_a\,p_b$ and $\tr_Yp_a\,p_b$ which give non-vanishing
expectation values in the molecular ion $1s\s_g$ state.

\vskip1cm

\subsection{SME Hamiltonian $H_{\rm SME}$ in the {\textrm {\textsf{EXP}}} frame.}\label{sect 5.1}

As discussed earlier, spectroscopic measurements are made in the \textsf{EXP} frame of reference
with the quantisation axis aligned with the externally applied magnetic field. In this frame, the molecule 
rotates and vibrates and is described, for Para-${\rm H}_2^{\,+}$, by the states $|v\, N\,J\,M_J\rangle$
or $|v\,N\,M_N\, M_J\rangle$ as described in section \ref{sect 2}. The electron spin quantum number
$M_S$ is referred to the 3-axis in the \textsf{EXP} frame, while ${\bf N}$ is the total molecular 
orbital angular momentum in this frame. We therefore need to transform the SME Hamiltonian
(\ref{e1}) from the \textsf{MOL} frame to the \textsf{EXP} frame in order to evaluate 
$V_{\rm SME}^e(\bf R)$.

The transformations under the rotation from \textsf{MOL} to \textsf{EXP} of the spin operators 
and SME couplings written as spherical tensors  are described by Wigner matrices as follows:
\begin{align}
\widehat{\s}_{s'} ~&=~ \s_s\, d_{s s'}^1(\theta) \,e^{-i s \phi} \ , \nonumber \\[8pt]
\wh{\cal V}_{njm'} ~&=~ {\cal V}_{njm}\, d_{mm'}^j(\theta)\, e^{i m\phi} \ , \nonumber \\[8pt]
\wh{\cal T}_{njm'} ~&=~ {\cal T}_{njm}\, d_{mm'}^j(\theta)\, e^{i m\phi} \ ,
\label{e2}
\end{align}
where $(\theta,\phi)$ are the standard spherical polar angles specifying the orientation of the
molecular axis in the \EXP frame.

It will be useful to record here some general properties of Wigner matrices which are used 
extensively below. They are defined in terms of angular momentum states $|j\,m\rangle$ by,
\begin{equation}
d_{m' m}(\theta) ~=~ \langle j\,m'|\,e^{-i\,\theta\, J_z}\,|j\,m\rangle \ ,
\label{e3}
\end{equation}
and are related to the spherical harmonics by,
\begin{equation}
Y_{jm}(\theta,\phi) ~=~ 
\sqrt{\frac{2j+1}{4\pi}} \,\, d_{m0}^j(\theta) \,e^{im\phi} \ .
\label{e4}
\end{equation}
They satisfy orthonormality relations,
\begin{equation}
\sum_k\, d_{m'k}^j \, d_{mk}^j ~=~ \d_{m' m} \ ,
\label{e5}
\end{equation}
and satisfy the useful identities,
\begin{equation}
d_{m' m}^j ~=~ (-1)^{m - m'}\, d_{m m'}^j ~=~ d_{-m,-m'}^j \ .
\label{e6}
\end{equation}
As reducible representations of the rotation group, products of the $d_{m' m}^j(\theta)$ may 
be expanded in terms of a sum of irreducible representations using Clebsch-Gordan coefficients,
\begin{equation}
d_{s' m'}^{j'}(\theta)\, d_{s m}^j(\theta) ~=~ \sum_J \, C_{j' \,s',\, j \,s}^{J \,(s'+s)}  ~ C_{j'\, m',\,j\,m}^{J\,(m+m')}~
d_{s'+s,\, m'+m}^J (\theta)  \ ,
\label{e7}
\end{equation}
with the sum over $J = |j' - j|, \ldots (j'+j)$.

\vskip0.2cm

First, consider the spin-independent couplings in (\ref{e1}). $\wh{\VV}_{000}^{\NR}$ and 
$\wh{\VV}_{200}^{\NR}$ are invariant under the rotation (\ref{e2}), while 
$\wh{\VV}_{220}^{\NR} \,=\, \VV_{22m}^{\NR}\,d_{m\,0}^2 \,e^{im\phi}$. So using (\ref{e4}) to
re-express in terms of spherical harmonics, and replacing the  electron
momentum factors with their expectation values, we find the contribution of the 
spin-independent couplings to the potential $V_{\rm SME}^e({\bf R})$:
\begin{equation}
V_{\rm SME}^{e\,\VV}({\bf R}) ~=~ - \sqrt{\frac{1}{4\pi}}\, \Big(\VV_{000}^{\NR} 
~+~ \tr\langle\,p_a\,p_b\,\rangle\, \VV_{200}^{\NR}\,\Big) 
~+~ \frac{1}{2}\,\tr_Y\langle\,p_a\,p_b\, \rangle\, Y_{2m}(\theta,\phi)\, \VV_{22m}^{\NR} \ .
\label{e8}
\end{equation}

For the spin-dependent terms, those where the spin and coupling indices are contracted
with $\d_{sm}$ are clearly invariant, since in that case both transform with Wigner matrices
with $j=1$ and we can use the orthonormality relation (\ref{e5}). 
The main technical problem is therefore to transform the terms of the form 
$\wh{\s}_s \,{\textsf{Y}}_{sm}\, \wh{\TT}_{2jm}^{\NR} $ and
$\wh{\s}_s \,{\textsf{T}}_{sm}\, \wh{\TT}_{2jm}^{\NR} $ in (\ref{e1}) to the \EXP frame.

First, we may write, for example,
\begin{equation}
\wh{\s}_s \,\textsf{Y}_{sm}\,\wh{\TT}_{21m}^{\NR\,(0B)} ~=~ 
\wh{\s}_m\,\wh{\TT}_{21m}^{\NR\,(0B)} ~-~ 3\, \wh{\s}_0\,\wh{\TT}_{210}^{\NR\,(0B)} \ ,
\label{e9}
\end{equation}
where the first term is invariant.
For the second term, using the product formula (\ref{e7}) and re-expressing in terms of spherical
harmonics, we find
\begin{align}
\wh{\s}_0 \,\wh{\TT}_{210}^{\NR\,(0B)} ~&=~ \s_s\,\big(e^{-is\phi}\, d_{s\,0}^1(\theta)\,d_{m\,0}^1(\theta) 
\,e^{im\phi}\big)\, \TT_{21m}^{\NR\,(0B)}  \nonumber \\[8pt]
&= \s_s\,\Big((-1)^s\,\sum_{J=0,1,2} \sqrt{\frac{4\pi}{2J+1}}\,\, C_{1\,0,\,1\,0}^{J\,0}\, C_{1\,-s,\,1\,m}^{J\,M}\,
Y_{JM}(\theta,\phi) \Big)\, \TT_{21m}^{\NR\,(0B)} \ ,
\label{e10}
\end{align}
with $M= -s + m$.  Here, we have used the identity $d_{s0}^1 \,=\,(-1)^s d_{-s,0}^1$ 
to recast the Wigner matrix product in a convenient form to simplify the subsequent calculation.
Then, since the Clebsch-Gordan coefficient $C_{1\,0,\,1\,0}^{1\,0} = 0$, only $J=0$ and $J=2$ contribute
to the sum in (\ref{e10}).  The $J=0$ contribution precisely cancels the first term on the r.h.s. of (\ref{e9})
leaving simply
\begin{equation}
\wh{\s}_s \,\textsf{Y}_{sm} \,\wh{\TT}_{21m}^{\NR\,(0B)} ~=~
- 3 \sqrt{\frac{4\pi}{5}}\sqrt{\frac{2}{3}} \,\,\, \s_s \Big((-1)^s \, \sum_M\, C_{1\,-s,\,1\,m}^{2\,M}
\, Y_{2M}(\theta,\phi) \Big)\, \TT_{21m}^{\NR\,(0B)} \ ,
\label{e11}
\end{equation}
where we have displayed the sum over $M = -2, \ldots 2$ (which is anyway implicit in the 
sums over $s,m$) simply for clarity. Substituting back into $\wh{H}_{\rm SME}$, we therefore find 
the contribution of the $(0B)$ and $(1B)$ couplings with $j=1$ to $V_{\rm SME}^e({\bf R})$ as
\begin{align}
V_{\rm SME}^{e\,(0B),(1B)} ~&=~ -\sqrt{\frac{3}{4\pi}} \, \Big[ \, \s_m\,\,\TT_{01m}^{\NR\,(0B)}
~+~ \frac{1}{3} \, \tr\langle \,p_a\,p_b\,\rangle\,\s_m\,\big(\TT_{21m}^{\NR\,(0B)} \,+\,
2\,\TT_{21m}^{\NR\,(1B)} \big)\,\Big]
\nonumber \\[8pt]
&-\sqrt{\frac{1}{10}}\,\tr_Y\langle \,p_a\,p_b\,\rangle\, \s_s\,\Big( (-1)^s \sum_M\, C_{1\,-s,\,1\,m}^{2\,M}
\,Y_{2M}(\theta,\phi)\,\Big)\,\big(\TT_{21m}^{\NR\,(0B)} \,-\, \TT_{21m}^{\NR\,(1B)} \big) \ .
\label{e12}
\end{align}

Next consider the $\TT_{22m}^{\NR\,(1E)}$ couplings. Here, we need to evaluate directly:
\begin{equation}
\wh{\s}_s \,\textsf{T}_{sm} \,i\wh{\TT}_{22m}^{\NR\,(1E)} ~=~
\s_s \,e^{-is\phi}\,(-1)^{s-1} \,\big(\, d_{-s,-1}^1\,d_{m, 1}^2 \,-\, d_{-s,1}^1\, d_{m,-1}^2\big)
e^{im\phi} \, i\TT_{22m}^{\NR\,(1E)} \ ,
\label{e13}
\end{equation}
where we again use a Wigner matrix identity, $d_{s,1}^1 = (-1)^{s-1} d_{-s,-1}^1$, to put the product
into a convenient form to re-express in spherical harmonics.  Then,
\begin{equation}
\big(\, d_{-s,-1}^1\,d_{m, 1}^2 \,-\, d_{-s,1}^1\, d_{m,-1}^2\big) ~=~
\sum_{J=1,2,3} \, C_{1\,-s,\,2\,m}^{J\,(-s+m)}
\,\, \big(C_{1\,-1,\,2\,1}^{J\,0}\,-\, C_{1\,1,\,2\,-1}^{J\,0}\big) \, d_{(-s+m),0}^J \ ,
\label{e14}
\end{equation}
and explicitly evaluating the Clebsch-Gordan coefficients in brackets shows that again only 
$J=2$ gives a non-vanishing contribution, leaving
\begin{equation}
\wh{\s}_s \,\textsf{T}_{sm}\, i\wh{\TT}_{22m}^{\NR\,(1E)} ~=~
\sqrt{2} \sqrt{\frac{4\pi}{5}}\, \s_s \Big((-1)^s \,\sum_M \, C_{1\,-s,\,2\,m}^{2M}\,Y_{2M}(\theta,\phi)\,\Big) 
\, i\TT_{22m}^{\NR\,(1E)}  \ .
\label{e15}
\end{equation}
The contribution of the $(1E)$ couplings to $V_{\rm SME}^e({\bf R})$ is therefore
\begin{equation}
V_{\rm SME}^{e\,(1E)} ~=~ -\frac{1}{\sqrt{2}} \, \tr_Y\langle \,p_a\,p_b\,\rangle \,\,
\s_s \,\Big( \,(-1)^s\,\sum_M\,C_{1\,-s,\,2\,m}^{2\,M} \, Y_{2M}(\theta,\phi)\Big) \, i\TT_{22m}^{\NR\,(1E)} \ .
\label{e16}
\end{equation}

This completes the transformation of the SME Hamiltonian from the \MOL frame to the
required expression in terms of \EXP frame spins and couplings. The expectation values of the 
electron momenta $\langle\, p_a\,p_b\,\rangle$ remain of course evaluated in the \MOL frame.
Finally, therefore, collecting the results (\ref{e8}), (\ref{e12}) and (\ref{e16}), we can write the full 
contribution to the potential $V_{\rm SME}^e({\bf R})$ as:
\begin{align}
V_{\rm SME}^e({\bf R}) ~&=~  - \sqrt{\frac{1}{4\pi}}\, \Big(\VV_{000}^{\NR} 
~+~ \tr\langle\,p_a\,p_b\,\rangle\, \VV_{200}^{\NR}\,\Big) 
~+~ \frac{1}{2}\,\tr_Y\langle\,p_a\,p_b\, \rangle\, Y_{2m}(\theta,\phi)\, \VV_{22m}^{\NR}  \nonumber\\[8pt]
&-~\sqrt{\frac{3}{4\pi}} \, \Big[ \, \s_m\,\,\TT_{01m}^{\NR\,(0B)} 
~+~ \frac{1}{3} \,\, \tr\langle \,p_a\,p_b\,\rangle\,\s_m\,\big(\TT_{21m}^{\NR\,(0B)} \,+\,
2\,\TT_{21m}^{\NR\,(1B)} \big)\,\Big]   \nonumber \\[8pt]
&+~\sqrt{\frac{1}{10}}\,\,\tr_Y\langle \,p_a\,p_b\,\rangle\, \s_s\,\Big( (-1)^s \sum_M\, C_{1\,-s,\,1\,m}^{2\,M}
\,Y_{2M}(\theta,\phi)\,\Big)\,\big(\TT_{21m}^{\NR\,(0B)} \,-\, \TT_{21m}^{\NR\,(1B)} \big) \nonumber \\
&-~\sqrt{\frac{1}{2}} \,\, \tr_Y\langle \,p_a\,p_b\,\rangle \,
\s_s \,\Big( \,(-1)^s\,\sum_M\,C_{1\,-s,\,2\,m}^{2\,M} \, Y_{2M}(\theta,\phi)\Big) \, i\TT_{22m}^{\NR\,(1E)} \ .
\label{e17}
\end{align}
As we have already seen for the spin-independent couplings in Paper I, and here in sections
\ref{sect 3} and \ref{sect 4},
the contributions proportional to $\tr_{\textsf{Y}}\langle p_a\,p_b\rangle$ involve a 
non-trivial dependence on the spherical harmonics $Y_{2M}(\theta,\phi)$, which capture the 
orientation of the molecular axis in the \EXP frame.

\vskip0.3cm
Now, as discussed in sections \ref{sect 3} and \ref{sect 4}, the next step in the Born-Oppenheimer 
analysis is to evaluate $V_{\rm SME}^e({\bf R})$ in the hyperfine states $|v\, N\,J\,M_J\rangle$ to
yield the modified inter-nucleon potential itself, $V_{\rm SME}^e(R)$.  These states are linear
combinations of the states $|v\,N\,M_N\,M_S\rangle$ in which the expectation values of
the spin operators in (\ref{e17}) are calculated. In general, this involves many contributions, 
but at this point we may use the simplification introduced in section \ref{sect 4}.

As we saw there, the hyperfine and Zeeman interactions do not mix states with different $M_J$.
Any terms in $V_{\rm SME}^e({\bf R})$ with matrix elements between states with $\D M_J\neq 0$ therefore
give rise to off-diagonal entries which are purely of $O({\rm SME})$ and therefore contribute
only at $O({\rm SME})^2$ to the energy eigenvalues. These terms should therefore be neglected.

In the present formalism, the requirement $\D M_J = \D M_N + \D M_S =0$ implies that
$M+s = 0$ in (\ref{e17}), restricting the spherical harmonics that can contribute at $O({\rm SME})$.
Then, since the Clebsch-Gordan coefficients impose $M=-s + m$, we see that only the
SME couplings with $m=0$ contribute at $O({\rm SME})$. This enormously simplifies the 
application of (\ref{e17}).

The required Clebsch-Gordan coeffcients in (\ref{e17}) with $m = 0$, $M=-s$ are 
$(-1)^s \,C_{1\,-s,\,1\,0}^{2\,-s} \,=\, -\sqrt{\tfrac{1}{2}},\,\,  -\sqrt{\tfrac{1}{2}},\,\, 
 \sqrt{\tfrac{2}{3}}$ and  
$(-1)^s \,C_{1\,-s,\,2\,0}^{2\,-s} \,=\, -\sqrt{\tfrac{1}{2}},\,\,   \sqrt{\tfrac{1}{2}},\,\, 0$
for the spin index $s \,=\, 1,\,\,-1,\,\, 0$ respectively. Keeping only these terms, (\ref{e17})
reduces to,
\begin{align}
V_{\rm SME}^e({\bf R}) ~&=~  - \sqrt{\frac{1}{4\pi}}\, \Big(\VV_{000}^{\NR} 
~+~ \tr\langle\,p_a\,p_b\,\rangle\, \VV_{200}^{\NR}\,\Big) 
~+~ \frac{1}{2}\,\tr_Y\langle\,p_a\,p_b\, \rangle\, Y_{20}(\theta,\phi)\, \VV_{220}^{\NR}  \nonumber\\[8pt]
&-~\sqrt{\frac{3}{4\pi}} \, \Big[ \, \s_0\,\,\TT_{010}^{\NR\,(0B)} 
~+~ \frac{1}{3} \,\, \tr\langle \,p_a\,p_b\,\rangle\,\s_0\,\big(\TT_{210}^{\NR\,(0B)} \,+\,
2\,\TT_{210}^{\NR\,(1B)} \big)\,\Big]   \nonumber \\[8pt]
&+~ \sqrt{\frac{1}{15}}\,\,\,\tr_Y\langle \,p_a\,p_b\,\rangle\,\s_0\,
\big(\TT_{210}^{\NR\,(0B)} \,-\, \TT_{210}^{\NR\,(1B)} \big) \, Y_{20}(\theta,\phi) \nonumber \\[8pt] 
&-\,\frac{1}{2}\,\sqrt{\frac{1}{5}}\,\,\,\tr_Y\langle \,p_a\,p_b\,\rangle \,\Big[\,
\s_1\,\big(\TT_{210}^{\NR\,(0B)} \,-\, \TT_{210}^{\NR\,(1B)} \,-\, \sqrt{5}\,i \TT_{220}^{\NR\,(1E)}\big)
\, Y_{2,\,-1}(\theta,\phi) \nonumber \\[8pt]
&~~~~~~~~~~~~~~~~~~~~~~~~~ +~ 
\s_{-1}\,\big(\TT_{210}^{\NR\,(0B)} \,-\, \TT_{210}^{\NR\,(1B)} \,+\, \sqrt{5}\,i \TT_{220}^{\NR\,(1E)}\big)
\, Y_{2,\,1}(\theta,\phi) \Big]  \ .
\label{e18}
\end{align}

This expression for $V_{\rm SME}^e({\bf R})$ in terms of the spherical tensor couplings
may now be compared directly with our original forms (\ref{c14}), (\ref{d1}) and (\ref{d9a}). 
We see that they coincide precisely, given the 
identifications:
\begin{equation}
\frac{1}{m^2}\,\tr\,E_{ij} ~=~ - 3\,\sqrt{\frac{1}{4\pi}}\,\, \VV_{200}^{\NR} \ ,  ~~~~~~~~~~~~
\frac{1}{m^2}\,\tr_YE_{ij} ~=~  3\,\sqrt{\frac{5}{4\pi}}\,\, \VV_{220}^{\NR}  \ , 
\label{e19}
\end{equation}
and also $A \,=\,-\sqrt{\tfrac{1}{4\pi}}\, \VV_{000}^{\NR}$ (though this is unobservable),
and for the spin-dependent couplings,
\begin{align}
B_3 ~&=~ - \sqrt{\frac{3}{4\pi}}\,\, \TT_{010}^{\NR\,(0B)}  \ ,  \nonumber \\[8pt]
\frac{1}{m^2}\,\tr\,F_{ij3}~&=~-\sqrt{\frac{3}{4\pi}}\,\, \big(\TT_{210}^{\NR\,(0B)} \,+\, 2\,\TT_{210}^{\NR\,(1B)} \big) \ ,
\nonumber \\[8pt]
\frac{1}{m^2}\,\tr_Y\,F_{ij3}~&=~\,\,2\,\sqrt{\frac{3}{4\pi}}\,\, \big(\TT_{210}^{\NR\,(0B)} \,-\, \,\TT_{210}^{\NR\,(1B)} \big) \ ,
\nonumber \\[8pt]
\frac{1}{m^2}\, F_{-3+} ~&=~ -\sqrt{\frac{3}{4\pi}}\,\, \big(\TT_{210}^{\NR\,(0B)} \,-\, \TT_{210}^{\NR\,(1B)} \,-\, 
\sqrt{5}\,i \TT_{220}^{\NR\,(1E)}\big) \ ,
\nonumber \\[8pt]
\frac{1}{m^2}\, F_{+3-} ~&=~ -\sqrt{\frac{3}{4\pi}}\,\, \big(\TT_{210}^{\NR\,(0B)} \,-\, \TT_{210}^{\NR\,(1B)} \,+\, 
\sqrt{5}\,i \TT_{220}^{\NR\,(1E)}\big) \ . 
\label{e20}
\end{align}

\subsection{Minimal SME \,--\, equivalence with Lagrangian couplings.}\label{sect 5.2}

We can verify these identifications in the minimal SME by expressing the spherical tensor couplings
in terms of those in the original SME Lagrangian and comparing with the equivalent formulae
for $E_{ij}$, $B_k$ and $F_{ijk}$ in sections \ref{sect 3} and \ref{sect 4}.

First, for the spin-independent terms 
\begin{equation}
\VV_{njm}^{\NR} ~=~ c_{njm}^{\NR} ~-~ a_{njm}^{\NR} \ ,
\label{e21}
\end{equation}
and we can check the equivalence directly with the results already stated in (\ref{c26}) and (\ref{c27}).

For the spin-dependent terms, the tensor operators $\TT_{njm}^{\NR}$ are defined in terms of
\textsf{CPT} odd, $g_{njm}^{\NR}$, and \textsf{CPT} even, $H_{njm}^{\NR}$, operators,
\begin{equation}
\TT_{njm}^{\NR} ~=~ g_{njm}^{\NR} ~-~ H_{njm}^{\NR} \ ,
\label{e22}
\end{equation}
for each of the types $(0B)$, $(1B)$ and $(1E)$. The required expressions for the $g_{njm}^{\NR}$
and $H_{njm}^{\NR}$ couplings in terms of $b_\m$, $g_{\m\n\l}$, $d_{\m\n}$ and $H_{\m\n}$ are given in 
\cite{Kostelecky:2013rta}, but in many cases require significant calculation to extract the results 
we need since the minimal SME identifications in \cite{Kostelecky:2013rta} are quoted in
terms of dual couplings and related to the spherical tensors before taking the 
non-relativistic limit (indicated by the subscript NR here) which mixes couplings with different
values of $n$.

For the spin-dependent $\TT_{010}^{\NR}$ couplings, we need
\begin{align}
\sqrt{\frac{3}{4\pi}}\,\, g_{010}^{\NR\,(0B)} ~&=~ \sqrt{\frac{3}{4\pi}}\, \,g_{010}^{\NR\,(1B)} 
~~=~~  \thalf (\tilde{b}_3 \,+\, \tilde{b}_3^* ) 
~=~ b_3 \,+\, m\,g_{120}  \ ,  \nonumber \\[8pt]
\sqrt{\frac{3}{4\pi}}\,\, H_{010}^{\NR\,(0B)} ~&=~ \sqrt{\frac{3}{4\pi}}\,\, H_{010}^{\NR\,(1B)} 
~~=~ - \thalf (\tilde{b}_3 \,- \, \tilde{b}_3^* ) 
~=~ m\,d_{30}\,+\,H_{12} \ ,
\label{e23}
\end{align}
consistent with the identification $B_3 = - \tilde{b}_3$ in section \ref{sect 4.3}.

Next, we have the $\TT_{210}^{\NR}$ relations familiar from applications in atomic
spectroscopy:
\begin{equation}
\sqrt{\frac{3}{4\pi}}\,\, g_{210}^{\NR\,(0B)} ~=~ 0 \ , ~~~~~~~~~~~~~~~~~~~~~~
\sqrt{\frac{3}{4\pi}}\,\, g_{210}^{\NR\,(1B)} ~=~ \frac{1}{2} \, \frac{1}{m^2}\,\,\tilde{g}_{D3} \ ,
\label{e24}
\end{equation}
and
\begin{equation}
\sqrt{\frac{3}{4\pi}}\, \,H_{210}^{\NR\,(0B)} ~=~ \frac{1}{m^2}\,\, \tilde{d}_3 \ , ~~~~~~~~~~~~~~
\sqrt{\frac{3}{4\pi}}\,\, H_{210}^{\NR\,(1B)} ~=~ 0 \ ,   ~~~~~~~~
\label{e25}
\end{equation}
which together with (\ref{d33}) confirms the identifications above for $\tr\,F_{ij3}$ and 
$\tr_YF_{ij3}$.  The identity (\ref{d34}) for the sum $(F_{+3-} + F_{-3+})$ also follows.

For the individual relations for $F_{+3-}$ and $F_{-3+}$ in (\ref{e20}), we also need the
$\TT_{220}^{\NR\,(1E)}$ coupling.  In this case, only $g_{220}^{\NR\,(1E)}$ contributes in the minimal SME,
and we find\footnote{There is a subtlety here since reading off the equivalent result from
\cite{Kostelecky:2013rta}, where results are quoted in terms of the four-dimensional duals
$\tilde{g}^{\m\n\l}$, there is an apparent discrepancy proportional to $\tr\,g_{i0j}$. However,
this can be re-expressed as $\big(\tilde{g}^{123}\,+\,\tilde{g}^{231} \,+\,\tilde{g}^{312}\big)$ which
is omitted in \cite{Kostelecky:2013rta} as being unobservable.}
\begin{equation}
i \,g_{220}^{\NR\,(1E)} ~=~ \sqrt{\frac{\pi}{15}}\,\,i\,\tr_Y g_{i0j} \ .
\label{e26}
\end{equation}
where the trace is over $i,j$, in agreement with  (\ref{d35}).
This completes the necessary identifications to establish the equivalence of $V_{\rm SME}^e({\bf R})$ 
in the spherical and Cartesian tensor frameworks. 

It is also interesting to verify directly the equivalence
of the SME Hamiltonians in the MOL frame. For this, we need to extend the above results to
include those with index $m\neq 1$. In particular, we need to verify the identities (\ref{app20})
given the expressions for $B_{\pm}$, $\tr\,F_{ij\pm}$ and $\tr_Y\,F_{ij\pm}$ in (\ref{d37}).

First, for $\TT_{01\pm1}^{\NR}$, we have 
\begin{align}
\sqrt{\frac{3}{4\pi}}\,\, g_{01\pm1}^{\NR\,(0B)} ~&=~ \sqrt{\frac{3}{4\pi}}\,\, g_{01\pm1}^{\NR\,(1B)} 
~~=~~  \mp \sqrt{\tfrac{1}{2}}\,\, (b_{\mp}\,\pm\,i \,m \,g_{\mp 30} )  \ ,  \nonumber \\[8pt]
\sqrt{\frac{3}{4\pi}}\,\, H_{01\pm1}^{\NR\,(0B)} ~&=~ \sqrt{\frac{3}{4\pi}}\,\, H_{01\pm1}^{\NR\,(1B)} 
~=~ \mp \sqrt{\tfrac{1}{2}}\,\,  (m\, d_{\mp 0} \,\pm\, i H_{\mp 3} )\ ,
\label{e27}
\end{align}
which confirms the required relation for $B_{\pm}$.

Next, we find
\begin{equation}
\sqrt{\frac{3}{4\pi}}\,\, g_{21\pm1}^{\NR\,(1B)} ~=~ \mp\, \frac{1}{2\sqrt{2}}\,\frac{1}{m^2}\,\tilde{g}_{D\mp} \ , 
~~~~~~~~~~~~
\sqrt{\frac{3}{4\pi}}\,\, H_{21\pm1}^{\NR\,(0B)} ~=~ \mp\, \frac{1}{\sqrt{2}}\,\frac{1}{m^2}\,\,\tilde{d}_{\mp} \ ,
\label{e28}
\end{equation}
as the natural generalisations of (\ref{e25}), noting that the vanishing of $g_{21m}^{\NR\,(0B)}$
and $H_{21m}^{\NR\,(1B)}$ holds in the minimal SME for all index $m$.
Then,
\begin{equation}
\sqrt{\frac{5}{4\pi}} \,i\, g_{22\pm 1}^{\NR\,(1E)} ~=~ \pm\,\frac{1}{2\sqrt{2}}\, \frac{i}{m} \, 
\big(g_{\mp 03} \,+\, g_{30\mp}\big) \ .
\label{e29}
\end{equation}
This closes the loop establishing the consistency of (\ref{app20}) with the expressions (\ref{d37}) for
$\tr\,F_{ij\pm}$ and $\tr_Y F_{ij\pm}$.

\vskip2.5cm

\section{Rovibrational spectrum of $\textbf{H}_2^{\,+}$ and 
$\overline{\textbf{H}}_2^{\,-}$}\label{sect 6}

In this final section, we use the analysis of Paper I to translate the results of sections \ref{sect 3}
and \ref{sect 4} for the inter-nucleon potentials $V_{\rm SME}^{e\,E}(R)$ and $V_{\rm SME}^{e\,F}(R)$,
and $\D E_{\rm SME}^n$,  into explicit expressions for the rovibrational energies in the 
hyperfine-Zeeman states.  We also comment briefly on how these results may influence an
experimental programme of testing Lorentz and \textsf{CPT} symmetry in high-precision
spectroscopy with ${\rm H}_2^{\,+}$ and $\overline{\rm H}_2^{\,-}$.
Further details will be presented elsewhere.

Restricting for clarity just to the leading terms in $(v+\thalf)$ and $N(N+1)$ in the expansion 
(\ref{bb13}), we consider here the rovibrational energies,
\begin{align}
E_{v N JM_J} ~&=~ {\cal E}_{\rm SME}^e\,+\, (1 + \d_{\rm SME}^e + \d_{\rm SME}^n)\,(v+\thalf) \,\w_0   \nonumber \\[8pt]
&~~~~~~~~~~~~~~~~~
+\, (B_0 + B_{\rm SME}^e + B_{\rm SME}^n) \,N(N+1) \,\w_0    ~~+~\ldots
\label{f1} 
\end{align} 
for Para-${\rm H}_2^{\,+}$.  The other terms in (\ref{bb13}) may be read off immediately
from Paper I given the results below.

The first term in (\ref{f1}) is defined as,
\begin{equation}
{\cal E}_{\rm SME}^e ~=~ \D E_{\rm SME}^{e\, B} \,+\, V_{\rm SME}^{e\,E} \,+\, V_{\rm SME}^{e\,F} \ ,
\label{f1a}
\end{equation}
where $V_{\rm SME}^{e\,E} \equiv V_{\rm SME}^{e\,E}(R_0)$ is evaluated from (\ref{c17}). 
Similarly for $V_{\rm SME}^{e\,F}$.
To determine the coefficients of $\w_0$, we first need the relations, 
\begin{equation}
\d_{\rm SME}^e \,=\, \frac{1}{2} \,\frac{1}{V_M^{''}}\, \Big[\, V_{\rm SME}^{e\,''}  \,-\, 
\frac{V_M^{'''}}{V_M^{''}}\,\, V_{\rm SME}^{e\,'}\Big] \ ,
~~~~~~~~~~~~~
B_{\rm SME}^e \,=\,  \l\, \,\frac{1}{V_M^{''}}\, \Big[\frac{1}{R_0}\,V_{\rm SME}^{e\,'} \,\Big] ~~+~\ldots\ ,
\label{f2}
\end{equation}
where the derivatives are evaluated at the mean bond length $R_0$. Also, 
\begin{equation}
\d_{\rm SME}^n ~=~ \frac{1}{2} \tilde{V}_{\rm SME}^n \ , 
~~~~~~~~~~~~~~~
B_{\rm SME}^n ~=~ B_0 \,\tilde{V}_{\rm SME}^n \ ,
\label{f3}
\end{equation}
with $B_0 = \l/2$, where recall $\l \simeq 0.027$ is a small parameter governing the hierarchy
of terms in the expansion of $E_{vNJM_J}$, and $\tilde{V}_{\rm SME}^n$ is given in (\ref{c25}).

$V_{\rm SME}^e(R)$ depends on the electron expectation values $\tr\,\langle\,p_a\,p_b\,\rangle$
and $\tr_Y\langle\,p_a\,p_b\,\rangle$.
Numerical values for these and $V_M(R)$ and their derivatives
were given in \cite{Muller:2004tc} and Paper I using a simple ansatz (which may be arbitrarily
improved as necessary) for the $1s\s_g$ wavefunction. Using these results, it was shown in
Paper I how to write $\d_{\rm SME}^e$ and $B_{\rm SME}^e$ (and the higher-order terms)
directly in terms of the coefficients of $\tr\,\langle\,p_a\,p_b\,\rangle$ and
$\tr_Y\langle\,p_a\,p_b\,\rangle$.

To extract the rovibrational energies from the inter-nucleon potential $V_{\rm SME\,\pm}^{e\,E}(R)$, 
we simply have to adapt these results for the hyperfine-Zeeman eigenstates, 
based on (\ref{c17}),
\begin{equation}
V_{{\rm SME}\,\pm}^{e\,E}(R)~=~ \frac{1}{3}\,\frac{1}{m_e^2}\,
\tr\,\langle\,p_a\,p_b\,\rangle\, \tr\,\tilde{E}_{ij}^e ~+~ 
\frac{1}{6} \,\frac{1}{m_e^2}\,\tr_Y\langle\,p_a\,p_b\,\rangle\, 
\tr_Y\tilde{E}_{ij}^e \,\, \wh{c}_{N M_J}^{\,\pm}(B)  \ .
\label{f3b}
\end{equation}
We find,\footnote{Recall that in the atomic units used here, the energy $E_{vNJM_J}$ is expressed
in units of the Rydberg constant, $R_H \simeq 13.6\, {\rm eV}$.  The vibration angular frequency
is $\w_0 = 0.020$. The SME coupling combinations $\tfrac{1}{m_e} \tr\,\tilde{E}_{ij}^e$ 
{\it etc.}~are dimensionless. In spectroscopic units where $h=1$,  $1\,{\rm eV} = 2.418 \times 10^{14}\,{\rm Hz}
= 8065.5\,{\rm cm}^{-1}$. See Paper I for details.}
\begin{align}
V_{\rm SME}^{e\,E} \,&=\, \,\big[\,0.782\,   \frac{1}{m_e}\,\tr\, \tilde{E}_{ij}^{e}   
\,+\, 0.120\, \wh{c}_{NM_J}^{\,\pm}\, \frac{1}{m_e}\,\tr_Y \tilde{E}_{ij}^{e} \,\big]   \nonumber  \\[10pt]
\d_{\rm SME}^{e\,E} \,&=\, \,\big[ -1.000\,   \frac{1}{m_e}\,\tr\, \tilde{E}_{ij}^{e}   
\,-\, 0.272\, \wh{c}_{NM_J}^{\,\pm}\, \frac{1}{m_e}\,\tr_Y \tilde{E}_{ij}^{e} \,\big]   \nonumber  \\[10pt]
B_{\rm SME}^{e\,E} \,&=\,\,\, \l\,\big[ -0.666 \, \frac{1}{m_e}\,\tr\, \tilde{E}_{ij}^{e}  
\,-\,0.112 \, \wh{c}_{NM_J}^{\,\pm}\, \frac{1}{m_e}\,\tr_Y \tilde{E}_{ij}^{e}  \,\big] \ ,
\label{f4}
\end{align}
using the appropriate $\wh{c}_{NM_J}^{\,\pm}(B)$ from (\ref{c18}), (\ref{c19})
for the mixed states with $J=N\pm\thalf$, and an identical result with $\wh{c}_{NM_J}$ 
from (\ref{c20}) for the unmixed states, with $J=N+\thalf, M_J=\pm(N+\thalf)$. 
Recall that the couplings are written in spherical tensor form as
\begin{equation}
\frac{1}{m}\,\tr\, \tilde{E}_{ij} \,=\, -3m\, \frac{1}{\sqrt{4\pi}}\,
\big(\tilde{c}_{200}^{\rm NR} \,-\, \tilde{a}_{200}^{\rm NR}\big) \ , 
~~~~~~~~~~~~
\frac{1}{m}\,\tr_Y\tilde {E}_{ij} \,=\, 3m\, \sqrt{\frac{5}{4\pi}}\,
\big(\tilde{c}_{220}^{\rm NR} \,-\, \tilde{a}_{220}^{\rm NR}\big)  \ .
~~~~\\[3pt]
\label{f5}
\end{equation}
and we maintain the notation where $\tilde{E}_{ij}^e$ contains both electron and proton
couplings, implying $\tilde{c}_{200}^{\NR\,e} = c_{200}^{\NR\,e} +\thalf c_{200}^{\NR\,p}$,
{\it etc.}

Similarly, from (\ref{c25}),
\begin{align}
\d_{\rm SME}^{n\,E} \,&=\,\,\, 0.333\,\big[\, \frac{1}{m_p}\,\tr\, {E}_{ij}^{p}   
\,-\, \wh{c}_{NM_J}^{\pm}(B)\, \frac{1}{m_p}\,\tr_Y {E}_{ij}^{p} \,\big]   \nonumber  \\[10pt]
B_{\rm SME}^{n\, E} \,&=\,\,0.333 \,\l\,\big[\, \frac{1}{m_p}\,\tr\, {E}_{ij}^{p}  
\,-\,\wh{c}_{NM_J}^{\pm}(B)\, \frac{1}{m_p}\,\tr_Y {E}_{ij}^{p}  \,\big]  \ ,
\label{f6}
\end{align}
with $\wh{c}_{NM_J}^{\pm}(B)$ and $\wh{c}_{NM_J}$ for the respective hyperfine-Zeeman eigenstates.

Next, for the $F_{ijk}^e$ couplings, recall from (\ref{d41}) that in the minimal SME where
$(F_{+3-} + F_{-3+}) = -\tr_Y F_{ijk}$, we can write $V_{\rm SME}^{e\,F}(R)$ in the form,
\begin{align}
V_{\rm SME\,\pm}^{e\,F}(R) ~&=~ \frac{1}{3} \,\frac{1}{m_e^2} \, \tr\,\langle\,p_a\,p_b\,\rangle \,
\tr\,F_{ij3}^e\,\, f_{NM_J}^\pm(B) 
\nonumber \\[8pt]
&~~~~~~~~~~~~~~~
-~\frac{1}{6} \,\frac{1}{m_e^2} \, \tr_Y\langle\,p_a\,p_b\,\rangle\,
\tr_Y F_{ij3}^e \,\,f_{NM_J}^{Y\,\pm}(B) \ ,
\label{f7}
\end{align}
for the mixed states with $J=N\pm\thalf$, 
where the $f_{NM_J}^{\pm}(B)$ and $f_{NM_J}^{Y\,\pm}$ are given in
(\ref{d42}) and (\ref{d43}), with a similar expression for the two unmixed states
with $f_{NM_J}$ from (\ref{d44}).   This now has a perfectly analogous form to $V_{\rm SME}^{e\,E}(R)$
in (\ref{f3b}), with different coefficients $f_{NM_J}$ and $f_{NM_J}^Y$ replacing the $\wh{c}_{NM_j}$. 
Notice that in this case both the $\tr\,\langle\,p_a\,p_b\,\rangle$ and 
$\tr_Y\langle\,p_a\,p_b\,\rangle$ terms have non-trivial coefficients.

The corresponding expressions for the rovibrational energies are then simply read off:
\begin{align}
V_{\rm SME}^{e\,F} \,&=\, \,\big[\, 0.782\,   f_{NM_J}^{\pm}(B)\,\,\frac{1}{m_e}\,\tr\,F_{ij3}^{e}   
\,-\, 0.120\, f_{NM_J}^{Y\,\pm}(B)\, \, \frac{1}{m_e}\,\tr_YF_{ij3}^{e} \,\big]   \nonumber  \\[10pt]
\d_{\rm SME}^{e\,F} \,&=\, \,\big[  -1.000\,   f_{NM_J}^{\pm}(B)\,\,\frac{1}{m_e}\,\tr\,F_{ij3}^{e}   
\,+\, 0.272\, f_{NM_J}^{Y\,\pm}(B)\, \, \frac{1}{m_e}\,\tr_YF_{ij3}^{e} \,\big]   \nonumber  \\[10pt]
B_{\rm SME}^{e\,F} \,&=\,\,\, \l\,\big[ -0.666 \, f_{NM_J}^{\pm}(B)\,\,\frac{1}{m_e}\,\tr\,F_{ij3}^{e}  
\,+\,0.112 \,   f_{NM_J}^{Y\,\pm}(B)\,\, \frac{1}{m_e}\,\tr_YF_{ij3}^{e}  \,\big] \ ,
\label{f8}
\end{align}
for the mixed states, and similarly with $f_{NM_J}$  and $f_{NM_J}^Y$ for the unmixed states.
Here, the couplings are (see (\ref{d33}) and (\ref{e24}), (\ref{e25})),
\begin{equation}
\frac{1}{m_e}\,\tr\,F_{ij3}^e ~=~ \frac{1}{m_e}\,\big(- \tilde{g}_{D3} \,+\, \tilde{d}_3 \big)\ , 
~~~~~~~~~~~~~~~~
\frac{1}{m_e} \tr_Y F_{ij3}^e ~=~ - \frac{1}{m_e}\,\big( \tilde{g}_{D3} \,+\, 2 \tilde{d}_3 \big) \ ,
\label{f9}
\end{equation}
with
\begin{equation}
\frac{1}{m_e} \, \tilde{g}_{D3} ~=~ 2m_e\, \sqrt{\frac{3}{4\pi}}\, \, g_{210}^{\NR\,(1B)} \ ,
~~~~~~~~~~~~~~~~
\frac{1}{m_e}\,\tilde{d}_3 ~=~ m_e\, \sqrt{\frac{3}{4\pi}} \,\, H_{210}^{\NR\,(0B)} \ .
\label{f10}
\end{equation}

Many of the important implications for ${\rm H}_2^{\,+}$ and $\overline{\rm H}_2^{\,-}$
rovibrational spectroscopy are encoded in the precise form of the various coefficients 
$\wh{c}_{NM_J}^\pm$, $f_{NM_J}^\pm$ and $f_{NM_J}^{Y\,\pm}$ determining the energies 
$E_{v N J M_J}$ through (\ref{f4}), (\ref{f6}) and (\ref{f8}). It is therefore useful to collect 
these coefficients together here, in both the zero and high magnetic field limits.

First, at zero $B$, and for the states $J= N\pm \thalf$,
\begin{align}
\wh{c}_{NM_J}^+ ~&=~ \frac{1}{(2N+1)(2N+3)}\,\big[ (N+\thalf)(N+ \tfrac{3}{2}) \,-\, 3 M_J^2\big] \ ,
\nonumber \\[10pt]
\wh{c}_{NM_J}^- ~&=~ \frac{1}{(2N-1)(2N+1)}\,\big[ (N-\thalf)(N+ \thalf) \,-\, 3 M_J^2\big] \ .
\label{f11}
\end{align}
The value $\wh{c}_{NM_J}$ for the unmixed states, $J=N+\thalf$, $M_J = \pm (N+\thalf)$
is read off from the first of these,
\begin{equation}
\wh{c}_{NM_J}~=~ - \frac{N}{2N+3} \ .
\label{f12}
\end{equation}
For the coefficients of the spin-dependent couplings, we have
\begin{align}
f_{NM_J}^+ ~&= ~\,\frac{2}{2N+1} \,M_J \ ,
~~~~~~~~~~~~~~~~~
f_{NM_J}^{Y\,+} ~=~ \frac{2N}{(2N+1)(2N+3)}\, M_J \ ,
\nonumber \\[10pt]
f_{NM_J}^- ~&=~ - \frac{2}{2N+1} \, M_J \ ,
~~~~~~~~~~~~~~
f_{NM_J}^{Y\,-} ~=~ - \frac{2(N+1)}{(2N-1)(2N+1)} \, M_J \ ,
\label{f13}
\end{align}
for the states $J=N\pm\thalf$, reducing to
\begin{equation}
f_{NM_J} ~=~ \pm 1 \ , ~~~~~~~~~~~~~~
f_{NM_J}^Y ~=~ \pm \frac{N}{2N+3} \ ,
\label{f14}
\end{equation}
for the unmixed states with $M_J = \pm (N+\thalf)$.

In the large $B$ limit, the states are labelled by $M_N = M_J - M_S$ and $M_S$.
The corresponding coefficients are,
\begin{equation}
{c}_{NM_N} ~=~ \frac{1}{(2N-1)(2N+3)} \,\big[N(N+1) - 3 M_N^2\big] \ ,
\label{f15}
\end{equation}
and
\begin{equation}
f_{NM_N}^{\pm} ~=~ \pm 1 \ ,   ~~~~~~~~~~~
f_{NM_N}^{Y\,\pm} ~=~ \pm {c}_{NM_N} 
\label{f16}
\end{equation}\\
for the states with $M_S = \pm \thalf$.

To complete the contributions to ${\cal E}_{\rm SME}^e$ in (\ref{f1a}), 
we also add the simple SME spin contribution
$\D E_{\rm SME}^{e\,B}$ given by (\ref{d4}), or (\ref{d5}) in the high magnetic field limit,
which is proportional to the coupling $\tilde{b}_3^e$.  At zero $B$, this is simply,
\begin{equation}
\D E_{\rm SME}^{e\,B} ~=~ \mp \,\tilde{b}_3^e\, \,\frac{2}{2N+1}\, M_J \ ,
\label{f17}
\end{equation}
for $J= N\pm \thalf$, reducing to $\pm \,\tilde{b}_3^e$ for the 
unmixed states, while in the large $B$ limit,
\begin{equation}
\D E_{\rm SME}^{e\,B} ~=~ - 2\,\tilde{b}_3^e\, M_S \ .
\label{f18}
\end{equation}

\vskip0.3cm
In Paper I, we discussed how the $N$, and $M_J$, dependence of the cefficients $\wh{c}_{NM_J}$
allow the spin-independent couplings to be individually determined from a relatively small
number of rovibrational transitions.
The results above show how this can be extended to the spin-dependent couplings $\tilde{b}_3$,
$\tilde{g}_{D3}$ and $\tilde{d}_3$ when transitions between hyperfine states identified by 
$J, M_J$, as in Fig.~\ref{FigHZ}, as well as $v, N$ are isolated. Already, in \cite{SAS2024},
the rovibrational transitions between $|v\,N\, J\rangle = |1\,0\,\thalf\rangle$ and 
$|3\,2\,\tfrac{3}{2}\rangle$ or $|3\,2\,\tfrac{5}{2}\rangle$ have recently been studied.
The specific $N$-dependence of the coefficients $f_{NM_J}^{\pm}$ and $f_{NM_J}^{Y\,\pm}$ 
is important in disentangling the individual SME couplings from measurements of several 
rovibrational transitions. Our results also describe the dependence on the SME couplings
of transitions between hyperfine states with the same rovibrational quantum numbers
$v, N$ \cite{Jefferts1969,SchillerCP}.

In the quest for high precision, one experimental strategy is to combine transitions so that
the linear  Zeeman effects proportional to $M_J$ cancel. This has been extended 
to sub-leading effects including the quadratic Zeeman and electric quadrupole shifts
in the comprehensive paper \cite{SchillerKorobov2018}. A concern might be that in cancelling 
the Zeeman contributions we may also lose sensitivity to the spin-dependent SME couplings. 
However, the $N$ and $M_J$ dependence of the $f_{NM_J}^{\pm}$ and $f_{NM_J}^{Y\,\pm}$
coefficients ensures that suitably chosen transitions will remain sensitive to most of the SME
couplings even when the Zeeman and sub-leading contributions are systematically cancelled.
Further details will be presented elsewhere. 

Throughout this paper we have carefully included the interplay of the SME couplings
with a background magnetic field.
The additional flexibility afforded by knowing the dependence of the rovibrational energies
on the magnetic field, encoded here in the $B$-dependence of the coefficients
$f_{NM_J}^{\pm}(B)$ and $f_{NM_J}^{Y\,\pm}(B)$ in (\ref{d42}) and (\ref{d43}), should also prove useful 
in isolating potential Lorentz and \textsf{CPT} violating effects from higher-order systematics 
and Zeeman and QED effects. 

In addition, it is probable that spectroscopy with $\overline {\rm H}_2^{\,-}$ will require
confinement in a trap with a magnetic field well into our high-$B$ regime.  Direct comparisons
could therefore require performing ${\rm H}_2^{\,+}$ spectroscopy in similar fields.

One of the most promising opportunities for an early discovery of Lorentz violation would be
the observation of annual variations of the transition frequencies in ${\rm H}_2^{\,+}$.
It is worth commenting in this context that the expected precision of rovibrational 
spectroscopy on the ${\rm H}_2^{\,+}$ molecular ion will far exceed that of current high-order
QED calculations, so Lorentz and \textsf{CPT} tests will require comparisons of different
spectroscopic measurements rather than a simple theory/experiment comparison
on a single transition.

In Paper I we gave an explicit formula for the annual variations implied by the spin-independent
couplings $c_{200}^{\NR}$ and $a_{200}^{\NR}$. In particular, it was shown how the variations
are sensitive to different components of the fundamental SME couplings than appear in
the basic transition frequencies themselves, though suppressed by $O(10^{-4})$ (the ratio of the
Earth's orbital velocity to the speed of light).  The same analysis can be applied to the
spin-dependent couplings $\tilde{b}_3$, $\tilde{g}_{D3}$ and $\tilde{d}_3$, with added
complexity due their non-isotropic nature.  In Appendix \ref{Appendix C} we give a brief
outline of some of the extra features needed to generalise the discussion of annual
variations to the spin-dependent couplings considered here.

A key point in Paper I was the observation that the dependence of the rovibrational energies
on the proton SME couplings arising from the direct contribution $\D E_{\rm SME}^n$ is
enhanced by $O(m_p/m_e)$ relative to the contribution from the inter-nucleon potential
$V_{\rm SME}^e(R)$. The latter gives the same parametric dependence as occurs in
atomic spectroscopy with ${\rm H}$ and $\overline{\rm H}$. This gives rovibrational
spectroscopy of the molecular ion a potential $O(10^3)$ increased sensitivity to
\textsf{CPT} violation in the proton sector given comparable experimental 
measurement precisions. 

However, to achieve this sensitivity gain for the spin-dependent proton couplings,
we need the analogue of $\D E_{\rm SME}^n$ which would arise from the term
proportional to $F_{ijk}^p\, I_k$ in the SME Hamiltonian, {\it i.e.}~depending on
the nucleon spin. This of course implies performing rovibrational spectroscopy 
with Ortho-${\rm H}_2^{\,+}$.  It is straightforward to extend all our results to 
Ortho-${\rm H}_2^{\,+}$ with no further issues of principle, simply involving the
extra complexity associated with the Clebsch-Gordan analysis required to
describe mixing amongst the $|v N F J M_J\rangle$ hyperfine states. 

In summary, the richness and extremely high precision of the spectrum of rovibrational
transitions make an experimental programme of ${\rm H}_2^{\,+}$, and in future
$\overline{\rm H}_2^{\,-}$, spectroscopy ideal for testing fundamental symmetries 
such as Lorentz and \textsf{CPT} invariance.
The detailed results presented here should help to guide this programme and ensure that 
the experimentally selected rovibrational transitions maintain maximum sensitivity
to potential Lorentz and \textsf{CPT} violating effects, wherever they may occur.

\vskip1cm
\noindent {\large{\textbf{Acknowledgements}}}
\vskip0.3cm
I am grateful to Stefan Eriksson for many helpful discussions in the course of this work
and to the Higgs Centre for Theoretical Physics at the University of Edinburgh for hospitality.

\newpage

\appendix{

\section{SME Hamiltonian $\wh{H}_{\rm SME}$ in the \textsf{MOL} frame.}\label{Appendix A}

The SME Hamiltonian for a single Dirac particle is given in terms of spherical tensor couplings
in ref.~\cite{Kostelecky:2013rta}, in a formalism which systematically includes higher-dimensional
operators. With particle momentum ${\bf p}$ with respect to the fixed frame in which the spherical tensor 
couplings are defined, the non-relativistic Hamiltonian is
\begin{align}
H_{\rm SME} ~&=~ - \sum_{njm} \, |{\bf p }|^n \, Y_{jm}(\hat{\bf p}) \, \VV_{njm}^{\NR}  \nonumber \\[6pt]
&~~~~~ -\,\s^r\, \sum_{njm}\, |{\bf p }|^n \, Y_{jm}(\hat{\bf p}) \, \TT_{njm}^{\NR\,(0B)} \nonumber \\[6pt]
&~~~~~ +\,\s^{\pm}\, \sum_{njm}\, |{\bf p }|^n \,\, {}_{\pm1}Y_{jm}(\hat{\bf p}) \, \big(\pm \TT_{njm}^{\NR\,(1B)}
\,+\, i\TT_{njm}^{\NR\,(1E)} \big)\ ,
\label{app1}
\end{align}
with
\begin{equation}
\VV_{njm}^{\NR} ~=~ c_{njm}^{\NR} ~-~ a_{njm}^{\NR} \ ,
\label{app2}
\end{equation}
and
\begin{equation}
\TT_{njm}^{\NR\,(0B)} ~=~ g_{njm}^{\NR\,(0B)} ~-~ H_{njm}^{\NR\,(0B)} \ ,
\label{app3}
\end{equation}
and similarly for the $(1B)$ and $(1E)$-type couplings.  For further details and the motivation
for (\ref{app1}), see ref.\cite{Kostelecky:2013rta}.

For our purposes, we first consider this to be the Hamiltonian for the electron, with the fixed frame
identified as the \MOL frame with whose $z$-axis is aligned with the molecular axis. 
The electron momentum ${\bf p}$, with components $p_a$, has direction $\hat{\bf p}$ specified by 
spherical polar angles $(\theta,\phi)$ in the \MOL frame.

In (\ref{app1}), the Pauli spin operators $\s^r$, $\s^{\pm}$ are defined in the `helicity' frame (\HEL)
whose $z$-axis is chosen to lie along the direction $\hat{\bf p}$ of the electron momentum.
The first step in adapting (\ref{app1}) to our problem is therefore to transform these 
spin operators to the \MOL frame so the SME Hamiltonian is written consistently in a single frame.
As in the main text, we identify the spin operators and couplings in the \MOL frame with
a circumflex accent, to distinguish from their components in the \EXP frame used later.

First, we need an explicit expression for the spin-weighted spherical harmonics ${}sY_{jm}(\hat{\bf p})$
in (\ref{app1}) in terms of the Wigner matrices $d_{sm}^j(\theta)$ introduced in the text, {\it viz.}
\begin{equation}
{}_sY_{jm}(\theta,\phi) ~=~ \sqrt{\frac{2j+1}{4\pi}}\,(-1)^m\, d_{-s\,m}^j(\theta) \, e^{im\phi} \ .
\label{app4}
\end{equation}

Then, writing $\s^r$, $\s^{\pm}$ in standard spherical tensor notation $\s_m$ $(m=1,-1,0)$ in the \HEL 
frame, we have $\s^+ = \s_{m=-1}^{\HEL}$, $\s^- = -\s_{m=1}^{\HEL}$ and $\s^r = \s_{m=0}^{\HEL}$
according to the definitions in \cite{Kostelecky:2013rta}. 
Rotating these spin operators from the \HEL to the \MOL frame using the Wigner matrices, as
described in the main text,
\begin{equation}
\s_{s'}^{\HEL} ~=~  \wh{\s}_s\, d_{s\,s'}^1(\theta)\, e^{-i s\phi} \ ,
\label{app5}
\end{equation}
and using the identity $d_{m'\,m}^j = (-1)^{m' -m} \,d_{m\,m'}^j$ to reorganise indices, we find
the SME Hamiltonian with all quantities in the \MOL frame:
\begin{align}
{\wh H}_{\rm SME} &= - \sum_{njm}\, |{\bf p}|^n\, \sqrt{\frac{2j+1}{4\pi}}\,\,
\bigg[\, e^{im\phi}\, d_{m\,0}^j\, \wh{\VV}_{njm}^{\NR}  \nonumber \\[6pt]
&~~~~~~~~~+~ e^{i (m-s)\phi}\,\,\Big( \wh{\s}_s \, d_{s\,0}^1\, d_{m\,0}^j\, \wh{\TT}_{njm}^{\NR\,(0B)} 
~+~ \wh{\s}_s\,\big(d_{s,-1}^1\,d_{m,-1}^j \,+\, d_{s\,1}^1\,d_{m\,1}^j \big) \, 
\wh{\TT}_{njm}^{\NR\,(1B)} \nonumber \\[6pt]
&~~~~~~~~~~~~~~~~~~~~~~~~~~~~~~~~~~~~~~~~~~~~~~
~+~ \wh{\s}_s\,\big(d_{s,-1}^1\,d_{m,-1}^j \,-\, d_{s\,1}^1\,d_{m\,1}^j \big) \, 
i \wh{\TT}_{njm}^{\NR\,(1E)}  \,\Big) \bigg]\ .
\label{app6}
\end{align}

The spherical tensor couplings are subject to a number of constraints on the allowed values of
$(n,j,m)$ which follow from their fundamental definition and the subsequent non-relativistic
expansion of the Hamiltonian in powers of $|{\bf p}|^n$, together with some identities which 
hold for low values of $n$ and $j$.

Consider first the contribution to (\ref{app6}) independent of momentum, so $n=0$. Here, the
index $j$ is resticted to $j=0$ for $\wh{\VV}_{0jm}^{\NR}$ and $j=1$ for $\wh{\TT}_{0jm}^{\NR\,(0B)}
\,=\,\wh{\TT}_{01m}^{\NR\,(1B)}$.
There is no corresponding $(1E)$ coupling with $n=0$.  Using orthonormality of the Wigner
matrices (see (\ref{e5})) and $d_{0\,0}^0 = \sqrt{4\pi} Y_{00} = 1$, the Hamiltonian simplifies to,
\begin{equation}
{\wh H}_{\rm SME}\big|_{p^2=0} ~~=~ -\, \sqrt{\frac{1}{4\pi}}\,\, \wh{\VV}_{000}^{\NR} 
~-~ \sqrt{\frac{3}{4\pi}}\,\, \wh{\s}_m\, \wh{\TT}_{01m}^{\NR\,(0B)} \ ,
\label{app7}
\end{equation}
where as always a sum over the index $m$ is assumed.

Next, consider the $O(|{\bf p}|^2)$ contribution. Here, for $n=2$, the permitted values for the 
spherical tensor couplings are $j=2, 0$ for $\wh{\VV}_{2jm}^{\NR}$, $j=3, 1$ for $\wh{\TT}_{2jm}^{\NR\,(0B)}$
and $\wh{\TT}_{2jm}^{\NR\, (1B)}$, and $j=2, 0$ for $\wh{\TT}_{2jm}^{\NR\,(1E)}$. So in general we have,
\begin{align}
{\wh H}_{\rm SME}\big|_{p^2} ~&=~ - \sqrt{\frac{1}{4\pi}} \,\,|{\bf p}|^2\, \sum_m\bigg[\, \wh{\VV}_{200}^{\NR} \,+\,
\sqrt{5}\,e^{im\phi}\, d_{m\,0}^2\, \wh{\VV}_{22m}^{\NR} \nonumber \\[6pt]
&+~ e^{i(m-s)\phi} \,\,\wh{\s}_s\,\bigg(\,  
\sqrt{3}\,d_{s\,0}^1\,d_{m\,0}^1\, \wh{\TT}_{21m}^{\NR\,(0B)}
~+~ \sqrt{7}\, d_{s\,0}^1\, d_{m\,0}^3\, \wh{\TT}_{23m}^{\NR\,(0B)} \nonumber \\[6pt]
&+~  \sqrt{3} \,\big(d_{s,-1}^1\, d_{m,-1}^1 \,+\, d_{s\,1}^1\,d_{m\,1}^1\big)\,\wh{\TT}_{21m}^{\NR\,(1B)}
~+~ \sqrt{7}\,\big(d_{s,-1}^1\, d_{m,-1}^3 \,+\, d_{s\,1}^1\,d_{m\,1}^3\big)\,\wh{\TT}_{23m}^{\NR\,(1B)}
\nonumber\\[6pt]
&+~ \sqrt{5} \,\big(d_{s,-1}^1\, d_{m,-1}^2 \,-\, d_{s,1}^1\,d_{m\,1}^2\big)\,i\wh{\TT}_{22m}^{\NR\,(1E)} 
\bigg)\,\bigg] \ .
\label{app8}
\end{align}

Evidently, the SME Hamiltonian in this full generality is complicated, with many couplings and 
angle-dependent coefficients. At this point, therefore, we specialise to the case of the molecular ion.
The first major simplification is to impose cylindrical symmetry, which requires the expectation values
$\langle\,p_a\,p_b\,\rangle = 0$ for $a\neq b$, and $\langle \,p_x^2\,\rangle = \langle\,p_y^2\,\rangle$.

Then, for example, the coefficient of $\wh{\VV}_{220}^{\NR}$ is 
$|{\bf p}|^2 \,d_{0\,0}^2 = -\thalf |{\bf p}|^2\, (1 - 3\cos^2\theta) \,=\, - \thalf \tr_Y \,p_a\,p_b$.
Examination of the $m\neq0$ contributions shows that they give vanishing expectation values,
{\it e.g.} $|{\bf p}|^2 \,d_{2\,0}^2 e^{2i\phi} \,=\, \tfrac{\sqrt{6}}{4}\, |{\bf p}|^2\, \sin^2\theta\, e^{2i\phi} \,=\, 
(p_x + i p_y)^2 \rightarrow 0$.  The $O(|{\bf p}|^2)$ Hamiltonian for the spin-independent couplings 
therefore reduces to,
\begin{equation}
{\wh H}_{\rm SME}^{\VV} \big|_{p^2} ~=~ - \sqrt{\frac{1}{4\pi}} \,\, \tr \,p_a\,p_b \, \,\wh{\VV}_{200}^{\NR}\,
\,+\,\frac{1}{2} \sqrt{\frac{5}{4\pi}} \,
\,\tr_Y \,p_a\,p_b \,\, \wh{\VV}_{220}^{\NR} ~+~ \ldots 
\label{app9}
\end{equation}
where $+\,\ldots$ indicate terms with zero expectation value for the molecular ion with the electron 
in the $1s\s_g$ ground state.

For the $(0B)$ and $(1B)$ couplings we need the products of Wigner matrices, which are evaluated 
in terms of Clebsch-Gordan coefficients using the formula (\ref{e7}) in the main text. It is convenient
here to first use the identity $d_{s\,0}^j = (-1)^s d_{-s\,0}^j$, then evaluate
\begin{equation}
d_{-s\,0}^1\, d_{m\,0}^j ~=~ \sum_{J=|j-1|}^{j+1}\,\, C_{1\,0,\,j\,0}^{J\,0}\,\, C_{1\,-s,\,j\,m}^{J\,(-s+m)} \,
d_{(-s+m)\,0}^J \ .
\label{app10}
\end{equation}
For $j=1$ only $J=0, 2$ are allowed, since $C_{1\,0,\,1\,0}^{1\,0} = 0$, and again keeping only terms with
non-vanishing expectation values when combined with $|{\bf p}|^2$, leaves
\begin{equation}
d_{-s\,0}^1\, d_{m\,0}^1 ~=~ -\sqrt{\frac{1}{3}} \,C_{1\,-s,\,1\,m}^{0\,0}\, d_{00}^0 ~+~ 
\sqrt{\frac{2}{3}} \,C_{1\,-s,\,1\,m}^{2\,0}\, d_{00}^2 ~+~ \ldots
\label{app11}
\end{equation}
Evaluating the Clebsch-Gordan coefficients then gives
\begin{equation}
(-1)^s \, d_{-s\,0}^1\, d_{m\,0}^1 ~=~ \frac{1}{3} \,\d_{sm}\, d_{0\,0}^0 ~-~ \sqrt{\frac{2}{3}}\,\sqrt{\frac{1}{6}}\,\,
\textsf{Y}_{sm}\, d_{0\,0}^2 ~+~ \ldots
\label{app12}
\end{equation}
with $\textsf{Y}_{sm} = \begin{pmatrix}
1 &~\,0&~0\\ 0&~\,1&~0 \\0&~\,0& -2\\
\end{pmatrix}$.
Finally, substituting back into the SME Hamiltonian, we find the contribution of the $(0B)$ couplings
with $j=1$ as,
\begin{equation}
{\wh H}_{\rm SME}^{\TT(0B)}\big|_{p^2} ~=~ \sqrt{\frac{3}{4\pi}} \,\, \wh{\s}_s\, \Big( \frac{1}{3}\, \tr \,p_a\,p_b\,\, \d_{sm}
~+~ \frac{1}{6} \tr_Y\,p_a\,p_b\, \textsf{Y}_{sm} \Big)\, \wh{\TT}_{21m}^{\NR\,(0B)} ~+~\ldots 
\label{app13}
\end{equation}

The same product of Wigner matrices, when combined with orthonormality, is sufficient to evaluate the 
coefficient of the $\wh{\TT}_{21m}^{\NR\,(1B)}$ couplings as well. For those with $j=3$, the same method can 
in principle be carried through, but since these couplings do not arise in the minimal SME since their
associated Lorentz and \textsf{CPT} violating operators are of higher dimension, we do not make further
use of them here.

This leaves the $(1E)$ couplings.  Here, since the products involve Wigner matrices with different $j$, the
orthonormality trick which simplifies the coefficient of the $\TT_{21m}^{\NR\,(1B)}$ is not available
and we must evaluate directly:
\begin{equation}
\big(d_{s,-1}^1\, d_{m,-1}^2 \,-\, d_{s\,1}^1\,d_{m\,1}^2\big) ~=
\sum_{J=2,3} \, \Big( C_{1\,-1,\,2\,-1}^{J\,-2}\,\, C_{1\,s,\,2\,m}^{J\,(s+m)} \, d_{(s+m)\,-2}^J 
~-~ C_{1\,1,\,2\,1}^{J\,2}\,\, C_{1\,s,\,2\,m}^{J\,(s+m)} \, d_{(s+m)\,2}^J \Big)
\label{app14}
\end{equation}
the sum being over $J=2,3$ only since $J=1$ has a vanishing Clebsch-Gordan coefficient.
The $J=2$ term contributes
\begin{equation}
\sqrt{\frac{1}{3}}\,\, C_{1\,s,\,2\,m}^{2\,(s+m)} \,\big( \,d_{(s+m), -2}^2 ~+~ d_{(s+m)\,2}^2\,\big) ~=~ 
- \frac{1}{6} (1 + \cos^2\theta) \, \textsf{T}_{sm} \ ,
\label{app15}
\end{equation}
restricting to $s=m$ by cylindrical symmetry, while after some remarkable simplification of the
Wigner matrices, the $J=3$ term gives
\begin{equation}
\sqrt{\frac{2}{3}}\,\, C_{1\,s,\,2\,m}^{3\,(s+m)} \, \big( \,d_{(s+m), -2}^3 ~-~ d_{(s+m)\,2}^3\,\big) ~=~ 
- \frac{2}{3} (1 - 2\cos^2\theta) \, \textsf{T}_{sm} \ ,
\label{app16}
\end{equation}
with $\textsf{T}_{sm} \,=\, \begin{pmatrix}
1 &~0&~0\\ 0&~-1&~0 \\0&~0& \,\,0\\
\end{pmatrix}$.
Combining the $J=2$ and $J=3$ contributions and evaluating with the momentum factor,
we find they once again conspire to give the familiar $\tr_Y \,p_a\,p_b$ combination,
leaving the $(1E)$ contribution to the SME Hamiltonian,
\begin{equation}
{\wh H}_{\rm SME}^{\TT(1E)}\big|_{p^2} ~=~ - \sqrt{\frac{5}{4\pi}}\,\, \frac{1}{2}\,\,\tr_Y\,p_a\,p_b\,\,
\wh{\s}_s\,\textsf{T}_{sm}\, i \wh{\TT}_{22m}^{\NR\,(1E)} ~~+~\ldots 
\label{app17}
\end{equation}

Finally, putting everything together, and keeping only the momentum factors that have
non-vanishing expectation values given the cylindrical symmetry of the molecular ion,
we find the following compact expression for the SME Hamiltonian with all quantities expressed 
in the \MOL frame:
\begin{align}
{\wh H}_{\rm SME} ~=~ &-\sqrt{\frac{1}{4\pi}}\, \bigg[\, \wh{\VV}_{000}^{\NR} ~+~ 
\tr\,p_a\,p_b\,\wh{\VV}_{200}^{\NR}
~-~ \sqrt{5}\,\frac{1}{2}\,\tr_Y p_a\,p_b\,\wh{\VV}_{220}^{\NR} \nonumber \\[8pt]
&+ \sqrt{3}\, \wh{\s}_s \d_{sm}\, \wh{\TT}_{01m}^{\NR\,(0B)} 
~+~ \sqrt{3}\, \wh{\s}_s \big( \,\frac{1}{3} \,\tr\,p_a\,p_b\, \d_{sm} ~+~ 
\frac{1}{6}\, \tr_Y p_a\,p_b\, \textsf{Y}_{sm} \big)\,\wh{\TT}_{21m}^{\NR\, (0B)} \nonumber \\[8pt]
&+~ \sqrt{3}\, \wh{\s}_s \big( \,\frac{2}{3} \,\tr\,p_a\,p_b\, \d_{sm} ~-~ 
\frac{1}{6}\, \tr_Y p_a\,p_b\, \textsf{Y}_{sm} \big)\,\wh{\TT}_{21m}^{\NR\, (1B)} \nonumber \\[8pt]
&+~ \sqrt{5}\, \frac{1}{2}\, \tr_Y p_a\,p_b\, \,\wh{\s}_s\,\textsf{T}_{sm}\, 
i \wh{\TT}_{22m}^{\NR\,(1E)} \, \bigg]\ .
\label{app18}
\end{align}
Recall we have also omitted the $\wh{\TT}_{23m}^{\NR\,(0B)}$ and $\wh{\TT}_{23m}^{\NR\,(1B)}$ 
couplings which do not occur in the minimal SME.

\vskip0.2cm

We can compare this with the equivalent expression for $\wh{H}_{\rm SME}$ written directly
from the original SME Lagrangian (\ref{a1}).  With all quantities defined in the \MOL frame,
and imposing the same cylindrical symmetry constraints on the momentum as in (\ref{app18}),
the spin-dependent terms are,
\begin{equation}
\wh{H}_{\rm SME} ~=~ \wh{B}_c \, \wh{\s}_c ~+~ 
\frac{1}{m^2}\,\bigg[\frac{1}{3}\, \tr\,p_a\,p_b\,\,\tr\,\wh{F}_{abc}\, \wh{\s}_c
~+~ \frac{1}{6}\, \tr_Yp_a\,p_b\,\,\tr_Y{\wh F}_{abc}\, \wh{\s}_c \, \bigg] \ .
\label{app19}
\end{equation}
Comparing (\ref{app18}) and (\ref{app19}), we see that they are equivalent provided the following
identities hold, in addition to those already quoted at the end of section \ref{sect 5.1}:
\begin{align}
\frac{1}{\sqrt{2}}\,\wh{B}_{\pm}~&=~ \mp\,\sqrt{\frac{3}{4\pi}}\,\, \wh{\TT}_{01\mp1}^{\NR\,(0B)} \ ,
\nonumber \\[2pt]
\frac{1}{\sqrt{2}}\,\frac{1}{m^2}\,\tr\,\wh{F}_{ab\pm} ~&=~ \mp\,\sqrt{\frac{3}{4\pi}}\,
\Big(\wh{\TT}_{21\mp1}^{\NR\,(0B)} \,+\,2\,\wh{\TT}_{21\mp1}^{\NR\,(1B)} \Big) \ ,
\nonumber \\[2pt]
\frac{1}{\sqrt{2}}\, \frac{1}{m^2}\,\tr_Y\wh{F}_{ab\pm} ~&=~ 
\mp\sqrt{\frac{3}{4\pi}}\, \Big( \wh{\TT}_{21\mp1}^{\NR\,(0B)}\,-\, \wh{T}_{21\mp1}^{\NR\,(1B)} \Big) 
~+~ \sqrt{\frac{5}{4\pi}}\,3\,i\, \wh{\TT}_{22\mp1}^{\NR\,(1E)} \ . 
\label{app20}
\end{align}

We verify these relations explicitly in the minimal SME in section \ref{sect 5.2}.

\newpage

\section{Clebsch-Gordan relations}\label{Appendix B}

We collect here a number of relations for weighted sums over Clebsch-Gordan coefficients
which are used in the main text in transforming results to the Para-${\rm H}_2^{\,+}$ hyperfine states
$|v\,N\,J\,M_J\rangle$.  First recall the relation between the basis states,
\begin{equation}
|v\,N\,J\,M_J\rangle  ~~=~~ \sum_{M_S} \, C_{N \,M_N,\,\thalf \,M_S}^{J \,M_J} \, |v\,N\,M_N\, M_S\rangle \ ,
\label{bpp1}
\end{equation}
with $M_N = M_J - M_S$, where for $J = N\pm \thalf$ we have the explicit forms:
\begin{align}
C_{N\,M_J\mp\thalf,\,\thalf\,\pm\thalf}^{N+\thalf\,\,M_J}  ~~&=~~~~ \frac{1}{\sqrt{2N+1}}\, \sqrt{N+\thalf \pm M_J}\ ,   
\nonumber \\[10pt]
C_{N\,M_J\mp\thalf,\,\thalf\,\pm\thalf}^{N-\thalf\,\,M_J}  ~~&=~~ \mp\,\frac{1}{\sqrt{2N+1}}\, \sqrt{N+\thalf \mp M_J} \ .
\label{bpp2}
\end{align}

First, for the unweighted sums, with $M_J$ fixed and  $J', J = N\pm \thalf$, we can explicitly verify,
\begin{equation}
\sum_{M_S} \, C_{N\,(M_J-M_S),\,\thalf\,M_S}^{J'\,M_J} \,\, C_{N\,(M_J-M_S),\,\thalf\,M_S}^{J\,M_J} ~=~  \d_{J' J} \ ,
\label{bpp3}
\end{equation}
which follows from the general orthonormality property of Clebsch-Gordan coefficients.

Next we need the corresponding sums weighted by the spin quantum number $M_S$:
\begin{align}
&\sum_{M_S} \,\Big(C_{N\,(M_J-M_S),\,\thalf\,M_S}^{N+\thalf\,M_J} \Big)^2 \,\,M_S~=~ \frac{1}{2N+1}\,M_J \ ,
\nonumber \\[10pt]
&\sum_{M_S} \, \Big(C_{N\,(M_J-M_S),\,\thalf\,M_S}^{N-\thalf\,M_J} \Big)^2 \,\, M_S~=~ - \frac{1}{2N+1}\,M_J \ ,
\nonumber \\[10pt]
&\sum_{M_S} \, C_{N\,(M_J-M_S),\,\thalf\,M_S}^{N+\thalf \,M_J} \,\, C_{N\,(M_J-M_S),\,\thalf\,M_S}^{N-\thalf\,M_J} \,\,M_S~=~ 
- \frac{1}{2N+1} \, \sqrt{(N+\thalf)^2 - M_J^2} \ ,
\label{bpp4}
\end{align}
which are required to determine the hyperfine-Zeeman energies and the effect of the SME coupling $B_3^e$
in sections \ref{sect 2} and \ref{sect 4}.

For the analysis in section \ref{sect 3} with the spin-independent couplings $E_{ij}$, we need these 
sums with the weight factor $c_{NM_N}$, defined below. Here,
\begin{align}
&\sum_{M_S}\, \Big(C_{N\,(M_J-M_S),\,\thalf\,M_S}^{N+\thalf\,M_J}\Big)^2\,\,c_{N (M_J-M_S)}~=~
\frac{1}{(2N+1)(2N+3)}\,\big[(N+\thalf)(N+\tfrac{3}{2}) - 3 M_J^2\big] \ ,  \nonumber \\[10pt]
&\sum_{M_S}\, \Big(C_{N\,(M_j-M_S),\,\thalf\,M_S}^{N-\thalf\,M_J}\Big)^2\,\,c_{N (M_J-M_S)}~=~
\frac{1}{(2N-1)(2N+1)}\,\big[(N-\thalf)(N+\thalf) - 3 M_J^2\big] \ ,   \nonumber \\[10pt]
&\sum_{M_S}\, C_{N\,(M_j-M_S),\,\thalf\,M_S}^{N+\thalf\,M_J} \,\,C_{N\,(M_j-M_S),\,\thalf\,M_S}^{N-\thalf\,M_J}
\,\,c_{N (M_J-M_S)}  \nonumber \\[4pt]
&~~~~~~~~~~~~~~~~~~~~~~~~~~~~~~~~~
=~ - 6\, \,\frac{1}{(2N-1)(2N+1)(2N+3)}\,\big[(N+\thalf)^2 - M_J^2\big]^{\thalf}\,\,\, M_J\ .
\label{bpp5}
\end{align}

Finally, for the momentum and spin-dependent couplings $F_{ijk}$ considered in section \ref{sect 4},
we need the corresponding sums with both $c_{NM_N}$ and $M_S$ as weight factors.
In this case, 
\begin{align}
&\sum_{M_S}\, \Big(C_{N\,(M_J-M_S),\,\thalf\,M_S}^{N+\thalf\,M_J}\Big)^2\,\,c_{N (M_J-M_S)}\,\,M_S
\nonumber \\
&~~~~~~~~~~~~~~~~~~~~~~~~~~~~~~~~~
=~ \frac{1}{(2N+1)(2N-1)(2N+3)}\,\big[N^2 + 4N + \tfrac{3}{4}  - 3
  M_J^2\big]\, M_J \ ,  \nonumber \\[10pt]
&\sum_{M_S}\, \Big(C_{N\,(M_j-M_S),\,\thalf\,M_S}^{N-\thalf\,M_J}\Big)^2\,\,c_{N (M_J-M_S)}\,\,M_S 
\nonumber \\
&~~~~~~~~~~~~~~~~~~~~~~~~~~~~~~~~~
=~ -\, \frac{1}{(2N+1)(2N-1)(2N+3)}\,\big[N^2 - 2N -\tfrac{9}{4} - 3 M_J^2\big]\,M_J \ ,   \nonumber \\[15pt]
&\sum_{M_S}\, C_{N\,(M_j-M_S),\,\thalf\,M_S}^{N+\thalf\,M_J} \,\,C_{N\,(M_j-M_S),\,\thalf\,M_S}^{N-\thalf\,M_J}
\,\,c_{N (M_J-M_S)} \,\,M_S \nonumber \\
&~~~~~~~~~~~~~~~~~~~~~~~~~~~~~~~~~
=~  -\,\frac{\sqrt{(N+\thalf)^2 - M_J^2}}{(2N-1)(2N+1)(2N+3)}\,
\big[N^2 + N - \tfrac{3}{4} - 3 M_J^2\big] \ .
\label{bpp6}
\end{align}

\vskip0,3cm
We also include here some other Clebsch-Gordan coefficients used in the evaluation of matrix elements
of the spherical harmonics in sections \ref{sect 3} and \ref{sect 4}:
\begin{align}
C_{N\,0,\,2\,0}^{N\,0} ~&=~ - \left(\frac{N(N+1)}{(2N-1)(2N+3)}\right)^{\thalf}  \ , \nonumber\\[8pt]
C_{N\,M_N,\,2\,0}^{N\,M_N} ~&=~ - \left(\frac{1}{N(N+1)(2N-1)(2N+3)}\right)^{\thalf}\,
\big(N(N+1)- 3 M_N^2\big)  \ , \nonumber \\[8pt]
C_{N\,M_N,\,\,2\,\pm 1}^{N\,(M_N\pm 1)} ~&=~ - \sqrt{\frac{3}{2}}\,\left(\frac{1}{N(N+1)(2N-1)(2N+3)}\right)^{\thalf}\,
\nonumber \\[8pt]
&~~~~~~~~~~~~~~~~~~~\times ~~\big((N+1\pm M_N)(N \mp M_N)\big)^{\thalf} \,(1 \pm2M_N) \ ,
\label{bpp7}
\end{align}
from which we find,
\begin{equation}
c_{N\,M_N} ~=~ C_{N\,M_N,\,2\,0}^{N\,M_N} \,\, C_{N\,0,\,2\,0}^{N\,0} ~=~ \frac{N(N+1) - 3 M_N^2}{(2N-1)(2N+3)} \ ,
\label{bpp8}
\end{equation}
which enters extensively in the calculations in Paper I, and here in sections \ref{sect 3} and \ref{sect 4}.

\vskip1.5cm

\section{Annual variations and spin-dependent couplings}\label{Appendix C}

To compare the constraints on the Lorentz violating SME couplings from different experiments,
it is necessary to refer them to a standard reference frame, generally taken to be a solar-centred,
or \textsf{SUN}, frame \cite{Kostelecky:2008ts,Kostelecky:2015nma}. 
A useful intermediary is provided by a standard laboratory,
\textsf{LAB}, frame with fixed orientation relative to the Earth. In order to study annual variations, we
therefore need to rotate our results from the \textsf{EXP} frame first to the \textsf{LAB} and then to
the \textsf{SUN} frames. We then perform a Lorentz boost to describe the Earth's orbital motion around
the Sun.

In the special case of the isotropic spin-independent couplings $c_{200}^{\NR}$ and $a_{200}^{\NR}$,
since they are invariant under rotations our results hold directly in the \textsf{SUN} frame, with
\begin{equation}
\sqrt{\frac{1}{4\pi}}\,\, c_{200}^{\NR} ~=~ \frac{1}{3 m} \big( c_{KK} \,+\, \tfrac{3}{2} c_{TT} \big) 
~=~ \frac{5}{6 m} c_{TT} \ ,
~~~~~~~~
\sqrt{\frac{1}{4\pi}}\,\, a_{200}^{\NR} ~=~  \big( a_{TKK} \,+\, a_{TTT}\big) \ ,
\label{cpp1}
\end{equation}
where we have again used the vanishing of the spacetime trace of the
$c_{\m\n}$ couplings \cite{Colladay:1998fq}.

The Lorentz boost depends on the Earth's orbital velocity ${\bf v}_\oplus$, given in \textsf{SUN} frame 
coordinates by,
\begin{equation}
{\bf v}_\oplus ~=~ v_\oplus\,\left(\sin\Omega_\oplus T\, {\bf e}_X ~-~ 
\cos\Omega_\oplus T \left(\cos\eta\,{\bf e}_Y\,+\, \sin\eta\, {\bf e}_Z\right)\,\right) \ ,
\label{cpp2}
\end{equation}
where $\Omega_\oplus$ is the orbital frequency and the tilt angle $\eta = 23.4^{\circ}$ 
is the angle between the Earth's equator and orbital plane.

To first order in $v_\oplus$, which is of $O(10^{-4})$, the boosts are simply,
\begin{align}
\d_\oplus\, c_{TT} ~&=~ 2\, v_\oplus^J\, c_{JT} \ ,  ~~~~~~~~~~~~~~~~~
\nonumber \\[8pt]
\d_\oplus\, a_{TTT} ~&=~ 3\, v_\oplus^J\, a_{JTT} \ , ~~~~~~~~~~~~~\,
\d_\oplus\, a_{TKK} ~=~ v_\oplus^J\,\big(a_{JKK} \,+\, 2 a_{JTT}\big) \ ,
\label{cpp3}
\end{align}
from which we find the annual variation,
\begin{align}
\tfrac{1}{\sqrt{4\pi}}\, \d_{\oplus} \big(c_{200}^{\rm NR} - a_{200}^{\rm NR} \big) 
\,&=\, v_{\oplus} \sin\Omega_{\oplus}T \Big(\tfrac{5}{3m} c_{XT} - 5 a_{XTT} - a_{XKK}\Big)
\nonumber \\[3pt]
&~~~- v_{\oplus}\cos\Omega_{\oplus}T 
\Big( \cos\eta \, \big( \tfrac{5}{3m} c_{YT} - 5 a_{YTT} - a_{YKK}\big)  \nonumber \\
&~~~~~~~~~~~~~~~~~~~~~~~
+\sin\eta \, \big(\tfrac{5}{3m} c_{ZT}  - 5 a_{ZTT} - a_{ZKK} \big)\Big) \ ,
\label{cpp4}
\end{align}
which implies the corresponding variation in the rovibrational transition frequencies.
Note that these variations are governed by the components $c_{JT}$, $a_{JTT}$ and $a_{JKK}$
of the SME Lagrangian couplings, different from those in the transition frequencies
themselves.

To generalise this to the spin-dependent couplings in the combinations
$\tilde{b}_3$, $\tilde{g}_{D3}$ and $\tilde{d}_3$, we need to rotate from the \textsf{EXP}
to the \textsf{SUN} frame. A significant simplification is to consider only 
sidereal averages,\footnote{In fact it is precisely these sidereal (daily) variations which
\cite{Vargas:2025efi} proposes to identify in order to constrain the SME couplings.
If we retain the sidereal variations in (\ref{cpp5}), we can show
\begin{align*}
\tilde{b}_3 ~&=~\big[\cos\theta \,\sin\b\,\cos\chi \,+\, \cos\b\,\sin\chi\big]
\,\big(\,\tilde{b}_X\,\cos\omega_{\oplus}T_{\oplus} \,+\, \tilde{b}_Y\, 
\sin\omega_{\oplus}T_{\oplus}\big)
\nonumber \\[5pt]
&~~+~ \big[\sin\theta \,\sin\b\big]\, 
\big(-\tilde{b}_X\,\sin\omega_{\oplus}T_{\oplus} \,+\, \tilde{b}_Y\, 
\cos\omega_{\oplus}T_{\oplus}\big)
\nonumber \\[5pt]
&~~+~\big[-\cos\theta \,\sin\b\,\sin\chi \,+\, \cos\b\,\cos\chi\big]\, \tilde{b}_Z \ ,
\end{align*}
where $\omega_{\oplus}$ is the sidereal angular frequency and $T_{\oplus}$ is a local 
sidereal time.  Identical results hold for $\tilde{g}_{D3}$ and $\tilde{d}_3$. 
Isolating the coefficients of $\cos\omega_{\oplus}T_{\oplus}$ and $\sin\omega_{\oplus}T_{\oplus}$
therefore enables constraints to be set on all three components, 
$\tilde{b}_X$ and $\tilde{b}_Y$ as well as $\tilde{b}_Z$.}
in which case we can show that the required rotation has non-vanishing components
only for $J=3$, for example,
\begin{equation}
\tilde{b}_3 ~=~ {\cal R}_{3J}\, \tilde{b}_J ~~~=~~~ {\cal R}_{3Z}\, \tilde{b}_Z \ .
\label{cpp5}
\end{equation}
Here, ${\cal R}_{3Z} = - \cos\theta\,\sin\b\,\sin\chi \, + \cos\b\,\cos\chi$,
where $\theta,\b$ describe the orientation of the \textsf{EXP} frame relative to \textsf{LAB}
({\it i.e.}~the direction of the applied magnetic field) and $\chi$ is the colatitude of the
laboratory. For a horizontal magnetic field, $\b = \pi/2$.
The same rotation also holds for $\tilde{g}_{D3}$ and $\tilde{d}_3$. 

The annual variations are then determined by the Lorentz boosts of 
the individual couplings in $\tilde{b}_Z$,  $\tilde{g}_{D3}$ and $\tilde{d}_3$. 
For example, for $b_Z$ itself  we have,
\begin{equation}
\d_\oplus\, b_Z ~=~ v_\oplus\, \cos\Omega_\oplus T \, \sin\eta\,\, b_T  \ .
\label{cpp6}
\end{equation}
The variations of the higher tensor couplings are defined in the same way,
with
\begin{align}
\d_\oplus\,b_Z ~&=~ v_\oplus^Z\, b_T  \ ,  \nonumber \\[8pt]
\d_\oplus\, d_{TZ} ~&=~ v_\oplus ^J\, d_{JZ} \,+\, v_\oplus^Z \, d_{TT} \ , \nonumber \\[8pt]
\d_\oplus\, H_{XY} ~&=~ v_\oplus^X\, H_{TY} \,+\, v_\oplus^Y \, H_{XT} \ , \nonumber \\[8pt]
\d_\oplus\, g_{XYT} ~&=~ v_\oplus^X\, g_{TYT} \,+\, v_\oplus^Y\, g_{XTT} \,+\, 
v_\oplus^J\, g_{XYJ} \ ,
\label{cpp7}
\end{align}
and so on for the remaining components, including the non-isotropic
spin-independent couplings,
$\sqrt{\tfrac{5}{4\pi}}\,c_{220}^{\rm NR}  = -\tfrac{1}{3m}\, \tr_Y c_{JK} \,\,$ 
and
$\sqrt{\tfrac{5}{4\pi}}\,a_{220}^{\rm NR} \,=\, - \tr_Ya_{TJK}$.

Evidently, combining all these terms to determine the annual variation of a particular
rovibrational transition frequency from the energy levels quoted in section \ref{sect 6}
gives a very complicated mix of couplings in the coefficients of $\cos\Omega_\oplus T$
and $\sin \Omega_\oplus T$. Nevertheless, the search for annual variations represents
one of the more promising routes to uncovering a signal for Lorentz and \textsf{CPT}
violation in ${\rm H}_2^{\,+}$ and $\overline{\rm H}_2^{\,-}$ molecular ion spectroscopy.

}

\end{document}